\documentclass[twocolumn,trackchanges]{aastex63}
\usepackage{graphicx}
\usepackage{float}
\usepackage{amsfonts,amsmath,amssymb} 
\usepackage{natbib}
\usepackage{url}
\usepackage{xcolor}
\usepackage{hyperref}
\usepackage[numbered]{bookmark} 
\usepackage{calculator}
\usepackage{placeins} 
\let\splitbox\relax
\usepackage{adjustbox} 
\usepackage{enumitem} 
\usepackage{xifthen}
\usepackage{chngcntr} 
\usepackage[normalem]{ulem} 

\usepackage{tabularx}

\usepackage{multirow}

\usepackage{nccmath}

\hypersetup{
    colorlinks,
    linkcolor={magenta!80!white},
    citecolor={blue!80!black},
    urlcolor={magenta!80!black},
}

\usepackage{listings} 
\usepackage{color}
\definecolor{lightgray}{gray}{0.9}

\usepackage{lmodern}
\usepackage{slantsc}

\usepackage{twoopt} 

\DeclareRobustCommand{\disambiguate}[3]{#2~#3} 

\newcommand{\TODO}[2][]{%
\ifthenelse{\isempty{#1}}%
{\textbf{\textcolor{red}{TODO: #2}}}%
{\textbf{\textcolor{red}{TODO (#1) #2}}}%
}

\def\gtsima{$\; \buildrel > \over \sim \;$}
\def\ltsima{$\; \buildrel < \over \sim \;$}
\def\prosima{$\; \buildrel \propto \over \sim \;$}
\def\gsim{\lower.5ex\hbox{\gtsima}}
\def\lsim{\lower.5ex\hbox{\ltsima}}
\def\simgt{\lower.5ex\hbox{\gtsima}}
\def\simlt{\lower.5ex\hbox{\ltsima}}
\def\simpr{\lower.5ex\hbox{\prosima}}

\def\h1{$h^{-1}$}
\def\beq{\begin{equation}}
\def\eeq{\end{equation}}
\def\8mu{8\,$\mu{\rm m}$}
\def\16mu{16\,$\mu{\rm m}$}
\def\24mu{24\,$\mu{\rm m}$}
\def\70mu{70\,$\mu{\rm m}$}

\def\SNR{\mathrm{S/N}}
\def\alphaCO{\alpha_{\mathrm{CO}}}

\def\deltaGDR{\delta_{\mathrm{GDR}}}

\def\deltaGas{\mu_{\mathrm{molgas}}}
\def\deltaMol{\mu_{\mathrm{mol\,frac}}}

\def\deltaMS{\delta{\mathrm{MS}}}
\def\DeltaMS{\Delta{\mathrm{MS}}}

\def\nH2{n_{\mathrm{H}_2}}
\def\H2{\mathrm{H}_2}
\def\HI{\mathrm{H}\textnormal{\textup{\uppercase\expandafter{\romannumeral 1}}}}

\def\metalZOH{12+\log_{10}(\mathrm{O/H})}
\def\Msun{\mathrm{M}_{\odot}}
\def\Msyr{\mathrm{M}_{\odot} \, \mathrm{yr}^{-1}}
\def\Lsun{\mathrm{L}_{\odot}}
\def\Kkmspc2{\mathrm{K} \, \mathrm{km} \, \mathrm{s}^{-1} \, \mathrm{pc}^{2}}
\def\SFR{\mathrm{SFR}}
\def\Mstar{M_{\star}}
\def\logMstar{\log_{10} M_{\star}}
\def\Mdust{M_{\mathrm{dust}}}
\def\Tdust{T_{\mathrm{dust}}}

\def\Mmolgas{M_{\mathrm{mol\,gas}}}
\def\Mtotalgas{M_{\mathrm{total\,gas}}}
\def\Matomicgas{M_{\mathrm{atomic\,gas}}}

\def\R52{R_{52}}

\def\fmol{f_{\mathrm{mol\,frac}}}

\def\tauDepl{\tau_{\mathrm{depl}}}
\def\fgas{f_{\mathrm{gas}}}
\def\fmolgas{f_{\mathrm{mol\,gas}}}

\def\d_{{\mathrm{d}}}

\def\L850{L_{\nu_{850\mu\mathrm{m,rest}}}}
\def\LRJ{L_{\nu_{\mathrm{RJ,rest}}}}
\def\vLv850{{\nu}L_{\nu_{850\mu\mathrm{m}}}}
\def\alphaRF850{\alpha_{850}}
\def\alphamolgasRF850{\alpha_{850,\mathrm{mol}}}
\def\alphaRJtot{\alpha_{\textsc{rj},\,\mathrm{tot}}}
\def\alphaRJmol{\alpha_{\textsc{rj},\,\mathrm{mol}}}
\def\alphaRJ_{\alpha_{\mathrm{RJ}}}
\def\alphaRJS17{\alpha_{850,\mathrm{S17}}}
\def\alphaRJH17{\alpha_{850,\mathrm{H17}}}
\def\alphaRJG15{\alpha_{160{\textendash}500,\mathrm{G15}}}

\def\LprimeCO{L^{\prime}_{\mathrm{CO}}}
\def\Lprime12CO10{L^{\prime}_{\mathrm{CO(1-0)}}}

\def\LnuRJ{L_{\nu_{\textsc{rj}}}}

\def\NumberOfComplementarySamples{20} 

\newcommand{\rom}[1]{{{\uppercase\expandafter{\romannumeral #1}}}}
\newcommand{\romup}[1]{{\textup{\uppercase\expandafter{\romannumeral #1}}}}

\newcommand{\incode}[1]{{\raggedright\lstinline|#1|}}
\newcommand{\incodep}[1]{{\raggedright\lstinline|"#1"|}}

\defcitealias{Scoville2017}{S17}
\defcitealias{Tacconi2018}{T18}
\defcitealias{Hughes2017}{H17}
\defcitealias{Groves2015}{G15}
\newcommand{\citealtalias}[1]{\citetalias{#1}}

\renewcommand{\figurename}{Fig.}

\newcommand{\inStringStartsWith}[4]{%
\ifnum\pdfmatch{#1}{#2}=1%
#3%
\else%
#4%
\fi%
}

\newcommand{\stkout}[1]{\ifmmode\text{\sout{\ensuremath{#1}}}\else\sout{#1}\fi}

\def\CurrentRevisionLabel{RefereeReport1} 
\definecolor{ColorForRevisedText}{RGB}{0,0,0}
\newcommandtwoopt{\REVISED}[3][][]{%
	\ifthenelse{\isempty{#3}}%
	{%
		\ifthenelse{\isempty{#2}}%
		{
			\inStringStartsWith{\CurrentRevisionLabel}{#1}{%
		        \textbf{\textcolor{ColorForRevisedText}{#1}}%
			}{%
			}%
		}%
		{
			\inStringStartsWith{\CurrentRevisionLabel}{#1}{%
			    \textcolor{gray}{\stkout{#2}}%
			}{%
			}%
		}%
	}%
	{
		\ifthenelse{\isempty{#2}}%
		{
			\inStringStartsWith{\CurrentRevisionLabel}{#1}{%
			    \textbf{\textcolor{ColorForRevisedText}{#3}}%
			}{%
			    {#3}%
			}%
		}%
		{
			\inStringStartsWith{\CurrentRevisionLabel}{#1}{%
			    \textcolor{gray}{\sout{#2}}%
			    \textbf{\textcolor{ColorForRevisedText}{#3}}%
			}{%
			    {#3}%
			}%
		}%
	}%
}
\revised{\today}
\usepackage{expl3}
\usepackage{varwidth}
\usepackage{tikz}
\usepackage{amsmath,amssymb}
\usepackage{bm}
\usetikzlibrary{arrows.meta}
\usetikzlibrary{shapes,arrows,chains}
\usetikzlibrary{calc}
\usetikzlibrary{positioning}
\makeatletter
\newcommand\XHUGE{\@setfontsize\Huge{29}{29}}
\newcommand\XXHUGE{\@setfontsize\Huge{38}{38}}
\makeatother

\definecolor{link(red)}{rgb}{1.0, 0.55, 0.12}
\definecolor{blue(ncs)}{rgb}{0.0, 0.53, 0.74}
\definecolor{blue(new)}{rgb}{0.0, 0.63, 0.84}

\dimendef\prevdepth=0 

\tikzstyle{StyleOfCatalog} = [rectangle, draw, color=black, fill=white, text=black, text centered, inner sep=0.5cm, line width=0.75mm, minimum width=8.0cm, execute at begin node={\begin{varwidth}{9.5cm}\centering\bf\XHUGE}, execute at end node={\end{varwidth}}, rounded corners]

\tikzstyle{StyleOfCatalog(OneLine)} = [rectangle, draw, color=black, fill=white, text=black, text centered, inner sep=0.45cm, line width=0.5mm, execute at begin node={\centering\bf\Huge}, execute at end node={}, rounded corners]

\tikzstyle{StyleOfEmptyNode} = [fill=white, inner sep=0cm, minimum width=0cm]

\tikzstyle{StyleOfProcess} = [diamond, draw, fill=cyan!15, text centered, rounded corners, inner sep=0.09cm, line width=0.5mm, execute at begin node={\begin{varwidth}{6.4cm}\centering\bf\Huge}, execute at end node={\end{varwidth}}]

\tikzstyle{StyleOfProcess(long)} = [diamond, draw, fill=cyan!15, text centered, rounded corners, inner sep=0.03cm, line width=0.5mm, minimum width=10.0cm, minimum height=10.0cm, execute at begin node={\begin{varwidth}{9.2cm}\centering\bf\Huge}, execute at end node={\end{varwidth}}]

\tikzstyle{StyleOfProcess(narrow)} = [diamond, draw, fill=cyan!15, text centered, rounded corners, inner sep=0.01cm, line width=0.5mm, execute at begin node={\begin{varwidth}{4.4cm}\centering\bf\Huge}, execute at end node={\end{varwidth}}]

\tikzstyle{StyleOfFlag} = [ellipse, draw, color=blue(ncs), fill=white, text=black, text centered, inner sep=0.09cm, line width=0.5mm, minimum width=5.4cm, minimum height=2.8cm, execute at begin node={\begin{varwidth}{5.8cm}\centering\bf\Huge}, execute at end node={\end{varwidth}}, rounded corners]

\tikzstyle{StyleOfFlag(long)} = [ellipse, draw, color=blue(ncs), fill=white, text=black, text centered, inner sep=0.01cm, line width=0.5mm, minimum width=8.2cm, minimum height=2.8cm, execute at begin node={\begin{varwidth}{8.2cm}\centering\bf\Huge}, execute at end node={\end{varwidth}}, rounded corners]

\tikzstyle{StyleOfFinalProduct} = [rectangle, draw, color=blue(ncs), fill=white, text=black, text centered, inner sep=0.50cm, minimum width=6.5cm, minimum height=4.5cm, line width=1.5mm, execute at begin node={\begin{varwidth}{8.5cm}\centering\bf\XXHUGE}, execute at end node={\end{varwidth}}, rounded corners]

\tikzstyle{StyleOfLine} = [draw, -{Latex[scale=3,length=5,width=3,angle'=25,open]}, line width=0.8pt, color=blue(ncs)]

\tikzstyle{StyleOfLine(thick)} = [draw, -{Latex[scale=3,length=5,width=3,angle'=25,open]}, line width=1.4pt, color=blue(ncs)]

\tikzstyle{StyleOfLine(thicker)} = [draw, -{Latex[scale=3,length=9pt,width=15pt,angle'=25,open]}, line width=2.8pt, color=blue(ncs)]

\tikzstyle{StyleOfLine(dotted)} = [draw, dotted, -latex', line width=2.0pt, color=blue(ncs)]

\tikzstyle{StyleOfLine(dashed)} = [draw, dashed, -latex', line width=2.5pt, color=blue(new)]
\shorttitle{A3COSMOS gas evolution}
\shortauthors{D. Liu et al.}

\begin{document}

\title{Automated Mining of the ALMA Archive in the COSMOS Field (A$^3$COSMOS): II. Cold Molecular Gas Evolution out to Redshift 6}

\email{dzliu@mpia.de}

\author[0000-0001-9773-7479]{Daizhong Liu}
\affiliation{Max-Planck-Institut f\"{u}r Astronomie, K\"{o}nigstuhl 17, D-69117 Heidelberg, Germany}

\author[0000-0002-3933-7677]{E. Schinnerer}
\affiliation{Max-Planck-Institut f\"{u}r Astronomie, K\"{o}nigstuhl 17, D-69117 Heidelberg, Germany}

\author[0000-0002-9768-0246]{B. Groves}
\affiliation{Research School of Astronomy and Astrophysics, Australian National University, Canberra ACT, 2611, Australia}
\affiliation{International Centre for Radio Astronomy Research, University of Western Australia, Crawley, Perth, Western Australia, 6009, Australia}

\author[0000-0002-6777-6490]{B. Magnelli}
\affiliation{Argelander-Institut f\"{u}r Astronomie, Universit\"{a}t Bonn, Auf dem H\"ugel 71, D-53121 Bonn, Germany}

\author[0000-0002-5681-3575]{P. Lang}
\affiliation{Max-Planck-Institut f\"{u}r Astronomie, K\"{o}nigstuhl 17, D-69117 Heidelberg, Germany}

\author[0000-0002-4826-8642]{S. Leslie}
\affiliation{Max-Planck-Institut f\"{u}r Astronomie, K\"{o}nigstuhl 17, D-69117 Heidelberg, Germany}

\author[0000-0002-2640-5917]{E. Jim\'{e}nez-Andrade}
\affiliation{Argelander-Institut f\"{u}r Astronomie, Universit\"{a}t Bonn, Auf dem H\"ugel 71, D-53121 Bonn, Germany}
\affiliation{National Radio Astronomy Observatory, 520 Edgemont Road, Charlottesville, VA 22903, USA}

\author[0000-0001-9585-1462]{D.~A. Riechers}
\affiliation{Department of Astronomy, Cornell University, Space Sciences Building, Ithaca, NY 14853, USA}
\affiliation{Max-Planck-Institut f\"{u}r Astronomie, K\"{o}nigstuhl 17, D-69117 Heidelberg, Germany}

\author[0000-0003-1151-4659]{G. Popping}
\affiliation{Max-Planck-Institut f\"{u}r Astronomie, K\"{o}nigstuhl 17, D-69117 Heidelberg, Germany}
\affiliation{European Southern Observatory, Karl-Schwarzschild-Strasse 2, D-85748, Garching, Germany}

\author[0000-0002-4872-2294]{Georgios E. Magdis}
\affiliation{Cosmic Dawn Center at the Niels Bohr Institute, University of Copenhagen and DTU-Space, Technical University of Denmark}
\affiliation{DTU Space, National Space Institute, Technical University of Denmark, Elektrovej 327, DK-2800 Kgs. Lyngby, Denmark}
\affiliation{Niels Bohr Institute, University of Copenhagen, DK-2100 Copenhagen $\O$}
\affiliation{Institute for Astronomy, Astrophysics, Space Applications and Remote Sensing, National Observatory of Athens, 15236, Athens, Greece}

\author[0000-0002-3331-9590]{E. Daddi}
\affiliation{CEA Saclay, Laboratoire AIM-CNRS-Universit\'{e} Paris Diderot, Irfu/SAp, Orme des Merisiers, 91191, Gif-sur-Yvette, France}

\author[0000-0003-1033-9684]{M. Sargent}
\affiliation{Astronomy Centre, Department of Physics and Astronomy, University of Sussex, Brighton BN1 9QH, UK}

\author[0000-0003-0007-2197]{Yu Gao}
\affiliation{Purple Mountain Observatory \& Key Laboratory for Radio Astronomy, Chinese Academy of Sciences, 10 Yuanhua Road, 
Nanjing 210033, PR China}

\author[0000-0001-7440-8832]{Y. Fudamoto}
\affiliation{Department of Astronomy, Universit\'e de Gen\`eve, Chemin des Maillettes 51, 1290 Versoix, Switzerland}

\author[0000-0001-5851-6649]{P. A. Oesch}
\affiliation{Department of Astronomy, Universit\'e de Gen\`eve, Chemin des Maillettes 51, 1290 Versoix, Switzerland}

\author{F. Bertoldi}
\affiliation{Argelander-Institut f\"{u}r Astronomie, Universit\"{a}t Bonn, Auf dem H\"ugel 71, D-53121 Bonn, Germany}

\begin{abstract}
We present new measurements of the cosmic cold molecular gas evolution out to redshift 6 based on systematic mining of the ALMA public archive in the COSMOS deep field (A$^3$COSMOS). 
Our A$^3$COSMOS dataset contains $\sim700$ galaxies ($0.3 \lesssim z \lesssim 6$) with high-confidence ALMA detections in the (sub-)millimeter continuum and multi-wavelength spectral energy distributions (SEDs). Multiple gas mass calibration methods are compared and biases in band conversions (from observed ALMA wavelength to rest-frame Rayleigh-Jeans(RJ)-tail continuum) have been tested. 
Combining our A$^3$COSMOS sample with $\sim1,000$ CO-observed galaxies at $0 \lesssim z \lesssim 4$ (75\% at $z < 0.1$), we parameterize galaxies' molecular gas depletion time ($\tauDepl$) and molecular gas to stellar mass ratio ($\deltaGas$) each as a function of the stellar mass ($\Mstar$), offset from the star-forming main sequence ($\DeltaMS$) and cosmic age (or redshift). 
Our proposed functional form provides a statistically better fit to current data (than functional forms in the literature), and implies a ``downsizing'' effect (i.e., more-massive galaxies evolve earlier than less-massive ones) and ``mass-quenching'' (gas consumption slows down with cosmic time for massive galaxies but speeds up for low-mass ones). 
Adopting galaxy stellar mass functions and applying our $\deltaGas$ function for gas mass calculation, we for the first time infer the cosmic cold molecular gas density evolution out to redshift 6 and find agreement with CO blind surveys as well as semi-analytic modeling. 
These together provide a coherent picture of cold molecular gas, SFR and stellar mass evolution in galaxies across cosmic time. 
\end{abstract}


\section{Introduction}
\label{Section_Introduction}

The interstellar medium (ISM), especially the cold molecular gas, is the fuel of star formation activity in galaxies. In recent years, our knowledge of the cosmic evolution of star formation and stellar mass growth has been obtained out to redshift $\sim5$ (e.g., see latest reviews by \citealt{Lutz2014Review} and \citealt{Madau2014Review}; see also \citealt{Davidzon2017}; \citealt{Liudz2017}; to name a few). However, the cosmic evolution of the cold molecular gas is much less well constrained and the validity of different tracers are debated (e.g., \citealt{Magdis2012SED}; \citealt{Santini2014}; \citealt{Genzel2015}; \citealt{Tacconi2018}; \citealt{Riechers2019}; \citealt{Decarli2019}). 

There are several widely used tracers of the molecular gas content in galaxies, including the commonly used carbon monoxide (CO) rotational transition lines, dust masses from dust spectral energy distribution (SED), and the cold dust continua at the Rayleigh-Jeans (RJ) tail of dust SED. We introduce each case below.

Observationally, CO lines at the rest-frame millimeter (mm) wavelengths have been established as the most-commonly used tracers of total molecular gas content in galaxies near and far since 1970s (e.g., see latest reviews by \citealt{Carilli2013Review}; \citealt{Combes2018Review}). At high-redshift, this method relies on galaxy samples with accurate spectroscopic redshifts and usually has uncertainties from the CO-to-H$_2$ conversion factor ($\alphaCO \equiv \Mmolgas/\LprimeCO$) and CO excitation. With this method, \cite{Genzel2010} and \cite{Tacconi2013,Tacconi2018} conducted the largest survey for individual galaxies (named PHIBSS) by observing hundreds of star-forming galaxies at $z\sim1-3$ to study the molecular gas scaling relation and evolution. Meanwhile, \cite{Walter2016} and \cite{Decarli2016,Decarli2019} have been conducting the largest blank-field survey (named ASPECS) by scanning a range of mm spectra within a fixed sky area to determine the CO luminosity function and thereby study the molecular gas mass density evolution. 

Alternatively, in the past few years, emission from dust grains located in the star-forming regions of galaxies has also been widely used as a proxy of the ISM. These dust grains absorb rest-frame ultra-violet (UV) photons from massive stars and re-emit thermal radiation in the infrared(IR)-to-mm wavelengths. By fitting a galaxy's full dust SED with models, e.g., modified blackbody models or multi-component physical models (e.g., \citealt{Draine2007SED}), the dust mass and dust temperature (or mean radiation field) can be obtained (e.g., \citealt{Santini2010,Santini2014}; \citealt{Magdis2011SED,Magdis2012SED}; \citealt{Magnelli2012,Magnelli2014}; \citealt{Saintonge2013}; \citealt{Sandstrom2013}; \citealt{Tan2014}; \citealt{Bethermin2015}; \citealt{Berta2016}; \citealt{Hunt2019}). The dust mass can then be converted to gas mass via the application of empirical gas-to-dust ratios ($\deltaGDR$).

However, a galaxy's full dust SED is a composite of a variety of dust components with different temperatures. Warmer dust exposed to strong radiation fields (e.g., photo-dominated regions; \citealt{Dale2001}; \citealt{Draine2007SED}) globally outshines the colder dust at shorter wavelengths of the SED, but the former is much less abundant (e.g., $<10\%$ in mass) and does not represent the bulk of dust in a galaxy. Thus obtaining reliable dust mass usually requires longer wavelength coverage that includes the RJ tail (e.g., $\lambda_{\mathrm{rest}} \gtrsim 250\,\mu$m). Also, different dust SED models can result in strong and not-easily-predictable systematic effects (\citealt{Berta2016}). Therefore, the RJ-tail method has been proposed by \cite{Scoville2013b,Scoville2014}, which directly uses the RJ-tail dust continuum to trace gas (yet the underlying physics of using dust mass to trace gas mass is the same as the in the dust SED method above).

The RJ-tail method has recently been proven to be as reliable as the CO method (e.g., \citealt{Scoville2014,Scoville2016}; \citealt{Groves2015}; \citealt{Hughes2017}; \citealt{Bertemes2018}; \citealt{Saintonge2018}; \citealt{Kaasinen2019}; and theoretical works, e.g., \citealt{Privon2018}) and is much more efficient in surveying large galaxy samples at high redshift. 
This method relies on the assumption that the dust grains providing most of the dust mass in galaxies are cold and mixed within the ISM. Their temperatures are likely always as cold as $\Tdust \approx 25 \, \mathrm{K}$ (see \citealt{Scoville2014,Scoville2016}), and hence they can trace the total gas content via a relatively stable gas-to-dust ratio ($\deltaGDR$; e.g., \citealt{Leroy2011GDR}; \citealt{RemyRuyer2014}).
Yet we bear in mind that metallicity, true dust temperature and mass distributions are all unsolved issue.

These studies have led to a rough picture of dust and gas evolution from redshift 3 to present, where: (a) the fraction of molecular gas mass to the total of molecular gas and stellar masses, 
{\setlength{\abovedisplayskip}{3pt}\setlength{\belowdisplayskip}{3pt}\noindent%
\begin{fleqn}%
\begin{equation}%
\begin{split}%
\quad & \fmolgas \equiv \Mmolgas/(\Mmolgas+\Mstar), \\
\quad & \textnormal{or} \ \deltaGas \equiv \Mmolgas/\Mstar, \\
\end{split}%
\label{Equation_mugas}
\end{equation}%
\end{fleqn}%
}%
decreases with cosmic age from $z\sim3$ to $z\sim0$, and depends on SFR and stellar mass; (b) the molecular gas depletion time, 
{\setlength{\abovedisplayskip}{3pt}\setlength{\belowdisplayskip}{3pt}\noindent%
\begin{fleqn}%
\begin{equation}%
\begin{split}%
\quad & \tauDepl \equiv \Mmolgas/\SFR, \\
\end{split}%
\end{equation}%
\end{fleqn}%
}%
increases from $z\sim3$ to present, and is significantly different between typical star-forming galaxies (which follow a tight $\Mstar-\SFR$ main sequence (hereafter MS) at each redshift; e.g., \citealt{Brinchmann2004}; \citealt{Noeske2007}; \citealt{Elbaz2007}; \citealt{Daddi2007}) and starbursts (i.e., located significantly above the MS; e.g., \citealt{Rodighiero2011,Rodighiero2014}). 
\cite{Genzel2015} first compiled a large sample of local and high-redshift ($0<z<3$) galaxies with both CO (500 galaxies) and dust SED (512 galaxies) methods. They studied the gas scaling relations by characterizing $\fmolgas$ and $\tauDepl$ as functions of $\Mstar$, SFR and redshift. More precisely, they found that gas fraction and depletion time are more strongly correlated with the SFR offset to the MS, 
{\setlength{\abovedisplayskip}{3pt}\setlength{\belowdisplayskip}{3pt}%
\begin{fleqn}%
\begin{equation}%
\begin{split}%
\quad & \deltaMS \equiv \SFR/\SFR_\mathrm{MS}, \\
\quad & \textnormal{or} \ \DeltaMS \equiv \log_{10}(\SFR/\SFR_\mathrm{MS}), \\
\end{split}%
\end{equation}%
\end{fleqn}%
}%
rather than the absolute SFR.

Utilizing the RJ-tail dust continuum method (at rest-frame 850\,$\mu$m), \citet[][hereafter \citetalias{Scoville2017}]{Scoville2017} studied the gas ($\fmolgas$ and $\tauDepl$) scaling relations with a large sample of 708 high-redshift \textit{Herschel} far-IR-selected galaxies ($0.3 < z < 4.5$), including a large number of public data in the Atacama Large Millimeter/submillimeter Array (ALMA) archive (at a 2.5--3$\,\sigma$ detection threshold), and characterized the $\fmolgas$ and $\tauDepl$ functional forms:
{\setlength{\abovedisplayskip}{2pt}\setlength{\belowdisplayskip}{3pt}\noindent%
\begin{fleqn}%
\begin{equation}%
\begin{split}%
&\deltaGas = 0.71 
\times (\deltaMS)^{+0.32} 
\times M_{\star,10}^{-0.70}
\times (1+z)^{+1.84} , \\[3pt]
&\tauDepl = 3.23 \; \mathrm{Gyr} 
\times (\deltaMS)^{-0.70} 
\times M_{\star,10}^{-0.01}
\times (1+z)^{-1.04} , 
\end{split}%
\label{Equation_Scoville2017}
\end{equation}%
\end{fleqn}%
}%
where $M_{\star,10}$ is $\Mstar/(10^{10}\;\Msun)$. 
With the same method but at rest-frame 250--500\,$\mu$m, \citet{Schinnerer2016} studied a smaller sample of optically-selected galaxies at $z=2.8-3.6$. However, discrepancies exist due to the slightly different methods and samples.

\citet[][hereafter \citetalias{Tacconi2018}]{Tacconi2018} expanded the work of \citet{Genzel2015} by obtaining nearly a hundred new CO detections in the PHIBSS2 survey and compiling more samples of local to high-redshift galaxies in the literature. They used all three methods for obtaining molecular gas measurements for 1,444 galaxies at $0<z<4$, and fitted them all together to derive the $\fmolgas$ and $\tauDepl$ functions:
{\setlength{\abovedisplayskip}{3pt}\setlength{\belowdisplayskip}{6pt}\noindent%
\begin{fleqn}%
\begin{align}%
\quad 
&\deltaGas = 2.32 
\times (\deltaMS)^{+0.53} 
\times M_{\star,10}^{-0.35} \notag\\
\quad 
&\hspace{3em}
\times 10^{-3.62 \times (\log_{10}(1+z)-0.66)^{2}}, \notag\\[3pt]
\quad 
&\tauDepl = 1.06 \ \mathrm{Gyr} 
\times (\deltaMS)^{-0.44} 
\times M_{\star,10}^{+0.09} \notag\\
\quad 
&\hspace{3em}
\times (1+z)^{-0.62},
\label{Equation_Tacconi2018}%
\end{align}%
\end{fleqn}%
}%
where we adopted their $\beta=2$ best-fit with the \cite{Speagle2014} MS and expressed their stellar mass in $M_{\star,10}$ to match Eq.~\ref{Equation_Scoville2017}.

Comparing Eqs.~\ref{Equation_Scoville2017} and \ref{Equation_Tacconi2018} at redshift 3 and $\Mstar=5\times10^{10}\;\Msun$ reveals a factor of 2.3 difference in $\deltaGas$ and a factor of 1.5 in $\tauDepl$. 
Such noticeable differences exist for other parameter values as well, raising concerns on the validity of the $\deltaGas$ and $\tauDepl$ functions and the predictability of $\deltaGas$ and $\tauDepl$ from a galaxy's redshift, stellar mass and SFR properties. In addition, previous works have constraints only for $z \lesssim 3-4$.

To solve the discrepancies and understand systematic bias especially for the latest RJ-tail dust method, a large, robust, galaxy sample from local to high redshift is needed to carry out the comprehensive analysis. Therefore, in this work, we present an independent study on the characterization of the molecular gas fraction ($\deltaGas$) and depletion time ($\tauDepl$) functional forms utilizing a large ($\sim700$), robust galaxy sample at $0.3 < z \lesssim 6$ in the 2\,deg$^2$ COSMOS field \citep{Scoville2007} from the A$^3$COSMOS project\,\footnote{\url{https://sites.google.com/view/a3cosmos}}, together with $\sim1,000$ CO-detected galaxies at $0 \lesssim z \lesssim 4$ (75\% at $z < 0.1$) from recent large surveys in the literature. All A$^3$COSMOS galaxies have robust (sub-)mm continuum detections from public ALMA archival data (release date up to Aug. 1st, 2018) with an expected spurious fraction close to zero and flux bias being corrected statistically (\citealt{Liudz2019}; hereafter paper~\romup{1}).

With such a combined large sample, we provide new molecular gas fraction ($\deltaGas$) and depletion time ($\tauDepl$) functional forms that are valid from redshift 0 to 6. We adopt galaxies' stellar mass functions and/or realistic galaxy modeling to analytically derive the cosmic molecular gas mass density evolution for the first time with such a large dataset out to redshift 6. The result supports a coherent picture of the evolution of galaxies' stellar mass, star formation and cold molecular gas.

This paper is organized as follows. Galaxy samples are presented in Sect.~\ref{Section_Sample_and_Data}, with the A$^3$COSMOS high-redshift sample in Sect.~\ref{Section_A3COSMOS_Pipelines} and complementary local-to-high-redshift samples from the literature in Sect.~\ref{Section_Complementary_Galaxy_Samples}. 
Molecular gas mass calculation and comparison are presented in  Sect.~\ref{Section_Molecular_Gas_Mass_Calculation} (dust SED method in Sect.~\ref{Section_Deriving_Gas_Mass_1}, RJ-tail method in Sect.~\ref{Section_Deriving_Gas_Mass_234}, and comparison in Sect.~\ref{Section_comparing_gas_mass_calibration}). 
The complexity and apparent correlations between $\deltaGas$, $\tauDepl$ and galaxies' redshifts, stellar masses and SFRs are discussed in Sect.~\ref{Section_Galaxy_Properties}. 
The characterization of the functional forms for $\deltaGas$ and $\tauDepl$ are presented in Sect.~\ref{Section_Fitting_Multi_Variate_Function}, and their implications are discussed in Sect.~\ref{Section: Predictions from Galaxy Molecular Gas Evolution Functions}. Finally, the cosmic evolution of molecular gas mass density is analytically obtained in Sect.~\ref{Section_Implication}, followed by the summary in Sect.~\ref{Section_Summary}.

In the Appendices, we thoroughly compare several important correlations related to our analysis: CO-to-H$_2$ conversion factor versus metallicity in Appx.~\ref{Section_ZACO}; gas-to-dust ratio versus metallicity in Appx.~\ref{Section_ZGDR}; molecular to total gas fraction versus stellar mass and/or metallicity in Appx.~\ref{Section_ZFMOL}; and stellar mass-metallicity relation in Appx.~\ref{Section_MZR}. 
These comparisons give useful insights into how different correlations impact the results presented in this work, as well as supporting our fiducial model in this work.

We adopt a flat $\Lambda$CDM cosmology with $H_0=70\;\mathrm{km\,s^{-1}\,Mpc^{-1}}$, $\Omega_M=0.3$ and $\Lambda_0=0.7$~\footnote{Same as those adopted by \citealtalias{Tacconi2018}.}, and a \cite{Chabrier2003} initial mass function (IMF).

\vspace{1truecm}
\section{Sample and Data}
\label{Section_Sample_and_Data}

\vspace{0.25truecm}
\subsection{The A$^3$COSMOS Galaxy Sample}
\label{Section_A3COSMOS_Pipelines}

In paper~\romup{1} we presented the A$^3$COSMOS project, i.e., an Automated ALMA Archive mining in the COSMOS field. We developed pipelines for producing continuum images using nearly all publicly available ALMA archival data in COSMOS (regardless of observing bands but discarded very high resolution (beam size $< 0.1''$) data; see paper~\romup{1}). We performed two major (sub-)mm continuum photometry extractions: one prior-based and one blind extraction, to make sure the photometries are robust and outliers are identified (see below). Both photometries are verified by extensive Monte Carlo simulations and corrected for flux bias and uncertainty. 
Additional photometry task using apertures following \citetalias{Scoville2017} show good consistency for isolated sources ($<20\%$ difference in average; but significant differ for blended or merger-like sources for which aperture photometry is not suitable). 

In order to obtain a most robust galaxy catalog from the initial (sub-)mm continuum detections, we applied very strict criteria to select ALMA detections: a peak flux to rms noise ratio of 5.40 for blind extraction and 4.35 for prior photometry, which correspond to an expected spurious source fraction of $\sim10\%$ (according to our statistical analysis). These spurious sources are statistically unavoidable in the initial photometry catalogs, but we developed a series of assessments to identify the most reliable detections. We hence removed ALMA detections which: 
(1) have inconsistent fluxes between blind- and prior-based (sub-)mm photometry (identified by the \incode{Flag_inconsistent_flux} in the A$^3$COSMOS catalog, which are about ten sources likely being mergers or blended sources and exhibit a $\gtrsim0.5\,$dex difference between blind-/prior-photometry fluxes; see examples in Appendix~B of paper~\romup{1}); (2) have a peculiar counterpart association quality (\incode{Flag_outlier_CPA}; which are likely because of chance alignment between a prior source and a noise peak); and/or (3) show an excess in ALMA flux relative to the galaxy SED (\incode{Flag_outlier_SED}; which are likely because of inconsistent photometric redshift, blended sources or noise). These criteria exclude sources that are either boosted by noise in the ALMA image or multiple galaxies co-aligned, plus other less-clear situations. For more details we refer the reader to paper~\romup{1}. 

After removing the spurious sources, our robust galaxy catalog from A$^3$COSMOS (version \incode{20180801}) contains 669 galaxies (36\% have spectroscopic redshifts mainly from the COSMOS spec-$z$ catalog compiled by M. Salvato; see references in paper~\romup{1}). Due to the strict additional selection criteria, the spurious fraction is reduced to close to zero according to our statistics in paper~\romup{1}. Yet this implies that we miss a significant number of low ALMA $\SNR$ sources which have a $<50\%$ chance of being real, faint galaxies. For comparison, \citetalias{Scoville2017} explored all ALMA Band\,6 and 7 data in the ALMA public archive and selected sources with total flux of $\SNR>2$. \cite{Betti2019} analyzed ALMA continuum data for 101 galaxies and selected 68 as detections with an aperture-based total flux of $\SNR>2$ or peak flux of $\SNR>3$. The data used in \cite{Betti2019} have been public in the ALMA archive before Aug. 2018, and are therefore in our catalog. 90 of their galaxies appear in our prior-fitting catalog without applying a $\SNR$ selection, however, only 8 galaxies have a peak flux $\SNR>4.35$, which is our selection criterion based on statistics (corresponding to a spurious fraction $\sim10\%$). This quick comparison demonstrates that our catalog has very strict constraints and only considers the statistically most-robust ALMA detections. Lowering the selection criterion for A$^3$COSMOS from a (peak flux) $\SNR$ of 4.35 to 3.0 doubles the A$^3$COSMOS galaxy sample, however 40\% of the sample will be spurious based on our simulation statistics. Given this trade-off between increased sample size and decreased reliability, we resort to the original robust galaxy catalog containing only highly-reliable sources from A$^3$COSMOS.

Galaxy properties in the A$^3$COSMOS galaxy catalog, including stellar mass ($\Mstar$), IR luminosity ($L_{\mathrm{IR}}$) and dust mass ($\Mdust$), are obtained from \textsc{MAGPHYS} (\citealt{daCunha2008,daCunha2015}) SED fitting to their optical-to-radio SEDs (see paper~\romup{1}). We compute the dust-obscured SFR from IR luminosity following the \cite{Kennicutt1998SFL} calibration and \cite{Chabrier2003} IMF: $\mathrm{SFR} = L_{\mathrm{IR}} \, / \, 1\times10^{10} \ \Msun \, \mathrm{yr^{-1}}$.

In addition, to understand whether using \textsc{MAGPHYS} SED fitting is biased due to the built-in SED templates or the assumption of energy balance, we performed two more independent SED fittings for each galaxy to fit the stellar (up to IRAC ch2) and near-IR-to-radio data points separately, with the \textsc{FAST} (\citealt{Kriek2009}) and ``\textsc{super-deblended}'' (\citealt{Liudz2017,Jin2018}) SED fitting tools, respectively. We find that the \textsc{MAGPHYS}-fitted stellar masses are systematically larger by about 0.25~dex than the \textsc{FAST} fitted values (with a scatter of 0.30~dex), while the dust-obscured SFRs are fully consistent between the \textsc{MAGPHYS} and \textsc{super-deblended} SED fitting. The systematic discrepancy in stellar mass has also been found by \cite{Battisti2019} and reproduced in SED modeling with various non-parametric star formation histories (e.g., \citealt{Leja2019}). 
Since this is not yet fully understood, we still adopt the \textsc{MAGPHYS} SED fitting results. We tested that using \textsc{FAST}-fitted stellar masses will not change our main results, but only alter the coefficients in our equations (by $\lesssim20\%$).


\begin{table*}[htb]
\caption{%
	Galaxy Samples Used for This Study
	\label{Table_galaxy_samples}
}
\begin{tabularx}{1.0\textwidth}{l l l l l}
\hline
\hline

Sample Name  & $z$ & $\log_{10} (\Mstar/\Msun)$ & 
$N_{\mathrm{det.}}$\,$^{a}$ & 
References \\

\hline

A$^3$COSMOS (Paper \romup{1}) & 0.29--5.667\,$^{b}$ & $\sim$10.--12. & 658\,$^{b}$ & \citet[][dataset \incode{20180801}]{Liudz2019} \\

\hline

DGS & 0--0.045 & 6.5--10.6 & 32 & \cite{RemyRuyer2014,RemyRuyer2015} \\

HRS & 0.0034--0.006 & 8.7--11.7 & 99 & \citet[][and refs. therein]{Andreani2018} \\

KINGFISH & 0.0005--0.006 & 6.3--10.7 & 28 & \citet[][and refs. therein]{Groves2015} \\

Saintonge+2017 (xCOLDGASS) & 0.01--0.05 & 9.0--11.4 & 330 & \cite{Saintonge2017} \\

Cicone+2017 (ALLSMOG) & 0.01--0.03 & 8.5--11.5 & 48 & \cite{Cicone2017} \\

Lisenfeld+2017 & 0.01--0.07 & 9.0--11.3 & 41 & \cite{Lisenfeld2017} \\

Cortzen+2019 & 0.03--0.29 & 8.9--11.5 & 51 & \cite{Cortzen2019} \\

Villanueva+2017 (VALES) & 0.03--0.33 & 10.1--11.3 & 49 & \cite{Villanueva2017} \\

Bertemes+2018 (Stripe82) & 0.03--0.20 & 10.0--11.3 & 78 & \cite{Bertemes2018} \\

Kirkpatrick+2014 (5MUSE) & 0.05--0.29 & 10.5--11.4 & 17 & \cite{Kirkpatrick2014} \\

Bauermeister+2013 (EGNOG) & 0.06--0.31 & 10.7--11.5 & 14 & \cite{Bauermeister2013} \\

Lee+2017 & 0.27--0.62 & 10.0--11.1 & 20 & \cite{Lee2017} \\

Spilker+2018 & 0.60--0.75 & $\sim 11.0$ & 4 & \cite{Spilker2018} \\

Combes+2013 (ULIRGs) & 0.61--0.97 & 9.3--12. & 12 (14) & \cite{Combes2013} \\

Magdis+2012a (BzK) & 0.51--1.60 & 10.5--11.0 & 9 & \cite{Magdis2012SED} \\

Tacconi+2018 (PHIBSS\,1\&2) & 0.50--2.49 & 9.8--11.6 & 148 & \cite{Tacconi2018} \\

Kaasinen+2019 & 1.78--2.93 & 10.6--11.7 & 10 & \cite{Kaasinen2019} \\

Magdis+2017 (LBG) & 2.8--2.9 & 11.28--11.38 & 1 & \cite{Magdis2017} \\

Magdis+2012b (LBG) & 2.9--3.2 & 11.0--11.3 & 1 & \cite{Magdis2012LBG} \\

Tan+2014 (GN20) & 4.05--4.06 & 10.6--11.0 & 3 & \cite{Tan2014} \\

\hline









\hline

\end{tabularx}
%
$^{a}$ 
    {We only include sources with $>3\sigma$ detections.}
    \\
$^{b}$
	The largest photometric redshift in the A$^3$COSMOS catalog is 5.54 based on the prior redshift information from \cite{Laigle2016} and/or \cite{Davidzon2017}, and 7.2 based on \cite{Jin2018}. The largest spectroscopic redshift is 5.667 based on the prior information from \cite{Capak2015}. There are 11 sources which have IR/mm photo-$z=5.7-7.2$ only from \cite{Jin2018} and are very uncertain. However, our test in Sect.~\ref{Section_Fitting_Multi_Variate_Function} shows that including or excluding them does not obviously alter our results. 
%
%
%
%
%
%
\end{table*}

\vspace{0.25truecm}
\subsection{Complementary Local-to-High-Redshift Galaxy Samples}
\label{Section_Complementary_Galaxy_Samples}

We include 
20 
samples of galaxies with CO observations and well-constrained stellar mass and SFR properties from the literature as complementary information to our analysis. The full list is presented in Table~\ref{Table_galaxy_samples} (starting from the second row). It encompasses most of the CO-observed samples analyzed by \citetalias{Tacconi2018}. Most of these samples are galaxies in the local Universe, and the largest sample is the xCOLD GASS survey sample \citep{Saintonge2017}.

\citet{Saintonge2017} applied a metallicity-dependent $\alphaCO$ according to \citet{Accurso2017} to convert their CO observations into molecular gas mass. A similar metallicity-dependent $\alphaCO$ is also adopted by the \cite{Bertemes2018} and \citetalias{Tacconi2018} samples (with slightly different equations; see Appx.~\ref{Section_ZACO}). 
While most other complementary samples either assume only a single $\alphaCO$ value, i.e., either a Galactic value or an Ultra-Luminous Infrared Galaxy (ULIRG) value (see Appx.~\ref{Section_ZACO}), or bimodal values depending on the galaxy type (e.g., \citealt{Villanueva2017}).

To homogenize the complementary sample, we recalculated all molecular gas masses from the CO line luminosities by applying the metallicity-dependent $\alphaCO$ following \citetalias{Tacconi2018}. 
We use metallicity to calculate $\alphaCO$ when available (mostly for $z \lesssim 0.3$ galaxies; where metallicity is from optical emission lines using the \cite{PP04} calibration, or converted to that calibration following \citealt{Kewley2008} where necessary). Otherwise we first estimate the metallicity using the mass-metallicity relation following \citet[][Eq.~12a; see also Appx.~\ref{Section_MZR}]{Genzel2015}, then calculates the $\alphaCO$. The re-computed molecular gas masses are within a factor of 2 ($\lesssim 0.36$\,dex in logarithm) from their originally obtained values.

\vspace{1truecm}
\section{Molecular Gas Mass Calculation}
\label{Section_Molecular_Gas_Mass_Calculation}

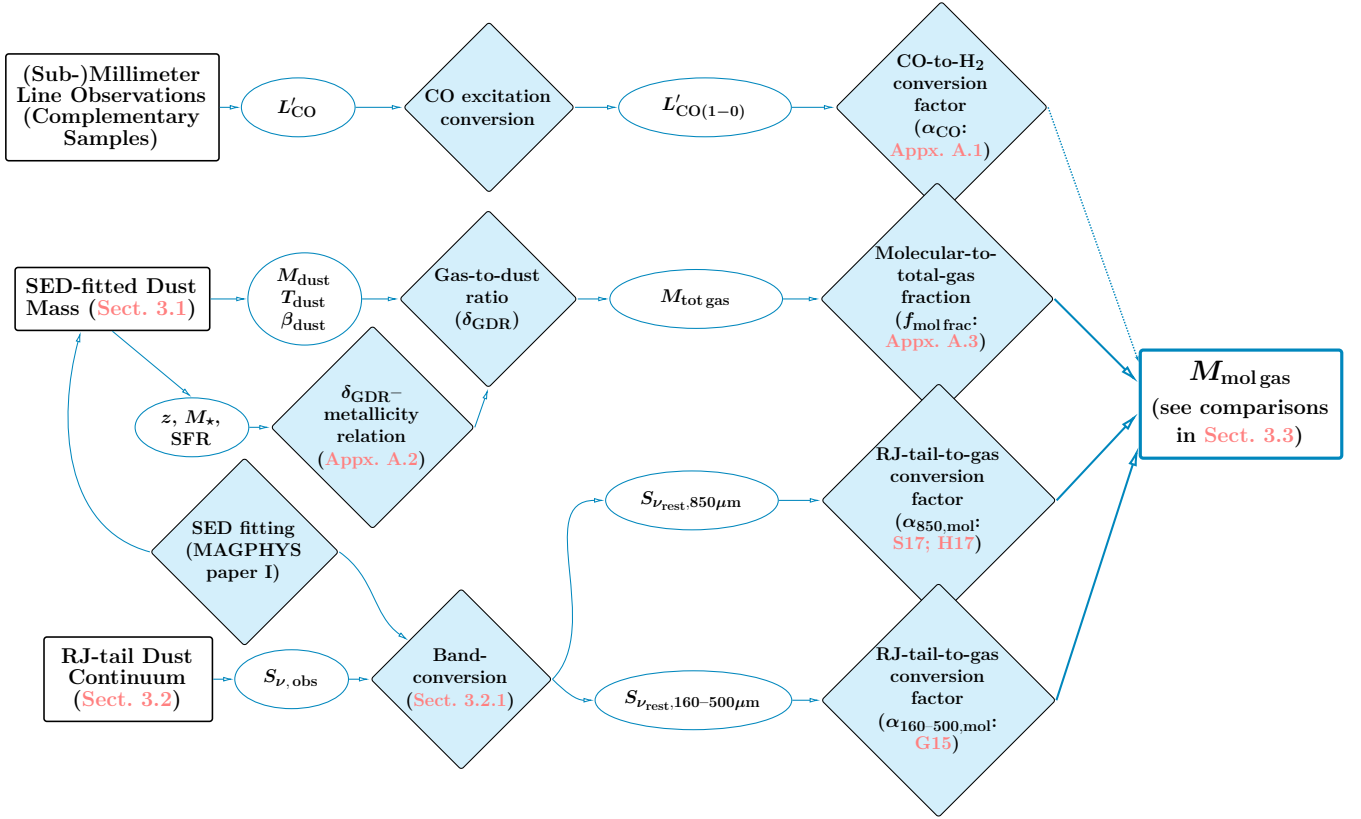
\begin{figure*}[ht]
\centering%
\begin{minipage}[b]{1.00\textwidth}
\resizebox{1.00\textwidth}{!}{


\def\hideLiteratureCatalogs{1}
\def\hideSpectralDataCubes{1}

\def\angleout{70}
\def\anglein{20}

\hypersetup{linkcolor=red!50!white, citecolor=red!50!white}

\begin{tikzpicture}[node distance = 2.5cm and 1.0cm, auto]

\node[StyleOfEmptyNode] (NodeOfCol1) {};

\node[StyleOfCatalog, below=of NodeOfCol1, xshift=0cm] (NodeOfMolecularLineObservations) {(Sub-)Millimeter Line Observations (Complementary Samples)};

\node[StyleOfFlag, right=of NodeOfMolecularLineObservations] (NodeOfMolecularLineProperties) {$\bm{L}^{\mathbf{\prime}}_{\mathbf{CO}}$};

\node[StyleOfProcess, right=of NodeOfMolecularLineProperties, xshift=1.25cm] (NodeOfExcitationConversion) {CO excitation conversion};

\node[StyleOfFlag(long), right=of NodeOfExcitationConversion, xshift=1.05cm] (NodeOfMolecularLineProperties2) {$\bm{L}^{\mathbf{\prime}}_{\mathbf{CO}\bm{(1-0)}}$};

\node[StyleOfProcess(long), right=of NodeOfMolecularLineProperties2, xshift=1.05cm] (NodeOfMolecularLineConversion) {%
CO-to-H$_{\bm{2}}$ \\
conversion \\
factor\\
($\bm{\alpha}_{\mathbf{CO}}$%
\ifx\isStandalone\undefined%
: \\%
\textcolor{red!50!white}{Appx.~\ref{Section_ZACO}}%
\fi%
)
};

\draw[StyleOfLine(thick)] 
(NodeOfMolecularLineObservations.east)
to
(NodeOfMolecularLineProperties.west);

\draw[StyleOfLine(thick)] 
(NodeOfMolecularLineProperties.east)
to
(NodeOfExcitationConversion.west);

\draw[StyleOfLine(thick)] 
(NodeOfExcitationConversion.east)
to
(NodeOfMolecularLineProperties2.west);

\draw[StyleOfLine(thick)] 
(NodeOfMolecularLineProperties2.east)
to
(NodeOfMolecularLineConversion.west);

\node[StyleOfCatalog, below=of NodeOfMolecularLineObservations, xshift=0cm, yshift=-2.5cm] (NodeOfSEDFittedDustMass) {%
SED-fitted Dust Mass
\ifx\isStandalone\undefined%
(%
\textcolor{red!50!white}{Sect.~\ref{Section_Deriving_Gas_Mass_1}}%
)%
\fi%
};

\node[StyleOfFlag, right=of NodeOfSEDFittedDustMass, xshift=0.75cm] (NodeOfMdust) {%
$\bm{M}_{\mathbf{dust}}$\\ 
$\bm{T}_{\mathbf{dust}}$\\ 
$\bm{\beta}_{\mathbf{dust}}$
};

\node[StyleOfProcess, right=of NodeOfMdust, xshift=0.75cm] (NodeOfGDR) {%
Gas-to-dust ratio\\
($\bm{\delta}_{\mathbf{GDR}}$%
)
};

\node[StyleOfFlag(long), right=of NodeOfGDR, xshift=0.50cm] (NodeOfMtot) {%
$\bm{M}_{\mathbf{tot\,gas}}$
};

\node[StyleOfProcess(long), right=of NodeOfMtot, xshift=0.75cm] (NodeOfMolecularFraction) {%
Molecular-to-\\
total-gas \\
fraction\\
($\bm{f}_{\mathbf{mol\,frac}}$%
\ifx\isStandalone\undefined%
: \\%
\textcolor{red!50!white}{Appx.~\ref{Section_ZFMOL}}%
\fi%
)
};

\draw[StyleOfLine(thick)] 
(NodeOfSEDFittedDustMass.east)
to
(NodeOfMdust.west);

\draw[StyleOfLine(thick)] 
(NodeOfMdust.east)
to
(NodeOfGDR.west);

\draw[StyleOfLine(thick)] 
(NodeOfGDR.east)
to
(NodeOfMtot.west);

\draw[StyleOfLine(thick)] 
(NodeOfMtot.east)
to
(NodeOfMolecularFraction.west);

\node[StyleOfCatalog, below=of NodeOfSEDFittedDustMass, xshift=0.75cm, yshift=-12cm] (NodeOfDustContinuumObservations) {%
RJ-tail Dust Continuum
\ifx\isStandalone\undefined%
(%
\textcolor{red!50!white}{Sect.~\ref{Section_Deriving_Gas_Mass_234}}%
)%
\fi%
};

\node[StyleOfFlag, right=of NodeOfDustContinuumObservations] (NodeOfDustContinuumFluxes) {%
$\bm{S}_{\bm{\nu},\,\mathbf{obs}}$
};

\node[StyleOfProcess, right=of NodeOfDustContinuumFluxes] (NodeOfBandConversion) {%
Band-conversion
\ifx\isStandalone\undefined%
(%
\textcolor{red!50!white}{Sect.~\ref{Section_Band_conversion}}%
)%
\fi%
};

\node[StyleOfFlag(long), right=of NodeOfBandConversion, xshift=+1.5cm,, yshift=+8.50cm] (NodeOfDustContinuumRF850) {%
$\bm{S}_{\bm{\nu}_{\mathbf{rest}},\bm{850{\mu}\mathrm{m}}}$
};

\node[StyleOfFlag(long), right=of NodeOfBandConversion, xshift=+1.0cm, yshift=-1.00cm] (NodeOfDustContinuumRF500) {%
$\bm{S}_{\bm{\nu}_{\mathbf{rest}},\bm{160{\textendash}500{\mu}\mathrm{m}}}$
};

\node[StyleOfProcess(long), right=of NodeOfDustContinuumRF850, xshift=+1.0cm, yshift=0cm] (NodeOfDustContinuumConversion) {%
RJ-tail-to-gas \\
conversion\\
factor\\
($\bm{\alpha}_{\mathbf{850,mol}}$%
\ifx\isStandalone\undefined%
: \\%
\textcolor{red!50!white}{%
\citetalias{Scoville2017}; 
\citetalias{Hughes2017}%
}%
\fi%
)
};

\node[StyleOfProcess(long), right=of NodeOfDustContinuumRF500, xshift=+0.35cm, yshift=0cm] (NodeOfDustContinuumConversion2) {%
RJ-tail-to-gas \\
conversion\\
factor\\
($\bm{\alpha}_{\mathbf{160{\textendash}500,mol}}$%
\ifx\isStandalone\undefined%
: \\%
\textcolor{red!50!white}{%
\citetalias{Groves2015}%
}%
\fi%
)
};

\draw[StyleOfLine(thick)] 
(NodeOfDustContinuumObservations.east)
to
(NodeOfDustContinuumFluxes.west);

\draw[StyleOfLine(thick)] 
(NodeOfDustContinuumFluxes.east)
to
(NodeOfBandConversion.west);

\draw[StyleOfLine(thick)] 
(NodeOfBandConversion.east)
to[out=+45,in=-180]
(NodeOfDustContinuumRF850.west);

\draw[StyleOfLine(thick)] 
(NodeOfBandConversion.east)
to[out=-45,in=-180]
(NodeOfDustContinuumRF500.west);

\draw[StyleOfLine(thick)] 
(NodeOfDustContinuumRF850.east)
to
(NodeOfDustContinuumConversion.west);



\draw[StyleOfLine(thick)] 
(NodeOfDustContinuumRF500.east)
to
(NodeOfDustContinuumConversion2.west);

\node[StyleOfFinalProduct, right=of NodeOfMolecularFraction, xshift=+3cm, yshift=-5cm] (NodeOfGasMass) {%
$\bm{M}_{\mathbf{mol\,gas}}$%
\ifx\isStandalone\undefined%
\\[0.5ex]
{\XHUGE(see comparisons in %
\textcolor{red!50!white}{Sect.~\ref{Section_comparing_gas_mass_calibration}}%
)}%
\fi%
};

\draw[StyleOfLine(dotted)] 
(NodeOfMolecularLineConversion.east)
to
([yshift=+2.0cm]NodeOfGasMass.west);

\draw[StyleOfLine(thicker)] 
(NodeOfMolecularFraction.east)
to
([yshift=+1.0cm]NodeOfGasMass.west);


\draw[StyleOfLine(thicker)] 
(NodeOfDustContinuumConversion.east)
to
([yshift=-0.4cm]NodeOfGasMass.west);


\draw[StyleOfLine(thicker)] 
(NodeOfDustContinuumConversion2.east)
to
([yshift=-2.0cm]NodeOfGasMass.west);


\node[StyleOfProcess, below=of NodeOfGDR, xshift=-5.5cm, yshift=5.5cm] (NodeOfMetallicity) {%
$\bm{\delta}_{\mathbf{GDR}}$--metallicity relation\\
\ifx\isStandalone\undefined%
(%
\textcolor{red!50!white}{Appx.~\ref{Section_ZGDR}}%
)%
\fi%
};

\node[StyleOfFlag, left=of NodeOfMetallicity] (NodeOfMstar) {%
$\bm{z}$, $\bm{M}_{\bm{\star}}$, \\
SFR
};

\node[StyleOfProcess, below=of NodeOfMstar, xshift=+2.5cm, yshift=+2.5cm] (NodeOfSEDFitting) {%
SED fitting \\
(\textsc{MAGPHYS} \\
paper~\romup{1})
};

\draw[StyleOfLine(thick)] 
(NodeOfSEDFittedDustMass.south)
to
(NodeOfMstar.north);

\draw[StyleOfLine(thick)] 
(NodeOfMstar.east)
to
(NodeOfMetallicity.west);

\draw[StyleOfLine(thick)] 
(NodeOfMetallicity.east)
to
(NodeOfGDR.south);


\draw[StyleOfLine(thick)] 
(NodeOfSEDFitting.west)
to[out=165,in=-105]
([xshift=-1.5cm]NodeOfSEDFittedDustMass.south);

\draw[StyleOfLine(thick)] 
(NodeOfSEDFitting.east)
to[out=-35,in=145]
([xshift=-2.5cm, yshift=-2.5cm]NodeOfBandConversion.north);


\ifdefined\isStandalone
\node[StyleOfEmptyNode, below=of NodeOfDustContinuumObservations, yshift=-4cm] (NodeOfEmpty001) {};
\fi

\end{tikzpicture}

\hypersetup{linkcolor=magenta!80!white, citecolor=blue}}
\vspace{-1.0ex}
\caption{%
Overview of popular molecular gas mass ($\Mmolgas$) calibration methods for high-redshift (i.e., $z>1$) galaxies sorted into three categories: (a) (sub-)millimeter emission line observations; (b) galaxy SED-fitted dust mass; and (c) RJ-tail dust continuum. 
See description in Sect.~\ref{Section_Molecular_Gas_Mass_Calculation}. 
The corresponding sections in this paper are labeled in the flow chart. 
The references for the $\alpha_{\mathrm{850,mol}}$ conversion factors are: \citealt{Scoville2017} (S17); \citealt{Hughes2017} (H17); and $\alpha_{160{\textendash}500,\mathrm{mol}}$: \citealt{Groves2015} (G15). 
\label{Figure_gas_mass_calibration_flow_chart}
}%
\end{minipage}
\vspace{2.5ex}
\end{figure*}

We summarize the three most commonly used molecular gas mass calibration methods for high-redshift galaxies in Fig.~\ref{Figure_gas_mass_calibration_flow_chart}. As mentioned in the introduction, they are (a) CO lines, (b) SED-fitted dust mass, and (c) RJ-tail dust continuum\,\footnote{The (b) and (c) methods have the same underlying physics, which is using dust mass to trace the total gas mass. Here we separate them because they have different technical steps and assumptions. See details in Sects.~\ref{Section_Deriving_Gas_Mass_1}~and~\ref{Section_Deriving_Gas_Mass_234}.}. 
The CO method infers the molecular gas mass via the $\alphaCO$ conversion factor, which relates to CO luminosity to H$_2$ gas mass and is correlated with metallicity (see details in Appx.~\ref{Section_ZACO}). When the observed CO line is not the ground transition ($J=1\to0$), an excitation ladder is needed to convert the higher-$J$ line luminosity to the $J=1\to0$ one (e.g., \citealt{Carilli2013Review}).

Given that the CO and dust RJ-tail 850\,$\mu$m-based gas mass calibrations have been extensively verified to be tightly correlated in a number of recent works at local and high redshift up to $z\sim2-3$ (e.g., \citealt{Scoville2014,Scoville2017,Hughes2017,Bertemes2018,Saintonge2018,Kaasinen2019}), we do not further discuss the CO method here, but focus on popular dust-based methods. 
In Sect.~\ref{Section_Deriving_Gas_Mass_1} we describe the use of SED-fitted dust mass to compute molecular gas mass, and in Sect.~\ref{Section_Deriving_Gas_Mass_234} we describe the use of the RJ-tail dust continuum for molecular gas mass calculation. 
There are multiple choices for calibration factors and wavelengths, thus we compare these methods thoroughly in Sect.~\ref{Section_comparing_gas_mass_calibration}.

Later we will combine CO- and dust-based samples together for our data fitting analysis (in Sects.~\ref{Section_Galaxy_Properties}~and~\ref{Section_Fitting_Multi_Variate_Function}). We assume that the consistency between CO- and (our adopted) dust-based gas mass calibration extends to all galaxies in our combined sample, which is at least supported by the aforementioned CO and dust calibration studies (but we also discussed the current caveats at the end of Sect.~\ref{Section_Fitting_Multi_Variate_Function}).

\vspace{0.25truecm}
\subsection{Molecular Gas Mass from SED-fitted Dust Mass}
\label{Section_Deriving_Gas_Mass_1}

In the SED-fitted dust mass method, we first obtain the dust mass ($\Mdust$), dust mean temperature ($\Tdust$) and dust emissivity ($\beta_{\mathrm{dust}}$; describing the dust opacity $\kappa$'s wavelength dependency; e.g., \citealt{Li2001}) from optical-to-mm SED fitting, then apply a gas-to-dust ratio, $\deltaGDR \equiv \Mtotalgas/\Mdust$, which relates total gas (molecular and atomic) mass to dust mass.

In the first step, different assumptions on dust grain models can lead to variations in the determined dust properties. Yet simulations (e.g., \citealt{Hayward2015}) and observations of local galaxies (e.g., \citealt{Hunt2019,Hayward2015}) indicate that SED fitting tools like \textsc{MAGPHYS} are able to reasonably recover galaxies' dust properties (at least for $L_{\mathrm{IR}}>10^{11}\;\Lsun$). 
For our work we ignore the systematic uncertainty introduced by different dust grain models (i.e., different SED fitting tools). This might not be entirely correct but further investigation of this topic requires a subsample of galaxies with well-sampled SEDs and accurate spectroscopic redshifts, and is beyond the scope of this paper.

In the second step, $\deltaGDR$ is found to strongly depend on metallicity (e.g., \citealt{Leroy2011GDR}; \citealt{RemyRuyer2014}; \citealt{DeVis2019}; see more details in Appx.~\ref{Section_ZGDR}), and the latter is correlated with the stellar mass (known as the mass--metallicity relation; see detailed discussion in Appx.~\ref{Section_MZR}). Differences exist among the empirical scaling relations in the literature, whereas our ALMA continuum observations preferentially select intensely star-forming galaxies with $\Mstar >  2 \times 10^{10} \; \Msun$, which exhibit a close-to-solar metallicity based on the mass-metallicity relation at $z\sim2.3$ of \cite{Erb2006}. Our analysis is, therefore, only affected by the relative small offset of 0.1--0.2~dex between the relations of \cite{Leroy2011GDR} and \cite{RemyRuyer2014} at $>0.5$ solar metallicity.

As A$^3$COSMOS galaxies do not have homogeneous metallicity measurements, we compute the metallicity based on redshift and stellar mass for each of the galaxies using the mass--metallicity relation of \citet[][Eq.~12a]{Genzel2015} and compute the $\deltaGDR$ using the \cite{RemyRuyer2014} prescription.  
Our detailed comparison of various forms of the mass--metallicity relation and the ``Fundamental Metallicity Relation'' (e.g., \citealt{Mannucci2010,Mannucci2011}; yet still debated) in Appx.~\ref{Section_MZR} shows that the Eq.~12a of \citet[][which is also used by \citealtalias{Tacconi2018}]{Genzel2015} provides the most plausible predictions for the metallicity of high-redshift $z>1$ galaxies. Here we adopt a slightly modified form of:
{\setlength{\abovedisplayskip}{1pt}\setlength{\belowdisplayskip}{2pt}%
\begin{fleqn}%
\begin{equation}
\begin{split}
&\metalZOH = \\
&\begin{cases}
a, \quad \quad \quad \quad \quad \quad \quad \quad \ \ \textnormal{if} \ \log_{10} (\Mstar/\Msun) \ge b(z), \\
a - 0.087 \times (\log_{10} (\Mstar/\Msun) - b(z))^2, \quad \quad \ \textnormal{else.}
\end{cases}\\
&\textnormal{where} \ a = 8.74 \ \textnormal{and} \
b(z) = 10.4 + 4.46 \times \log_{10}(1+z) \\& - 1.78 \times (\log_{10}(1+z))^{2}.
\end{split}
\label{Equation_metallicity}
\end{equation}
\end{fleqn}%
}%
The modification (under the $\log_{10} (\Mstar/\Msun) \ge b(z)$ condition) prevents a drop in metallicity for very massive galaxies at $z<1$ (see Fig.~\ref{Plot_metalZ_Mstar_SFR_z_bins}).

Finally, we consider that our A$^3$COSMOS high-redshift galaxies are molecular-rich (same as assumed by \citetalias{Tacconi2018} at $z>0.4$), i.e., the molecular-to-total-gas ratio $\fmol$ is unity. In this way, we obtain the dust-SED-based $\Mmolgas$ by multiplying $\Mdust$ with the mass--metallicity-derived $\deltaGDR$ and ignore the contribution from atomic gas. 
Hereafter we refer to this method as the ``$\delta_{\mathrm{GDR},\,Z}$'' method.

We caution that, as discussed in Appx.~\ref{Section_ZFMOL}, observations of local galaxies actually indicate that $\fmol$ is usually below 50\% even for a galaxy with $\Mstar\sim 1 \times 10^{11} \; \Msun$. Applying an actual $\fmol$, e.g., based on the \citet{KMT09} correlation with stellar mass or metallicity, will lead to a lower $\Mmolgas$. Based on our next comparison of $\Mmolgas$ calibrations (see Sect.~\ref{Section_comparing_gas_mass_calibration}), this will cause even larger difference to the RJ-tail dust continuum methods where atomic gas is also not considered. Therefore, here we choose to not account for the atomic gas, and leave the consideration of an actual $\fmol$ to future work.

\vspace{0.25truecm}
\subsection{Molecular Gas Mass from RJ-tail Dust Continuum}
\label{Section_Deriving_Gas_Mass_234}

Recent studies show that dust continuum luminosity at rest-frame RJ-tail wavelengths tightly correlates with gas mass or CO line luminosity across two orders of magnitude in local and high-redshift galaxies (e.g., \citealt{Bourne2013}; \citealt{Scoville2014,Scoville2016}; \citealt{Groves2015}; \citealt{Hughes2017}; \citealt{Saintonge2017}; \citealt{Bertemes2018}; \citealt{Kaasinen2019}). 
\cite{Scoville2014} found a constant ratio between dust continuum luminosity and $\Mtotalgas$: 
{\setlength{\abovedisplayskip}{3pt}\setlength{\belowdisplayskip}{3pt}%
\begin{fleqn}%
\begin{equation}
\quad \alphaRJtot \equiv \frac{%
L_{\nu_{\textsc{rj}}} \ / \ [\mathrm{erg\,s^{-1}\,Hz^{-1}}]
}{%
\Mtotalgas \ / \ [\Msun]
}
\label{Equation_alphaRJtot}
\end{equation}
\end{fleqn}
}%
where they calibrated $\alphaRJtot$ to be $1.0 \pm 0.23 \times 10^{20}$ $[\mathrm{erg\,s^{-1}\,Hz^{-1}\,\Msun^{-1}}]$ at rest-frame 850\,$\mu$m with a small sample of 12 local galaxies. 
Meanwhile, \citet{Groves2015} studied the atomic, molecular gas and dust continuum at rest-frame 70, 100, 160, 250, 350 and 500\,$\mu$m in 36 local spiral galaxies. They found a mean $\Mtotalgas/\nu L_{\nu,500} = 28.5$ for near-solar metallicity galaxies, corresponding to $\alpha_{\mathrm{500,tot}} = 2.2 \times 10^{20}$ $[\mathrm{erg\,s^{-1}\,Hz^{-1}\,\Msun^{-1}}]$, and a factor of ten lower values for much more metal-poor galaxies. 
According to Eq.~9 of \citet{Scoville2014}, $\alphaRJtot$ is proportional to dust opacity $\kappa_{\nu}$, which scales with frequency by $\kappa_{\nu} \propto \nu^{1.7-2.0}$ (\citealt{Li2001}), thus it is expected that $\alphaRJtot$ is a factor of 2.5 higher at 500\,$\mu$m than at 850\,$\mu$m. Therefore the calibrations are consistent between \cite{Groves2015} and \cite{Scoville2014}. 
Meanwhile, the variation from metal-rich to metal-poor galaxies can also be explained by a dramatic change in $\deltaGDR$ (Appx.~\ref{Section_ZGDR}).

Focusing on molecular gas only, \cite{Scoville2016} calibrated the ratio between the RJ-tail dust continuum luminosity and $\Mmolgas$: 
{\setlength{\abovedisplayskip}{3pt}\setlength{\belowdisplayskip}{3pt}%
\begin{fleqn}%
\begin{equation}
\quad \alphaRJmol \equiv \frac{%
L_{\nu_{\textsc{rj}}} \ / \ [\mathrm{erg\,s^{-1}\,Hz^{-1}}]
}{%
\Mmolgas \ / \ [\Msun]
}
\label{Equation_alphaRJmol}
\end{equation}
\end{fleqn}
}%
to be $6.7 \pm 1.7 \times 10^{19}$ $[\mathrm{erg\,s^{-1}\,Hz^{-1}\,\Msun^{-1}}]$ at rest-frame 850\,$\mu$m for a few tens of local spirals, ULIRGs and $z\sim2$ submillimeter galaxies (SMGs). Later studies with larger samples of CO and RJ-tail continuum observations found slightly non-linear correlations, i.e., $\alphaRJmol$ has a dependency on $\LnuRJ$ or $\LprimeCO$ (e.g., \citealt{Hughes2017}, \citealt{Bertemes2018} and \citealt{Saintonge2018}). 
As their samples span a wide range of stellar mass from $10^{9}$ to $10^{12}\;\Msun$, and CO $J=1\to0$ line luminosity $L^{\prime}_{\mathrm{CO(1-0)}}$ from $10^{7}$ to $10^{12}\;\mathrm{K\,km\,s^{-1}\,pc^{2}}$, galaxies have significantly varied metallicity, $\fmol$ and $\deltaGDR$. A simple explanation for the variations is that $\alphaRJmol$ scales with $\fmol^{-1}$ and $\deltaGDR^{-1}$, which both relate to metallicity (see Appx.~\ref{Section_ZFMOL} and \ref{Section_ZGDR} respectively).

Since the literature on the calibration of $\alphaRJmol$ is already very rich, we do not further discuss it here. In the following we will adopt the three calibrations from \citetalias{Scoville2017}, \cite{Hughes2017} and \cite{Groves2015}, referred to as the ``$\alphaRJS17$'', ``$\alphaRJH17$'' and ``$\alphaRJG15$'' method, respectively. 
For the ``$\alphaRJG15$'' method, we use the calibration factors for the $\log_{10}(\Mstar/\Msun)>9$ galaxies in the Table~5 of \cite{Groves2015}, and assume that our A$^3$COSMOS galaxies have negligible atomic gas contribution.
These works directly calibrate the ratio between $\LnuRJ$ and $\Mmolgas$ (and ``$\alphaRJH17$'' and ``$\alphaRJG15$'' include a luminosity dependency), therefore the need for a calibration of the underlying $\fmol$ and $\deltaGDR$ is bypassed.

\vspace{0.25truecm}
\subsubsection{Band conversion from observed-frame to rest-frame RJ-tail}
\label{Section_Band_conversion}

The good agreement between RJ-tail dust continuum-to-gas mass calibrations and the overwhelming observational efficiency compared to (sub-)mm line observations make the RJ-tail dust method very favorable and promising for large surveys at high-redshift.

\begin{figure}[htb]
\centering
\includegraphics[width=0.99\linewidth]{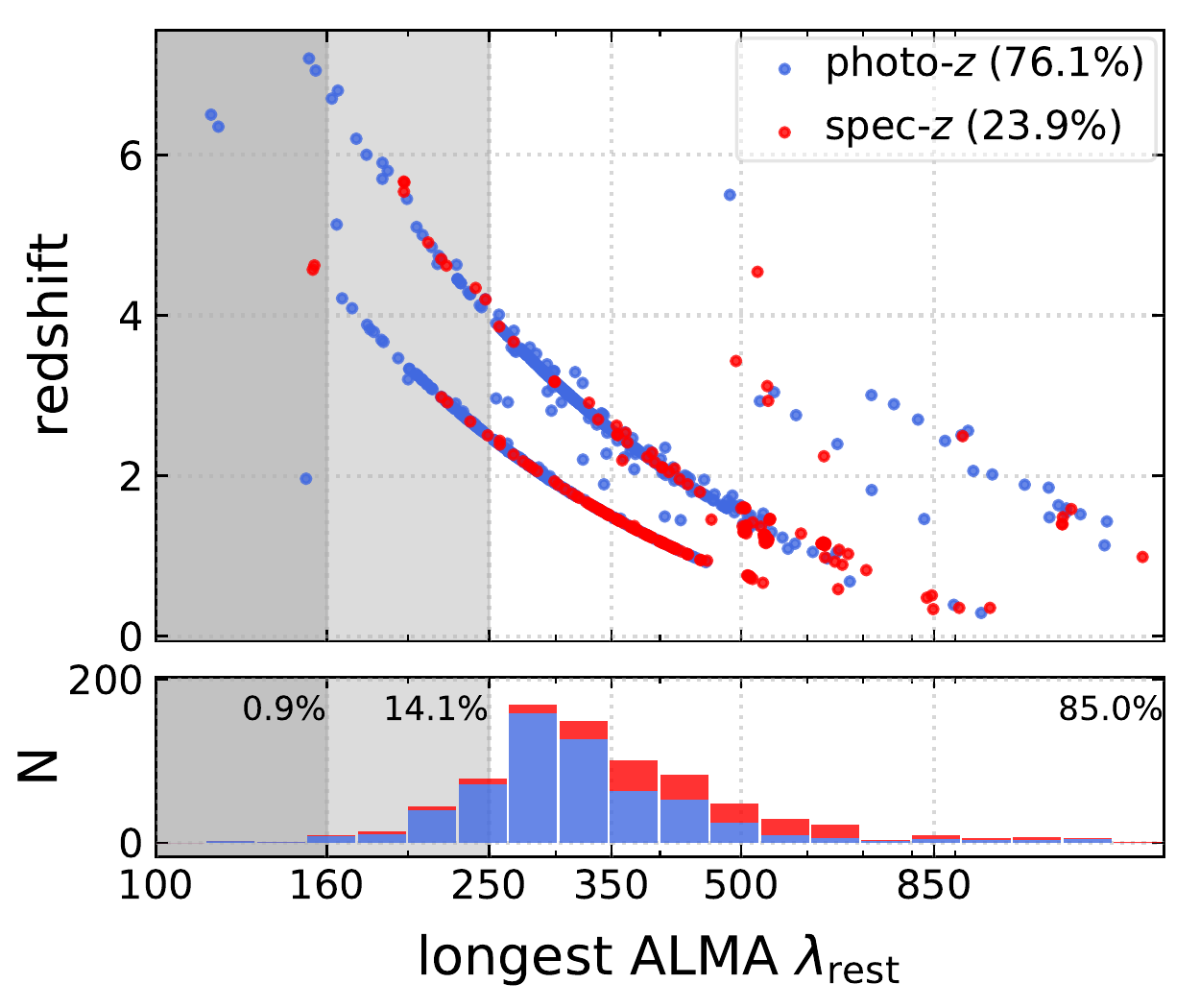}
\caption{%
{\it Upper panel:} Redshift versus the longest rest-frame wavelengths ($\lambda_{\mathrm{rest}}$) of the available ALMA data for each galaxy in our sample. Color indicates whether the source has a spectroscopic redshift (spec-$z$; red) or a photometric redshift (photo-$z$; blue). The dark-gray and light-gray shading represents $\lambda_{\mathrm{rest}} < 160\,\mu\mathrm{m}$ and $160 \le \lambda_{\mathrm{rest}} < 250\,\mu\mathrm{m}$ respectively.
{\it Lower panel:} Histograms of the longest ALMA $\lambda_{\mathrm{rest}}$. Color indicates the same subsample as above. The three labels are the percentages of sources with $\lambda_{\mathrm{rest}} < 160\,\mu\mathrm{m}$, $160 \le \lambda_{\mathrm{rest}} < 250\,\mu\mathrm{m}$, and $\lambda_{\mathrm{rest}} \ge 250\,\mu\mathrm{m}$, respectively. 
}
\label{Plot_rest_frame_wavelength_vs_redshift}
\end{figure}

Our high-redshift galaxies are most commonly observed in ALMA Band 6 and 7, which correspond to rest-frame $\lesssim 250\,\mu$m and $\lesssim 160\,\mu$m, respectively, for galaxies at $z\gtrsim4$. 
In Fig.~\ref{Plot_rest_frame_wavelength_vs_redshift}, we show the longest rest-frame wavelengths of the available ALMA data for each galaxy in our sample. 85\% of our sources have $\lambda_{\mathrm{rest}} \ge 250\,\mu\mathrm{m}$, while the rest only probe shorter-wavelength dust continua.

In order to apply the $\alphaRJmol$ conversion from dust continuum to molecular gas mass, a ``band conversion''\,\footnote{This means first applying the $K$-correction (\citealt{Humason1956,Oke1968}) to the best-fit SED, then interpolating/extrapolating to certain calibration wavelengths, then scale the observed ALMA flux accordingly, see details afterwards.}
is needed to obtain the corresponding flux density at the calibrated rest-frame wavelength, i.e., rest-frame 850\,$\mu$m for applying the $\alphaRJS17$ and $\alphaRJH17$ methods, and either 160, 250, 350 or 500\,$\mu$m for applying the $\alphaRJG15$ method.

We use our \textsc{MAGPHYS} SED fitting for the band conversion, i.e., predicting longer-wavelength flux density with an SED covering only shorter wavelengths. 
\textsc{MAGPHYS} fits the dust SED with two dust components, one associated with actual star-forming birth clouds and the other exposed to the ambient interstellar radiation field. The former dust usually has a high temperature and dominates the short-wavelength (e.g., $\lambda_{\mathrm{rest}}<60\,\mu\mathrm{m}$) flux density, while the latter dust is constrained to have a temperature in the range of 15--25~K (\citealt{daCunha2008,daCunha2015}) and dominates the long-wavelength flux density. 
A similar idea of composite dust models is also adopted by \cite{Draine2007SED} and used in fitting local star-forming galaxies (e.g., \citealt{Draine2007SINGS}; \citealt{Aniano2012}) and high-redshift galaxies (e.g., \citealt{Magdis2012SED,Magdis2014,Magdis2017}). 

Using such composite-model SED fitting for band conversion has a large advantage over using a single-temperature modified blackbody, as it is much less biased toward the luminosity-weighted dust temperature. \cite{Privon2018} studied the systematic bias of the band conversion using their zoomed-in cosmological simulations, finding that assuming a single-temperature modified blackbody SED for conversion leads to a more than 0.5\,dex overestimation in $\L850$ when the true dust temperature is a factor of two different than assumed (see their Fig.~5).

Whereas \textsc{MAGPHYS} performs well in fitting the dust SED shape, the sampling of the dust SED is usually limited by the available data for $z>4$ sources (as shown in Fig.~\ref{Plot_rest_frame_wavelength_vs_redshift}). 
In Appx.~\ref{Section_Appendix_SED_band_conversion} we perform a test to estimate the bias of lacking long-wavelength data in predicting longer wavelength flux density. We find that when having only $\lambda_{\mathrm{rest}} \le 160\,\mu\mathrm{m}$ data points, \textsc{MAGPHYS} under-predicts the rest-frame 850\,$\mu$m flux density by up to 0.8\,dex (on average 0.4\,dex) when the dust continua photometries have a quadratic-added mean $\SNR \lesssim 15$. Meanwhile, the worse case of having only the rest-frame 160\,$\mu$m data point available over the 8\,$\mu$m to 3\,mm range causes a similar bias by \textsc{MAGPHYS}.

To apply the band conversion, we first compute the ratio between the SED-predicted flux densities at $850\times(1+z)\,\mu\mathrm{m}$ and the observed wavelength: 
{\setlength{\abovedisplayskip}{3pt}\setlength{\belowdisplayskip}{3pt}\noindent%
\begin{fleqn}%
\begin{equation}
\begin{split}
&\Gamma^{\mathrm{SED}} \equiv {S_{\nu_{\mathrm{850\times(1+z)\,\mu\mathrm{m}}}}^{\;\mathrm{SED}}} / {S_{\nu_{\mathrm{obs}}}^{\;\mathrm{SED}}},
\end{split}
\end{equation}%
\end{fleqn}%
}%
then we scale the observed ALMA flux density by $\Gamma^{\mathrm{SED}}$ and compute the luminosity:
{\setlength{\abovedisplayskip}{3pt}\setlength{\belowdisplayskip}{3 pt}\noindent%
\begin{fleqn}%
\begin{equation}
\begin{split}
&\L850 = 4 \pi d_{\mathrm{L}}^2 \times S_{\nu_{\mathrm{obs}}}^{\mathrm{ALMA}} \times \Gamma^{\mathrm{SED}} \,/\, (1+z). \\
\end{split}
\end{equation}%
\end{fleqn}%
}%
In principle we can also directly take the SED-predicted rest-frame 850\,$\mu$m flux density $S_{\nu_\mathrm{850\times(1+z)\,\mu\mathrm{m}}}^{\;\mathrm{SED}}$. 
But this would lead to under-predicted scatter in our analysis due to the degeneracy within SED models. 

Finally, we divide the luminosity $\L850$ by $\alphaRF850$ derived from \citetalias{Scoville2017} and \cite{Hughes2017} to obtain the ``$\alphaRJS17$'' and ``$\alphaRJH17$'' molecular gas masses. 
In the ``$\alphaRJG15$'' method, as \cite{Groves2015} provided calibrations at six calibration wavelengths (70, 100, 160, 250, 350 and 500\,$\mu$m), we perform the band conversion from the longest-wavelength ALMA data to its nearest calibration wavelength.

\begin{figure*}[ht]
\centering
\includegraphics[width=0.8\textwidth]{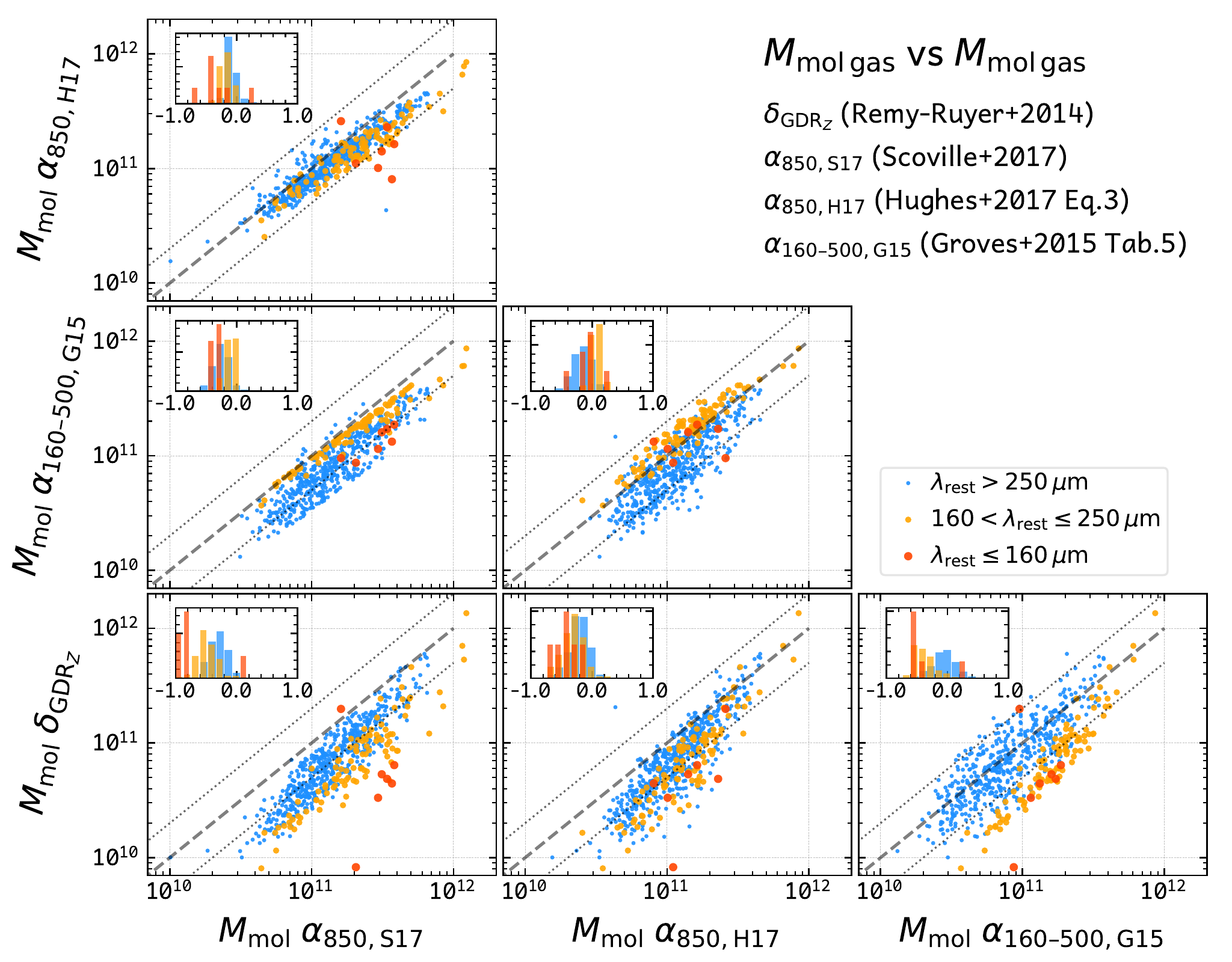}
\vspace{0.5ex}
\caption{%
Comparisons between four methods of gas mass calibration based on dust SED and/or RJ-tail continuum as presented in Sect.~\ref{Section_Molecular_Gas_Mass_Calculation} and labeled at top-right. 
We divide the sources into three categories based on their longest available rest-frame ALMA wavelength (denoted as $\lambda_{\mathrm{rest}}$): $\lambda_{\mathrm{rest}} \le 160 \,\mu\mathrm{m}$ (red), $160 < \lambda_{\mathrm{rest}} \le 250 \,\mu\mathrm{m}$ (orange) and $\lambda_{\mathrm{rest}} > 250 \,\mu\mathrm{m}$ (blue), which also correspond to the three same-color-shaded areas in Fig.~\ref{Plot_rest_frame_wavelength_vs_redshift}, respectively. 
In each panel, the dashed line is a one-to-one line, and the parallel dotted lines indicate a factor of two variation. 
The embedded histogram plotted in each Y-versus-X scatter plot is the normalized distribution of $\log_{10} (\mathrm{Y/X})$. 
See discussion in Sect.~\ref{Section_comparing_gas_mass_calibration}.
}
\label{Plot_Mmolgas_vs_Mmolgas}
\vspace{3ex}
\end{figure*}

\vspace{0.25truecm}
\subsection{Comparing gas mass calibrations}
\label{Section_comparing_gas_mass_calibration}

In Fig.~\ref{Plot_Mmolgas_vs_Mmolgas} we compare the molecular gas masses estimated from the above mentioned ``$\delta_{\mathrm{GDR},\,Z}$'', ``$\alphaRJS17$'', ``$\alphaRJH17$'' and ``$\alphaRJG15$'' methods. 
As shown in the bottom row of the figure, ``$\delta_{\mathrm{GDR},\,Z}$'' leads to systematically lower gas masses than the other three RJ-tail continuum methods. The bias is stronger for sources which do not have long-wavelength ($\lambda_{\mathrm{rest}}>250\,\mu\mathrm{m}$) coverage. This is closely related to the \textsc{MAGPHYS} SED fitting feature, where missing long-wavelength data seems to lead to an underestimation of the cold, ambient dust which dominates the total dust mass (consistent with the tests in Appx.~\ref{Section_Appendix_SED_band_conversion}).

In the first-row panel, ``$\alphaRJS17$'' and ``$\alphaRJH17$'' methods agree within 0.1~dex for sources with long-wavelength coverage (but up to about 0.3~dex for sources lacking $>160\,\mu\mathrm{m}$ data). However, a systematic offset of about 0.1~dex exists, which is likely because ``$\alphaRJS17$'' uses a single conversion factor while ``$\alphaRJH17$'' uses a luminosity-dependent conversion factor. The latter has been confirmed by many other works (e.g., \citealt{Bertemes2018}; \citealt{Saintonge2018}) and therefore is more reliable.

For panels in the second row, the gas masses based on the ``$\alphaRJS17$'' and ``$\alphaRJH17$'' methods are compared to those using the ``$\alphaRJG15$'' method. The ``$\alphaRJG15$'' method leads to 0.25~dex lower molecular gas masses than ``$\alphaRJS17$'', or 0.15~dex lower than ``$\alphaRJH17$'' for the majority of sources. A small number of sources with poor long-wavelength coverage, however, have smaller differences. 
This is probably due to the smaller $\Mstar>10^{9}\;\Msun$ sample in \cite{Groves2015} and the intrinsic variation in $\LRJ/\Mmolgas$.

To summarize, we find that the gas mass calibrations are: $\Mmolgas \, _{(\delta_{\mathrm{GDR},\,Z})} \lesssim \Mmolgas \, _{(\alphaRJG15)} < \Mmolgas \, _{(\alphaRJH17)} < \Mmolgas \, _{(\alphaRJS17)}$. The systematic offsets are about 0.15--0.25~dex, but are comparable to the scatter of the data. 
Considering the relatively better agreement of the ``$\alphaRJH17$'' method to other methods as well as recent observations (\citealt{Bertemes2018}; \citealt{Saintonge2018}), we choose the ``$\alphaRJH17$'' method as our final gas mass calculation for the A$^3$COSMOS galaxies. 
We also tested our full analysis with other gas mass calibrations in Sect.~\ref{Section_Fitting_Multi_Variate_Function}, finding that our results are not obviously altered.

\vspace{1truecm}
\section{Galaxy Molecular Gas and Star Formation Properties}
\label{Section_Galaxy_Properties}

\begin{figure*}[htb]
\centering
\includegraphics[trim=0 0 0 0, width=0.75\textwidth]{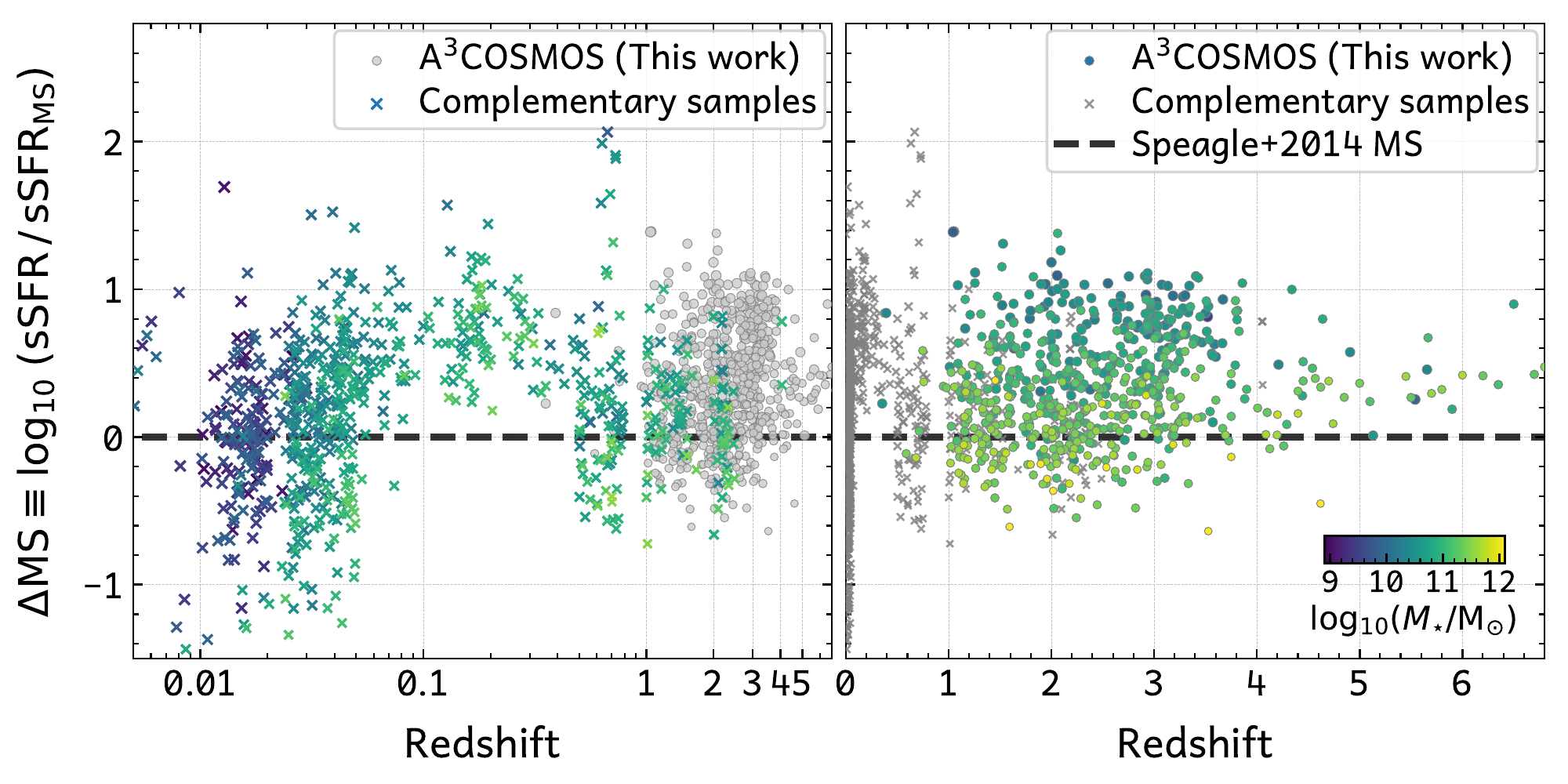}
\vspace{0.5ex}
\caption{%
Specific star formation rate ($\mathrm{sSFR} \equiv \SFR/\Mstar$) versus redshift distribution of our galaxies (see Table~\ref{Table_galaxy_samples}) normalized by the star-forming MS SFR at each redshift. The $\mathrm{Y}=0$ dashed line means exactly on the \cite{Speagle2014} MS. Data points are the same in the {\it left} and {\it right panels}, except that the redshift axis is logarithmic in the {\it left panel} for better illustrating the complementary samples (with A$^3$COSMOS data points in gray) and is linear in the {\it right panel} for illustrating our A$^3$COSMOS sample (with all complementary samples shown in gray). 
\label{Plot_z_DeltaMS}
}
\end{figure*}

After the calculation of molecular gas mass for A$^3$COSMOS galaxies, we combine them with our complementary galaxy samples listed in Table~\ref{Table_galaxy_samples}, allowing us to study galaxy molecular gas and star formation scaling relations and gas evolution in the following sections. In total, we have 
1,663
galaxies with redshift, SFR, stellar mass and molecular gas mass measurements. All complementary galaxies are selected to have CO detections and their molecular gas masses are homogenized with metallicity-dependent $\alphaCO$ as detailed in Sect.~\ref{Section_Complementary_Galaxy_Samples}. 
Such a combined sample is the largest, most-robust individually-detected sample so far, yet it still exhibits certain incompleteness in the parameter space of redshift, stellar mass and star formation due to sample selection biases. Therefore, before analyzing the gas scaling relations and the resulting gas evolution, we first provide detailed inspections below to constrain potential sample selection biases.

\begin{figure*}[htb]
\centering
\includegraphics[trim=0 0 0 0, width=1\linewidth]{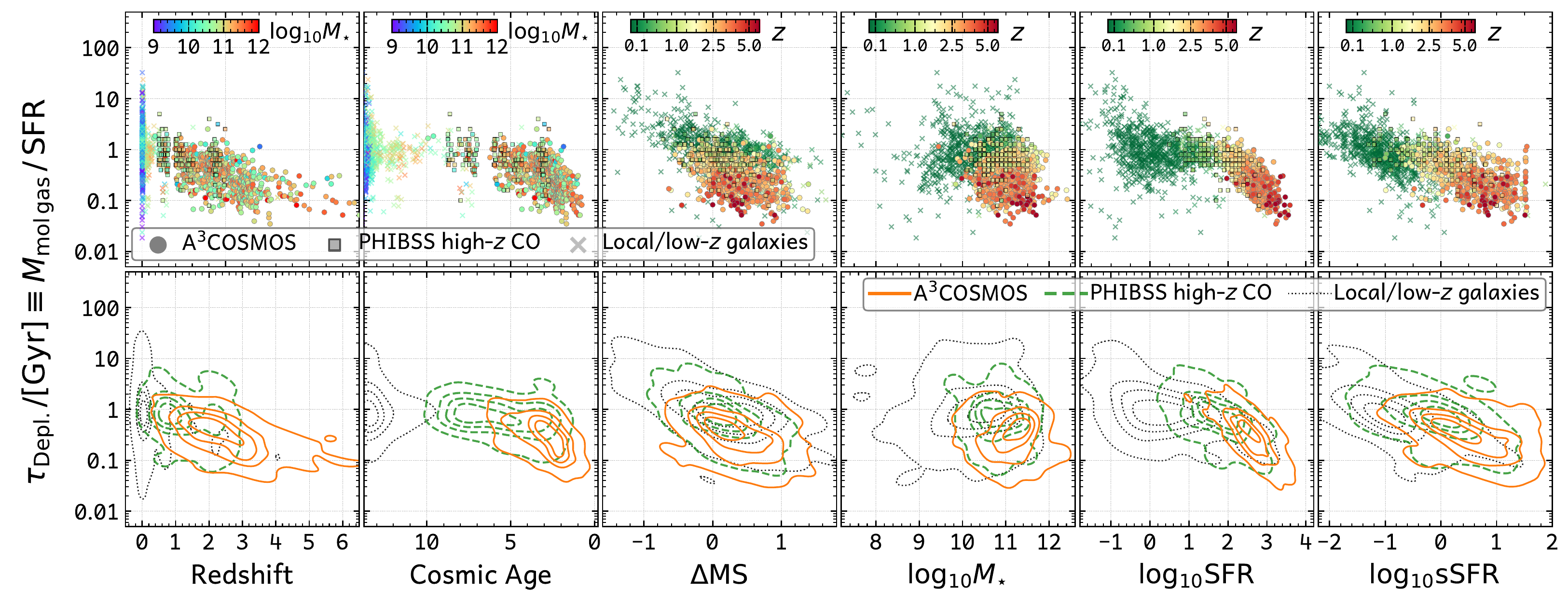}
\vspace{-2.5ex}
\caption{%
Molecular gas depletion time $\tauDepl$ versus various galaxy properties ({\em from left to right}): redshift, cosmic age, $\DeltaMS$, $\Mstar$, $\SFR$ and sSFR, respectively (see Sect.~\ref{Section_Galaxy_Properties}). 
For each distribution we show two vertically-adjacent panels: the upper panel shows the scatter plot and the lower panel provides density contours. 
In the scatter plots, data points are color-coded by $\Mstar$ in the first two top panels and by redshift (in $\log_{10}(1+z)$ scale but tick labels are $z$) in the remaining panels. 
In the contour plots, we divide the sample into three main subsamples: orange solid contours represent the A$^3$COSMOS galaxies ($0.5 \lesssim z \lesssim 6$), green dashed contours the PHIBSS\,1\&2 galaxies ($0.5 \lesssim z \lesssim 2$)
and gray dotted contours are all other local/low-$z$ galaxies. 
\label{Plot_z_tauDepl_multi_panel}
}
\vspace{2ex}
\includegraphics[trim=0 0 0 0, width=1\linewidth]{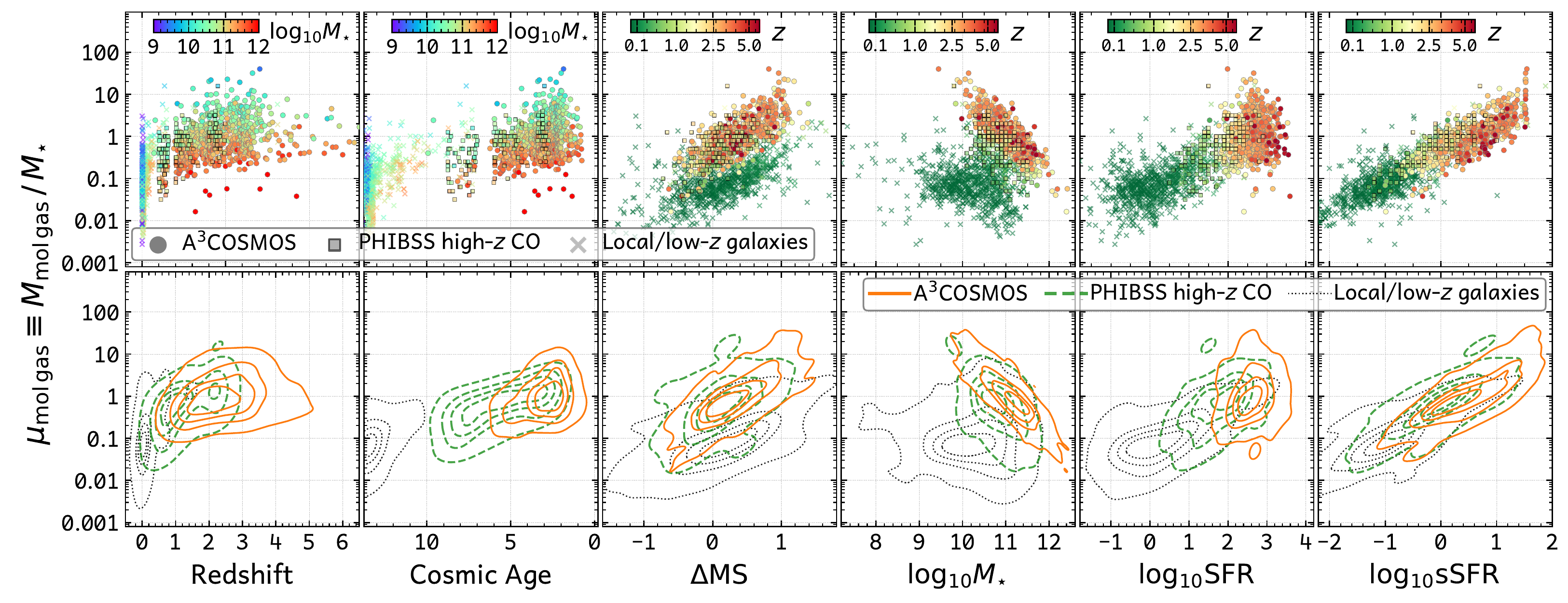}
\vspace{-2.5ex}
\caption{%
Molecular-gas-to-stellar-mass ratio $\deltaGas$ as a function of various galaxy properties. See Fig.~\ref{Plot_z_tauDepl_multi_panel} caption for the description of data points and contours. 
\label{Plot_z_deltaGas_multi_panel}
}
\vspace{2ex}
\end{figure*}

\vspace{0.25truecm}
\subsection{Sample distribution across the MS}
\label{Section_Results_z_DeltaMS}

In Fig.~\ref{Plot_z_DeltaMS} we show the specific star formation rate ($\mathrm{sSFR}$) versus redshift distribution, where sSFR is normalized by the MS sSFR of \citet{Speagle2014} at each redshift. The left and right panels have the same data points but have different X-axis scales and color schemes: the X-axis (redshift) is logarithmic in the left panel and only the complementary samples are color coded, while the redshift is linear in the right panel and only A$^3$COSMOS data points are color coded. Whereas our A$^3$COSMOS sample primely populates the $z>1$ regime, the complementary samples provide coverage at $z<1$. However, we do notice that the MS is not well sampled at $0.1 \lesssim z \lesssim 0.5$ and $z>1$. Only the most massive A$^3$COSMOS galaxies ($\log_{10}(\Mstar/\Msun)\gtrsim11.5$) sample well the MS, while less massive ones lie above. 

The majority of $1 \lesssim z \lesssim 2$ complementary sample sources are from the PHIBSS\,1\&2 surveys (\citetalias{Tacconi2018}) and \cite{Kaasinen2019}. Compared to the A$^3$COSMOS galaxies, they are slightly less massive (see Table~\ref{Table_galaxy_samples}), thus \citetalias{Tacconi2018} sources are able to represent the $\log_{10}(\Mstar/\Msun)\sim10-11$ MS while the A$^3$COSMOS galaxies are probing the $\log_{10}(\Mstar/\Msun)\sim11-12$ MS at $z>1$. 

We also notice that only very low redshift ($z \lesssim 0.03$) galaxies cover the $\log_{10}(\Mstar/\Msun)\sim9-10$ parameter space. In this low stellar mass range, the metallicity-dependent $\alphaCO$ might be more uncertain and so are the estimated molecular gas masses. However, this regime is important in understanding molecular gas scaling relations as shown in latter sections, thus here we still fit these galaxies from the complementary samples.

\begin{figure*}[htb]
\centering
\begin{interactive}{js}{Plot_z_Mstar_deltaGas_DeltaMS_in_3D.x3d}
\includegraphics[width=0.75\textwidth]{{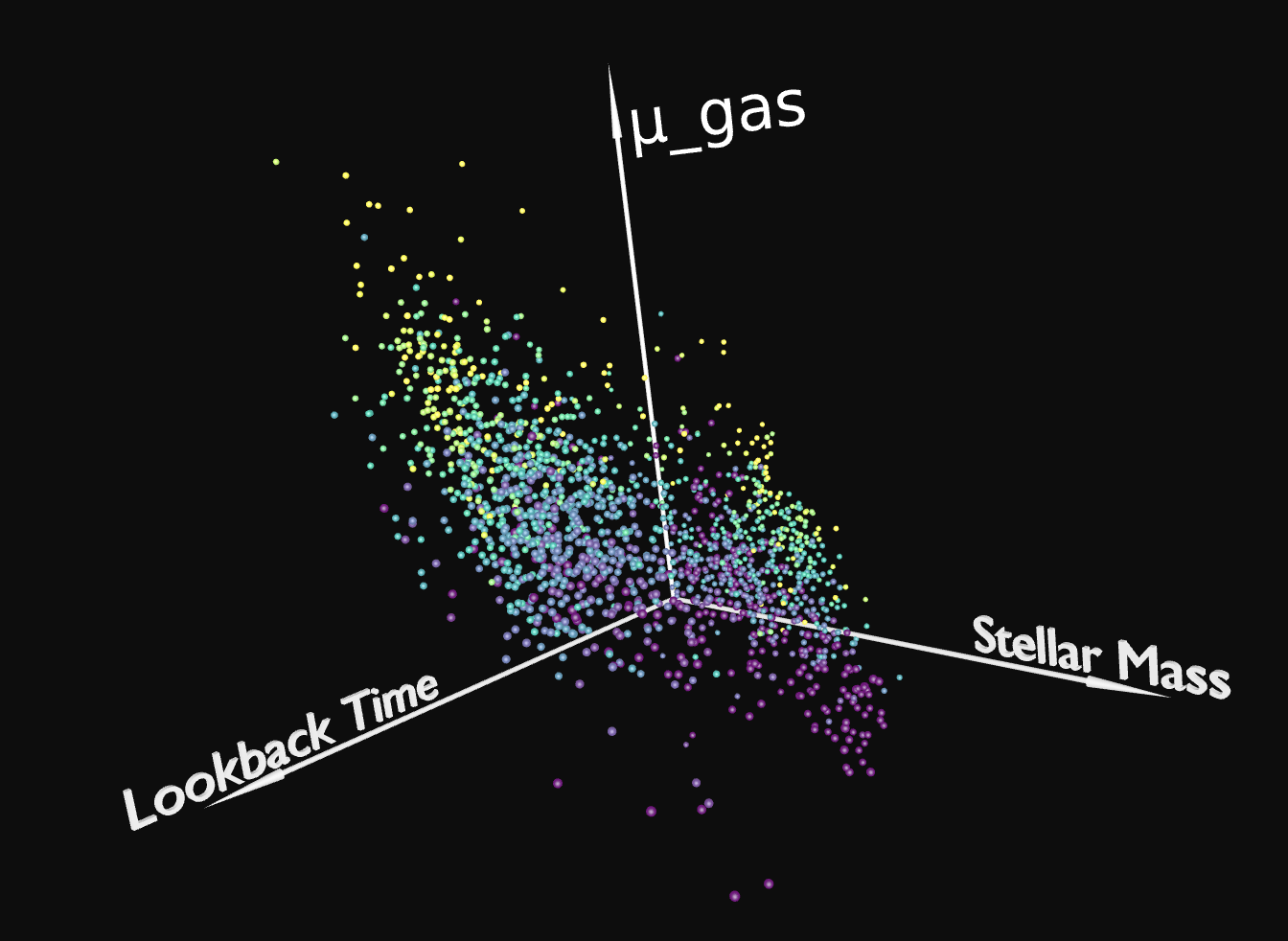}}
\end{interactive}
\vspace{2.0ex}
\caption{%
Three-dimensional view of the molecular gas to stellar mass ratio $\deltaGas$ as a function of lookback time and stellar mass. Data points are colored by their offsets from the MS ($\DeltaMS$): purple is below the MS, green is on the MS and yellow is above the MS. 
{\it (This interactive figure is only available in the online journal, where the reader can rotate/zoom/shift the view to see how the data points are distributed in 3D.)}
\label{Plot_z_deltaGas_3D}
}
\end{figure*}

\vspace{0.25truecm}
\subsection{Correlating molecular gas fraction and depletion time to galaxy stellar mass and star formation properties}
\label{Section_Results_deltaGas_tauDepl}

Here we study the scaling relations for the two most important molecular gas properties: molecular gas depletion time, $\tauDepl \equiv \Mmolgas / \SFR$, and molecular gas to stellar mass ratio, $\deltaGas \equiv \Mmolgas / \Mstar$. 
In Figs.~\ref{Plot_z_tauDepl_multi_panel}~and~\ref{Plot_z_deltaGas_multi_panel}, we show their distributions versus other galaxy properties, i.e., redshift, cosmic age, offset to the MS ($\DeltaMS$; using the MS of \citealt{Speagle2014}), $\Mstar$, SFR and sSFR. 
Two diagrams are shown for each distribution: a scatter plot ({\it upper panels}) and a contour plot ({\it lower panels}). 
In the contour plot, we show three sets of contours representing the data densities of A$^3$COSMOS galaxies (orange), PHIBSS\,1\&2 $0.5 \lesssim z \lesssim 2$ galaxies (green) and all other local/low-redshift galaxies (gray), respectively. 

The molecular gas depletion time $\tauDepl$ spans about one order of magnitude in the high-redshift range from $z\sim1$ to 6, but has more than two orders of magnitude variation at $z\sim0$. The latter can be due to the strong correlations with either $\DeltaMS$, SFR and/or sSFR, as shown in the corresponding scatter plots in Fig.~\ref{Plot_z_tauDepl_multi_panel}. 

However, as the SFR and sSFR have redshift dependency, and the $\Mstar$ distribution is biased differently from low to high redshift, it is unclear from just this figure which galaxy property mostly determines $\tauDepl$. There is even a break or turn-over feature in the cosmic age and SFR versus $\tauDepl$ panels, which is likely caused by the selection bias at $z>3$ where we only cover the most massive galaxies ($\log_{10} \Mstar \sim 11-12$). In the intermediate redshift range ($1 \lesssim z \lesssim 3$), the A$^3$COSMOS and PHIBSS\,1\&2 surveys' galaxies have very similar distributions as can be seen in the contour plots. 

Similar plots are shown in Fig.~\ref{Plot_z_deltaGas_multi_panel} for the molecular-gas-to-stellar mass ratio, $\deltaGas$. Compared to gas depletion time distributions, $\deltaGas$ has a nearly three orders of magnitude variation from local to high redshift, and even at $z>1$ the variation is still as large as two orders of magnitude. From the first two panels, we see a moderate redshift evolution and a strong dependency on stellar mass, respectively. $\deltaGas$ also exhibit a strong dependency on $\DeltaMS$, but local galaxies are systematically offset from the high-redshift ones by nearly one dex. 

As shown in the $\Mstar$--$\deltaGas$ panel, local galaxies and high-redshift galaxies seem to follow different distributions: local galaxies have similar $\deltaGas$ across different stellar mass, while high-redshift ones exhibit a steep slope. However, we caution that this is likely an artifact of the high-redshift sample selection using submm data, as the submm selection is similar to an SFR-selection or a dust-mass-selection, picking up massive MS galaxies and less massive but starbursty galaxies (see Fig.~\ref{Plot_z_DeltaMS}). 

The last two columns of Fig.~\ref{Plot_z_deltaGas_multi_panel} show clear and tight correlations between $\deltaGas$ and SFR and sSFR. Local, PHIBSS\,1\&2 intermediate-redshift and A$^3$COSMOS higher-redshift galaxies form a contiguous distribution from SFR~$\sim0.1$ to $>$1000$\;\Msyr$ and sSFR~$\sim0.01$ to $\sim50\;\mathrm{Gyr}^{-1}$. This is likely a combined effect of the SFR or sSFR evolution and the evolution of molecular gas content. 

The last scatter plot, when canceling out the $\Mstar$ term in both axes, is equivalent to the $\Mmolgas$ versus SFR correlation, i.e., Kennicutt-Schmidt law (\citealt{Schmidt1959,Kennicutt1998SFL}). A log-log space linear fitting gives a slope of $\sim0.7$ with a scatter of 0.3\,dex, consistent with \cite{Kennicutt1998SFL}, as well as the slope of $\sim0.8$ as measured by \cite{Sargent2014} for MS and strong starbursts separately (see also Sect.~\ref{Section_Results_SF_Law}).

\begin{figure*}[htb]
\centering%
\includegraphics[width=0.75\textwidth]{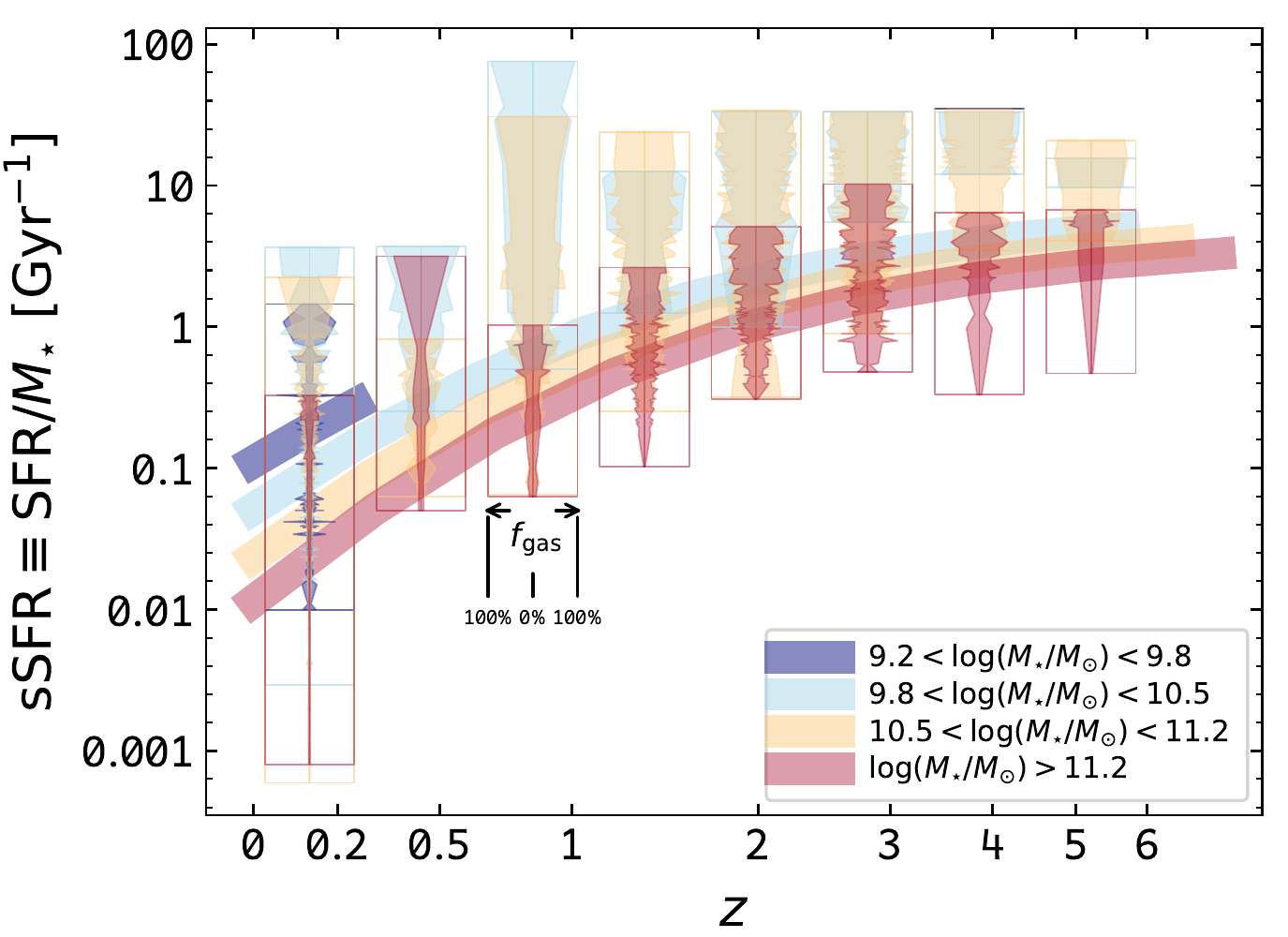}
\caption{%
Galaxies' redshift and specific star formation rate (sSFR $\equiv \SFR / \Mstar$) and gas fraction ($\fmolgas$; Eq.~\ref{Equation_mugas}). We divide our sample into four stellar mass ranges as labeled at the bottom-right corner. 
The colored curves represent the evolution of the galaxy MS sSFR (using \citet{Speagle2014} MS) for each subsample. 
Data points of each subsample are binned by redshift, and in each bin (i.e., each box in the figure) the sSFR versus $\fmolgas$ distribution is shown in a symmetric vertical histogram style (i.e., a ``spindle'' diagram). The horizontal axis indicates $\fmolgas$ in the following way: $\fmolgas=0\%$ when the width of the histogram is zero (i.e., a thin line at the bin center), and $\fmolgas=100\%$ when the width equals the box width. The vertical position in each box shares the same Y-axis of the whole figure, i.e., corresponds to sSFR. 
See discussion in Sect.~\ref{Section_Results_z_sSFR}. 
\textit{The complete figure set (5 images) is available in the online journal, where each stellar mass bin is shown individually for better readability.}
\label{Plot_z_sSFR_gas_evolution}%
}
\vspace{2ex}
\end{figure*}

Again, while Figs.~\ref{Plot_z_tauDepl_multi_panel}~and~\ref{Plot_z_deltaGas_multi_panel} show that observational data from a variety of samples span a wide range of parameters space and are consistent where they overlap, we caution that not all parameters are independent in these plots and the apparent correlations have degeneracies. Therefore, a high-dimensional-space fitting to the data is important to characterize the relative contribution of each key parameters to the observed gas fraction and depletion time. 
In Fig.~\ref{Plot_z_deltaGas_3D} (online-only), we show our data points in three-dimensional (3D) space to better illustrate the complexity. 
And we perform such a high-dimensional-space function fitting in the next section.

\begin{figure*}[htbp]
\centering%
\includegraphics[width=0.55\textwidth]{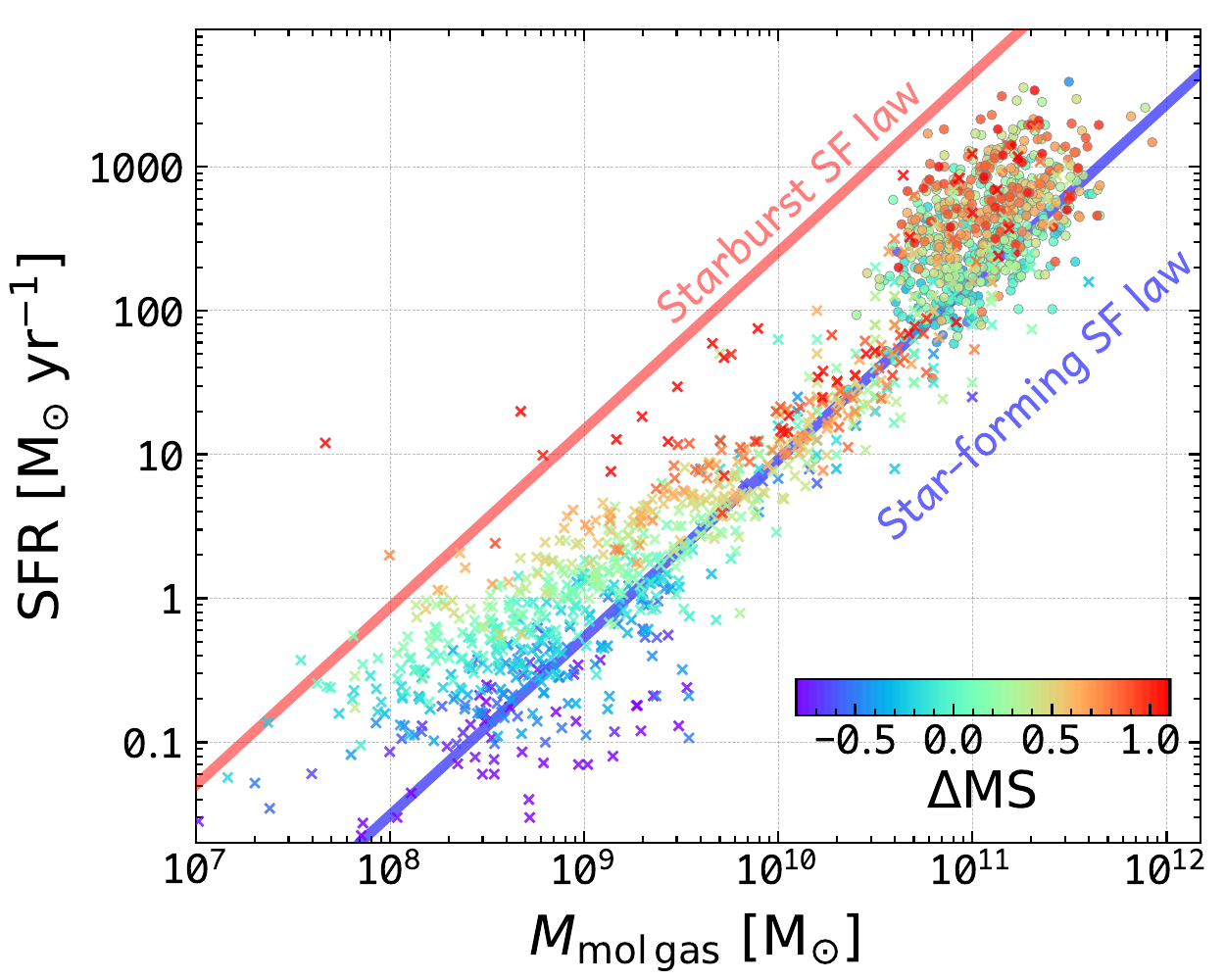}
\vspace{0.5ex}
\caption{%
Galaxy $\Mmolgas$--SFR relation, i.e., star formation law. Data points are color-coded by galaxies' offsets from the $\Mstar$--SFR MS ($\DeltaMS$; using \citealt{Speagle2014} MS). Solid circles are A$^3$COSMOS galaxies and crosses represent all other galaxies from the complementary samples. The blue and red lines represent, respectively, the star formation laws from \cite{Sargent2014} for normal star-forming galaxies and extreme starbursts with a factor $\sim$15 offset. 
See discussion in Sect.~\ref{Section_Results_SF_Law}. 
\label{Plot_SF_Law}%
}
\vspace{2ex}
\end{figure*}

\vspace{0.25truecm}
\subsection{Composite view of galaxy gas fraction and MS evolution}
\label{Section_Results_z_sSFR}

We show in Fig.~\ref{Plot_z_sSFR_gas_evolution} the composite view of the distributions among the four parameters: gas fraction, redshift, stellar mass and SFR. The first-level information in the figure is the sSFR evolution of our galaxies binned by redshift and stellar mass (curves are the \citet{Speagle2014} MS). The second-level information is that in each redshift and stellar mass bin (the boxes in the figure), the horizontal spanning represents the gas fraction $\fmolgas$ (Eq.~\ref{Equation_mugas}; 0\% to 100\% from bin center to edges; shown symmetric for illustration purpose) and the Y position is still sSFR as indicated by the global Y-axis. We can see that in the high-redshift bins if a galaxy has a high sSFR in each box, it spans more, meaning a higher gas fraction.  

Since all data in these boxes share the same Y-axis, the sSFR of the data can be directly read off from the figure and compared to the MS curves. The inhomogeneity of our sample is obvious in the less-massive galaxy bins (i.e., blue and yellow boxes) extending one to two dex above their corresponding MS, while the most-massive galaxies (i.e., the red boxes) merely extend more than one dex above the MS. 

To summarize, with this ``spindle'' diagram, we can more clearly see that:
\begin{enumerate}[label=(\alph*), topsep=0pt, noitemsep]
\item At a fixed redshift, more-massive galaxies have both lower sSFR and gas fraction than less-massive ones.
\item At a fixed redshift, galaxies that lie further above the MS exhibit higher gas fractions ($\fmolgas$ approaching 100\% for the ones with lowest mass and highest sSFR). 
(Yet such a trend is debated for individual or small ($\sim10$) samples of strong starburst galaxies ($\DeltaMS>0.6$), e.g., \citealt{Silverman2015CO,Silverman2018CO}.)
\item Similar to the sSFR evolution, gas fraction evolves with redshift: galaxies of similar stellar mass and distance to the MS but at an earlier cosmic time tend to have a higher gas fraction.
(These trends have been known anecdotally for more than ten years, e.g., \citealt{Daddi2008BzK,Daddi2010BzK}, \citealt{Tacconi2008,Tacconi2010,Tacconi2013}, \citealt{Genzel2010}, \citealt{Magdis2012LBG}, to name a few, but they have only been quantified recently with sufficiently large samples as presented here.)
\end{enumerate}

\begin{figure*}[htb]
\centering
\includegraphics[width=1\linewidth]{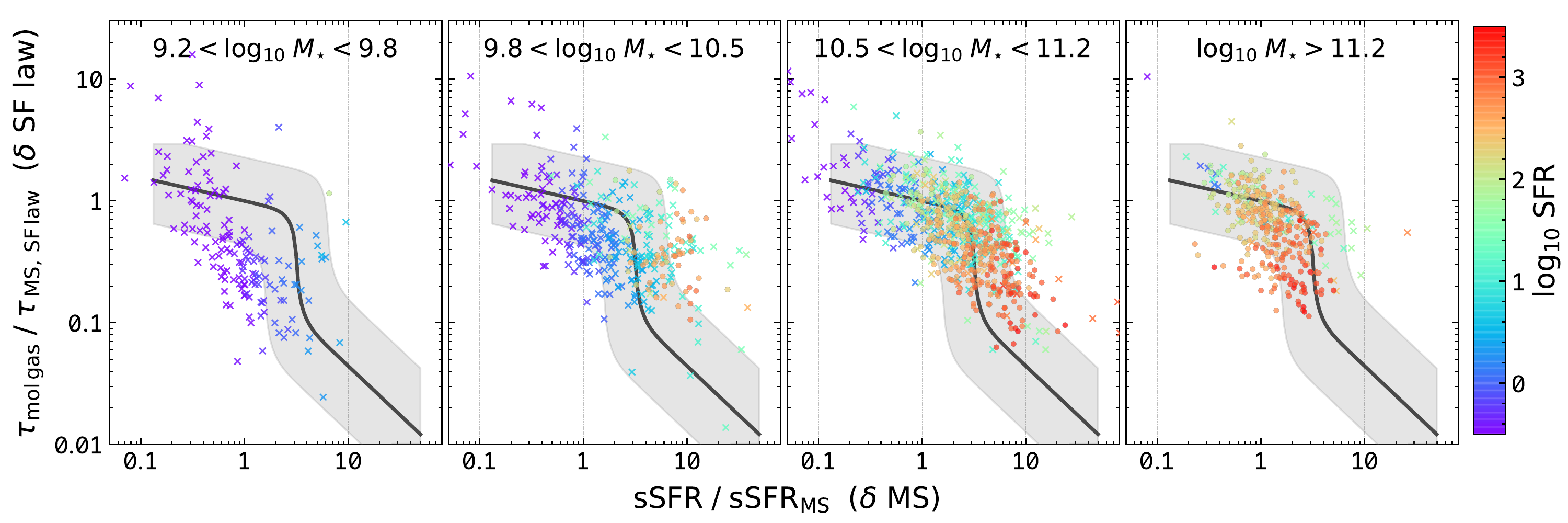}
\vspace{-2.5ex}
\caption{%
Correlation between galaxies' offsets from the star formation law (as indicated by the Y-axis $\tau_{\mathrm{mol\,gas}} / \tau_{\mathrm{mol\,gas,\,MS\,SF\,law}}$) and from the galaxy MS (as indicated by the X-axis $\mathrm{sSFR} / \mathrm{sSFR_{MS}}$) for four stellar mass bins as labeled. We adopt the \cite{Sargent2014} SF law and \cite{Speagle2014} MS (adopting a different MS from \citealt{Sargent2014} will only result in small shift of $-0.16$, $-0.24$, $-0.17$ and $-0.05$~dex in X direction (and smaller offsets in Y direction) for the four panels, respectively). 
Symbol shapes are the same as in Fig.~\ref{Plot_SF_Law}, and are color-coded by $\mathrm{log}_{10}\,\mathrm{SFR}$. 
The black line is the predicted median trend from the two-star-formation-mode (2-SFM) framework by \cite{Sargent2014}. The gray shaded area indicates a 0.3~dex scatter in both X and Y. 
\label{Plot_DeltaMS_Delta_SF_Law}
}
\vspace{2ex}
\end{figure*}

\vspace{0.25truecm}
\subsection{Linking to the galaxy star formation law}
\label{Section_Results_SF_Law}

The star formation law (or Kennicutt-Schmidt law; \citealt{Schmidt1959}; \citealt{Kennicutt1998SFL}; hereafter SF law) describes the correlation between molecular gas mass and star formation rate and has an empirical form of $\SFR = A \times \Mmolgas^{N}$, with a slope $N \approx 1.4$ in the log-log space (\citealt{Kennicutt1998SFL}; \citealt{Gao2004b}). It is physically motivated by the fact that star formation is fueled by molecular gas. However, galaxies show a large scatter in the $\Mmolgas$--$\SFR$ plane, and some galaxies like local ultra-luminous infrared galaxies (ULIRGs; e.g., \citealt{Sanders2003}) and bright high-redshift submm galaxies (SMGs; e.g., \citealt{Smail1997,Blain2002}) are more than a one dex offset from ``normal'' star-forming galaxies. This is also referred to as the bimodal SF law (e.g., \citealt{Daddi2010SFL}; \citealt{Genzel2010} for a strictly bimodal scenario; and \citealt{Sargent2014} for a continuous dichotomy between ``normal'' star-forming and starburst galaxies). 
However, why these galaxies are offset from the normal star-forming SF law and whether they are also starbursts in the MS relations is still poorly explored. 
Given the popular assumption (or intense debate) on the main-sequence/starburst dichotomy and bimodality of SF laws, e.g., in analytic galaxy modeling (\citealt{Sargent2012,Sargent2014,Bethermin2012,Bethermin2017Model}), and observational studies (\citealt{Daddi2010SFL,Genzel2010,Silverman2015CO,Silverman2018CO,Elbaz2018,Cibinel2019}), we investigate these two topics with our large A$^3$COSMOS and compiled sample and present how current models are fitting the data.

Fig.~\ref{Plot_SF_Law} shows the correlation between SFR and $\Mmolgas$ for all galaxies in this work. Data points are color-coded by $\DeltaMS$, and the A$^3$COSMOS and complementary samples are distinguished by different symbols (circle and cross, respectively).  
For A$^3$COSMOS galaxies with large SFR ($\sim100-3000\;\Msyr$) and $\Mmolgas$ ($\sim5\times10^{10}-5\times10^{11}\;\Msun$), a higher $\DeltaMS$ means more deviation from the normal star-forming SF law (see the blue line in Fig.~\ref{Plot_SF_Law}; adopted from \citealt{Sargent2014}). The strongest starbursts with more than one dex offset from the MS show a $0.43$~dex (median) offset from the star-forming SF law, while the offset for MS galaxies ($\DeltaMS<0.5$) is only $-0.12$~dex (median). Considering that the A$^3$COSMOS sample does not sample well the below-MS region, the $0.12$~dex offset does not prevent us from drawing the conclusion that MS galaxies also follow the normal star-forming SF law.

However, the starbursts which lie significantly above the MS ($\DeltaMS \sim 1$) seem to behave differently between the high-redshift and low-redshift/local samples. High-redshift starbursts do not show large enough offsets to reach the starburst SF law as indicated by the red line in Fig.~\ref{Plot_SF_Law}, which is offset by about one dex from the star-forming galaxies' SF law. This is also recently found by CO observations of a small sample of 12 strong starbursts at $z\sim1.5$ by \cite{Silverman2015CO,Silverman2018CO}. 
Meanwhile, some low-redshift/local MS starbursts with $\DeltaMS\sim0.5$ are able to reach the red line, and the trend between $\DeltaMS$ and the offset to the star-forming SF law is more clear there. 

In Fig.~\ref{Plot_DeltaMS_Delta_SF_Law}, we more clearly illustrate the correlation between galaxies' offsets to MS and SF law. The X-axis, $\mathrm{sSFR} / \mathrm{sSFR_{MS}}$, represents the offset to the MS, with $\mathrm{sSFR_{MS}}$ computed following \cite{Speagle2014}. The Y-axis, $\tau_{\mathrm{mol\,gas}} / \tau_{\mathrm{mol\,gas,\,MS\,SF\,law}}$, represents the offset to the star-forming galaxies' SF law, where $\tau_{\mathrm{mol\,gas,\,MS\,SF\,law}} \equiv M_{\mathrm{mol\,gas,\,SF\,law}} / \SFR_{\mathrm{MS}} = \alpha \times \SFR^{\beta} / \SFR_{\mathrm{MS}}$, and the $\alpha$ and $\beta$ coefficients are taken from \cite{Sargent2014}. 
Galaxies are binned into four panels by their stellar masses in Fig.~\ref{Plot_DeltaMS_Delta_SF_Law}. Data points are color-coded by SFR. Model-predicted curves from \cite{Sargent2014} are shown for comparison. Their model, named the two-star-formation-mode (2-SFM) model, assumes that galaxies have two modes of star formation --- a MS mode and a starburst mode. MS galaxies (e.g., $\mathrm{sSFR} \lesssim 3\times\mathrm{sSFR_{MS}}$) obey the SF law with a Galactic-like $\alphaCO$, while starbursts with sSFR above the MS (e.g., $\mathrm{sSFR} \gtrsim 3\times\mathrm{sSFR_{MS}}$) are shifted toward the starbursts' SF law and they also have a much lower $\alphaCO$. The shift in the SF law plane happens most rapidly when the sSFR increases from $\sim3$ to $\sim4\times$ the MS's sSFR (see Fig.~9 of \citealt{Sargent2014}), thus causing the steep model turnover seen in Fig.~\ref{Plot_DeltaMS_Delta_SF_Law}. 

The data are more complicated than what the 2-SFM model predicts. Galaxies in the lowest mass bin ($\logMstar<9.8$) are below the model-predicted curve, while in the mid-stellar-mass bins ($\logMstar\sim9.8-11.2$) some galaxies are above it. The turnover is likely seen in the two higher mass bins ($\logMstar>10.5$ whereas it is less obvious in the two lower mass bins. The difference can not be explained by the calibration of the MS because of the reasonably good agreement between MS calibrations (see Fig.~\ref{Plot_DeltaMS_Delta_SF_Law} caption). The molecular gas masses for the lowest-mass galaxies, which are mostly from complementary samples, are calculated via a metallicity-dependent $\alphaCO$ (see Sect.~\ref{Section_Complementary_Galaxy_Samples}), therefore, $\alphaCO$ seems to be not strong biased. 
While their SFRs are derived using optical photometry and lack far-IR data, they intrinsically have a low metallicity and are dust poor, thus the lack of far-IR/mm should not introduce a significant bias. 
Unfortunately, observational evidence is still scarce. 
\cite{Coogan2019} presented CO non-detections for five low-mass ($\langle\logMstar\rangle=9.8$) galaxies at $z\sim2$, resulting in an upper limit on their gas depletion times of $<0.8\,$Gyr, or $\tau_{\mathrm{mol\,gas}}/\tau_{\mathrm{MS,\,SF\,law}}<0.6$ (assuming a Galactic $\alphaCO$), in agreement with our findings. 
If the difference between data and model is truly significant, then it implies that low-mass ($\logMstar \lesssim 10.0$) MS galaxies might follow a different SF law with $3\times$ faster molecular gas depletion than higher-mass MS galaxies. But this is yet to confirm with more observations. 

In the other three higher-mass bins, from MS to starburst regime, we find good agreements between the data and model for galaxies close to and below the MS, meaning again that MS galaxies also obey the SF law. 
Nevertheless, a number of strong starburst galaxies show slower gas depletion (longer gas depletion times) than they should have according to the model. 
The majority of these strongest starburst outliers with long gas depletion times are from the complementary samples (e.g., \citealt{Villanueva2017} and \citealt{Combes2013}) with CO observations but without metallicity information. In this case, their $\alphaCO$ values are indirectly inferred from their stellar masses and SFRs (see Appx.~\ref{Section_MZR}). 
\cite{Silverman2015CO,Silverman2018CO} found that a different choice of $\alphaCO$ alters the $\tau_{\mathrm{mol\,gas}}/\tau_{\mathrm{MS,\,SF\,law}}$ ratio from close to one to 0.2 for a starburst galaxy with $\DeltaMS\sim1$~dex (see their Fig.~8). 
Therefore it is still unclear how well the molecular gas masses (or stellar mass) can be constrained in these strongest starbursts (more detailed multi-line gas studies are needed, e.g., with RJ-tail dust continuum plus multi-$J$ CO [e.g., \citealt{Liudz2015}] plus other tracers, to settle this issue). 

We also caution that our high-redshift, intermediate-mass ($\log_{10}\Mstar \sim 10 - 11$) sample has a strong bias toward a higher $\DeltaMS$, thus we sample better the region above the model curve than below it. This sample bias is less significant for the most massive bin ($\log_{10}\Mstar \gtrsim 11$) where we most clearly see the turnover. 

To summarize the link between the MS and the SF law, we find that: 
{\it (a)} massive ($\log_{10}\Mstar \sim 10 - 12$) MS galaxies obey the star-forming galaxies' SF law; 
{\it (b)} from the MS to $\DeltaMS\gtrsim1$, galaxies start to deviate from the star-forming galaxies' SF law toward the starbursts' SF law, with a rapid change at $\DeltaMS\sim0.4-0.6$~dex roughly in agreement with the 2-SFM prediction; 
{\it (c)} low-mass ($\log_{10}\Mstar \lesssim 10$) galaxies appears to have systematically shorter gas depletion times and even the MS ones do not obey the star-forming galaxies' SF law. 

These details will likely stimulate further refinement of the popular models and observing strategies.

\vspace{0.25truecm}
\subsection{Intermediate summary on the advantage and caveats of this sample}

In the previous sections, we illustrated the wide dynamical range of our sample. Such a data set is the largest sample for the study of gas scaling relation and its evolution to-date, and will grow with future processing of the ALMA archive in the COSMOS deep field under A$^3$COSMOS. The distribution of our sample in the ($z$,$\Mstar$,$\DeltaMS$) high-dimensional space is roughly contiguous from $(z,\,\logMstar,\,\DeltaMS) \sim (0.0,\,9.5,\,-1)$ to $\sim(5.0,\,12.0,\,+1)$. The gas mass calibrations for the CO and dust sub-samples are in good agreement where they overlap in the parameter space.

Nevertheless, the comprehensive presentation of our data set in previous figures also reveals that the sample is non-uniformly distributed and only partially covers the full parameter space. Our sample is biased toward submm-detected (i.e. IR-bright), massive high-redshift galaxies, as well as CO-detected (gas-rich), massive local/low-redshift galaxies. The impact of such sample biases on the results are hard to quantify with the current dataset.  
Stacking ALMA data for sufficient numbers of faint galaxies with similar properties can help to cover additional portions of the parameter space and will be presented in future work. 
Meanwhile, low-$J$ CO and RJ-tail dust observations toward samples covering the less-probed areas of the parameter space hold the key to further improve such studies.

\vspace{1truecm}
\section{Characterizing Galaxy Molecular Gas Scaling Relation via Functional Fitting}
\label{Section_Fitting_Multi_Variate_Function}

Through our previous discussion of galaxy properties (molecular gas to stellar mass ratio $\deltaGas$, molecular gas depletion time $\tauDepl$, stellar mass $\Mstar$, star formation rate $\SFR$, and redshift $z$ or the corresponding cosmic age $t_{\mathrm{cosmic age}}$), we can already see the complexity inherent in the their scaling relations. 
In this section, we provide high-dimensional functional fittings to simultaneously quantify the underlying dependencies of $\deltaGas$ and $\tauDepl$ on $z$ (or $t_{\mathrm{cosmic age}}$), $\Mstar$ and $\SFR$.

\begin{figure*}[htbp]
\centering
\includegraphics[width=\textwidth]{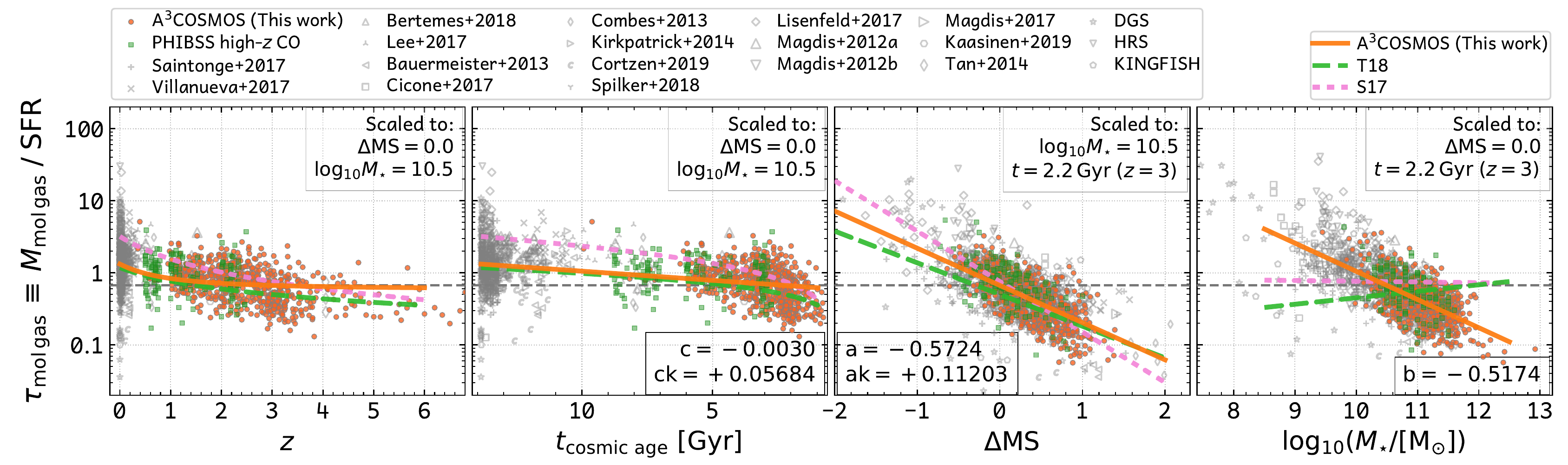}
\includegraphics[width=\textwidth, trim=0 0 0 22mm, clip]{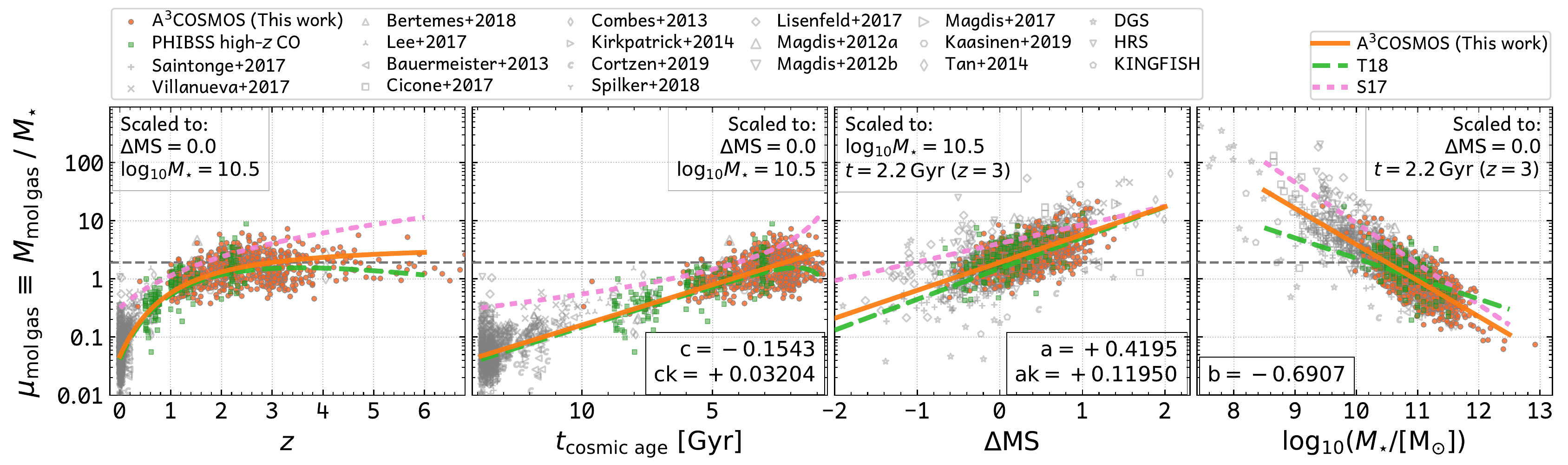}
\vspace{-2.2ex}
\caption{%
Characterizing molecular gas depletion time $\tauDepl$ ({\it upper panels}) and molecular gas to stellar mass ratio $\deltaGas$ ({\it lower panels}) in the functional form of Eq.~\ref{Equation_tauDepl_deltaGas}. From {\it left} to {\it right}, we show $\tauDepl$ versus redshift, $t_{\mathrm{cosmic\,age}}$, $\DeltaMS$ and $\Mstar$, respectively. Data points in each panel are re-scaled using the best-fit function so as to remove the dependency on other parameters and leave only the correlation with the current X-axis parameter (with coefficient(s) labeled at the bottom of each panel). Orange data points are from A$^3$COSMOS, while green ones are from the PHIBSS\,1\&2 surveys (\citetalias{Tacconi2018}) and gray ones are from the literature as listed in Table~\ref{Table_galaxy_samples} and at the top. We distinguish these samples by different symbols in order to better reveal outliers and sample biases against each parameter after removing other parameter-dependencies.
Our best-fit function is shown as the orange solid line in each panel, while the functions from \citetalias{Tacconi2018} (see Eq.~\ref{Equation_Tacconi2018}) and \citetalias{Scoville2017} (see Eq.~\ref{Equation_Scoville2017}) are shown as green dashed and pink dotted lines, respectively. 
\label{Plot_multi_variate_function_fitting}
}
\vspace{2ex}
\end{figure*}

We propose a new functional form which accounts for the different behaviors of galaxies due to their stellar masses seen in the previous figures: 
\begin{equation}
\begin{split}
\log_{10} \tauDepl 
&= (\mathsf{a} + \mathsf{ak} \times \log_{10} (\Mstar/10^{10})) \times \DeltaMS \\
& \; + \mathsf{b} \times \log_{10} (\Mstar/10^{10}) \\
& \; + (\mathsf{c} + \mathsf{ck} \times \log_{10} (\Mstar/10^{10})) \times t_{\mathrm{cosmic\,age}} \\
& \; + \mathsf{d} \\[0.5ex]
(\textnormal{best-fit:} \ 
& \mathsf{a}  = -0.5724, 
\ \mathsf{b}  = -0.5174, \ \\
& \mathsf{c}  = -0.002997, 
\ \mathsf{d}  = +0.02964, \ \\
& \mathsf{ak} = +0.1120 \ \textnormal{and} 
\ \mathsf{ck} = +0.0568) 
\\[1.25ex]
\log_{10} \deltaGas 
&= (\mathsf{a} + \mathsf{ak} \times \log_{10} (\Mstar/10^{10})) \times \DeltaMS \\
& \; + \mathsf{b} \times \log_{10} (\Mstar/10^{10}) \\
& \; + (\mathsf{c} + \mathsf{ck} \times \log_{10} (\Mstar/10^{10})) \times t_{\mathrm{cosmic\,age}} \\
& \; + \mathsf{d} \\[0.5ex]
(\textnormal{best-fit} \ 
& \mathsf{a}  = +0.4195, 
\ \mathsf{b}  = -0.6906, \ \\
& \mathsf{c}  = -0.1543, 
\ \mathsf{d}  = +0.9339, \ \\
& \mathsf{ak} = +0.1195 \ \textnormal{and} 
\ \mathsf{ck} = +0.0320) 
\end{split}
\label{Equation_tauDepl_deltaGas}
\end{equation}
where $t_{\mathrm{cosmic\,age}}$ and $\Mstar$ are in units of Gyr and $\Msun$ respectively.%
\,\footnote{%
See Appx.~\ref{Section_Appendix_MCMC_Corner_Diagram} for the probability distributions of the fitted coefficients.
We also provide a \incode{Python} package for the calculation with our functions: \url{https://ascl.net/code/v/2377}.
}

Here we adopt the MS function from the \#49 fitting of Table~7 of \cite{Speagle2014}, which is their preferred fit (see their abstract and Table~9). This functional form also uses cosmic age as a free parameter (as in our function).%
\,\footnote{%
We remind the reader that in \citet{Speagle2014} the functional fitting with redshift ($\log_{10} (1+z)$) in their Table~8 is not suitable to use at high redshifts, e.g., $z\gtrsim4$. Those fits are much different from the functions with cosmic age. For example, with the same \#49 fitting set, the two MS functions agree only at $z\lesssim 1.5$, while the difference can be 0.5~dex at $z\sim4.3$ (with the MS function with $\log_{10} (1+z)$ being higher).
}
We compared various MS in the literature (e.g., \citealt{Whitaker2014}; \citealt{Sargent2014}; \citealt{Bethermin2015}; \citealt{Schreiber2015}; \citealt{Lee2015}; \citealt{Tomczak2016}; \citealt{Pearson2018}), finding that the \cite{Speagle2014} cosmic age MS function provides the most reasonable fitting (see Appx.~\ref{Section_MS}). 
It can be rewritten in the same style as the above functions as: 
\begin{equation}
\begin{split}
\log_{10} \SFR_{\mathrm{MS}} 
&= \mathsf{b} \times \log_{10} (\Mstar/10^{10}) \\
& \; + (\mathsf{c} + \mathsf{ck} \times \log_{10} (\Mstar/10^{10}) ) \times t_{\mathrm{cosmic\,age}} \\
& \; + \mathsf{d} \\[0.5ex]
\textnormal{where} 
\ & \mathsf{b} = +0.84, 
\   \mathsf{c} = -0.15, 
\   \mathsf{d} = +1.89, \ \textnormal{and} \\
\ & \mathsf{ck} = -0.026, 
\end{split}
\label{Equation_Speagle2014}
\end{equation}

We fit our new functional form to the combined sample in this work, as well as re-fitted both the \citetalias{Scoville2017} and \citetalias{Tacconi2018} functional forms, i.e., described in Eq.~\ref{Equation_Scoville2017} and Eq.~\ref{Equation_Tacconi2018}, respectively. 
We use the \incode{Python} packages \incode{pymc3} and \incode{scipy.optimize.curve_fit} for the fitting\,\footnote{\incode{pymc3} documentation: \url{https://docs.pymc.io/}; and \incode{scipy.optimize.curve_fit}: \url{https://docs.scipy.org/doc/scipy/reference/generated/scipy.optimize.curve_fit.html}.}. 
The former package performs Markov chain Monte Carlo (MCMC) fitting to calculate the probability distribution of the fitting, while the latter one performs least-chi-square minimization to find the best fit. The two algorithms agree very well, and the former one provides better uncertainty estimates for the fitted parameters. 

We list our best-fit parameters in Table~\ref{Table_functions} as well as in Eq.~\ref{Equation_tauDepl_deltaGas}. The parameters fitted by \citetalias{Scoville2017} and \citetalias{Tacconi2018} for their own functional forms are also provided in Table~\ref{Table_functions} for comparison. 

Our re-fitting of the \citetalias{Tacconi2018} function agrees with their original fitting: only the redshift coefficient is slightly changed by about 10\%, which implies that the two fittings are consistent ($<30\%$) at $z<2$ and slightly discrepant at $z\sim3-5$ where our fitting predicts about 30\%-50\% lower gas fractions and shorter depletion times, mainly driven by the new data coverage from this work (their data only covers $z\sim0-3$).

For our new function, the fitted dependencies of gas fraction and depletion time on each parameter are presented in Fig.~\ref{Plot_multi_variate_function_fitting}. We show in each panel the best-fit function curve and the data points with a rescaling to remove the dependencies on other parameters than the current one presented by the X-axis of that panel. This rescaling uses our best-fit result, for example, for the rescaling in the first panel, the gas-to-stellar mass ratio $\deltaGas$ of a galaxy with $\DeltaMS=1$ and $\logMstar=10.5$ will be scaled by $-0.4123\times\DeltaMS$~dex, bringing it down to the MS galaxy level. In this way, each panel only indicates the dependency of our function fitting on the parameter presented by the X-axis.

We also show the original best-fits of \citetalias{Tacconi2018} and \citetalias{Scoville2017} (to their own functional forms, i.e., Eqs.~\ref{Equation_Tacconi2018}~and~\ref{Equation_Scoville2017}, respectively) in Fig.~\ref{Plot_multi_variate_function_fitting}. 
In comparison, our new functional form has a log-linear dependency on cosmic age, therefore $\deltaGas$ and $\tauDepl$ almost flatten beyond redshift $\sim4$ for the same stellar mass and $\DeltaMS$ galaxies. The \citetalias{Tacconi2018} best-fit function predicts a drop at $z\sim4$ in $\deltaGas$, while the \citetalias{Scoville2017} best-fit function predicts $\deltaGas$ to continue increasing with redshift. In the $\DeltaMS$ and $\Mstar$ panels, we also see certain differences, but our function is in between the \citetalias{Tacconi2018} and \citetalias{Scoville2017} ones.

The current fitting still has some minor caveats. For example, in the redshift panel, the limited number of $z>4.5$ data points are mostly below our best-fit function. But as we discussed in Sect.~\ref{Section_Band_conversion}, the band conversion for rest-frame R-J tail dust continuum has a large uncertainty when there are no long-wavelength data, which is the case for $z>4$ galaxies. The test in Appx.~\ref{Section_Appendix_SED_band_conversion} shows that our SED fitting tends to underestimate their true R-J tail dust continuum by a factor of 2--6.

In addition, we tested the stability of our fitting for subsamples of galaxies: (a) only $z<4$ data, and (b) without $z>1$ CO (which are mostly from the PHIBSS 1\&2 surveys from \citetalias{Tacconi2018}), i.e., only using A$^3$COSMOS dust-based data at $z>1$. The tests show that the $z>4$ data and $z>1$ CO data do not statistically bias our fitting results, likely because their numbers are not large enough compared to the full sample. 
The $\chi^2$ information of these test fittings are listed in Table~\ref{Table_function_fittings}, which shows that our proposed functional form in Eq.~\ref{Equation_tauDepl_deltaGas} gives statistically better fits to the data in this work than both the \citetalias{Tacconi2018} (Eq.~\ref{Equation_Tacconi2018}) and \citetalias{Scoville2017} (Eq.~\ref{Equation_Scoville2017}) functions, and that the \citetalias{Tacconi2018} one is better than the \citetalias{Scoville2017} one. This is likely because our function (Eq.~\ref{Equation_tauDepl_deltaGas}) has one more free parameter than the \citetalias{Tacconi2018} function, which further has one more degree of freedom than the \citetalias{Scoville2017} function.

Moreover, we have run our full fitting process for other gas mass calibration methods. 
We find that using the \citetalias{Scoville2017} gas mass calibration, which slightly overpredicts gas masses compared to the H17 calibration, leads to $\lesssim11\%$ changes in the coefficients in Eq.~\ref{Equation_tauDepl_deltaGas}, and results in a $11\%$ shallower $\deltaGas$ versus stellar mass (negative) dependency. This in turn increases the prediction of $\deltaGas$ for main-sequence, $\logMstar\sim10.5$ galaxies by a small amount of about $20\%$ at $z\sim6$. 
On the other hand, using the $\delta_{\mathrm{GDR},\,Z}$ gas mass calibration which tends to underestimate the gas masses, the coefficients change by $\lesssim40\%$. This results in a $\sim40\%$ steeper $\deltaGas$ versus stellar mass (negative) dependency, and consequently lower $\deltaGas$ for main-sequence, $\logMstar\sim10.5$ galaxies at $z\sim6$ by about $50\%$. The scatters in the diagnostic plots similar to those in Fig.~\ref{Plot_multi_variate_function_fitting} are also larger by about 0.06\,dex (e.g., the scatters around the best-fit lines in Fig.~\ref{Plot_multi_variate_function_fitting} are about 0.28--0.30\,dex with the $\alphaRJH17$ calibration, while they are 0.33--0.36\,dex with the $\delta_{\mathrm{GDR},\,Z}$ calibration). 
We also note that the slope of the $\deltaGas$ versus $\DeltaMS$ correlation is much less obviously affected ($\mathsf{a}\sim0.39$--0.41), and the fits close to $z\sim0$ are not obviously affected due to the large number of local/low-$z$ galaxies in our sample with CO-based gas masses. 
These tests show that the choice of gas mass calibration method is not significantly altering our result --- by at most a factor of two at $z\sim6$ and $\logMstar\sim10.5$, and less at lower redshifts.

Finally, we emphasize that, despite the fact that nearly all gas masses for our high-redshift ($z>2.5$) galaxies are dust-based and similarly those for local/low-redshift ($z<1$) galaxies are CO-based, we verified that our results are not significantly biased. 
Specificially we have: (a) excluded all CO-based galaxies and (b) all dust-based galaxies from our fitting. We find that the slopes of both the $\deltaGas$ versus $\DeltaMS$ and $\deltaGas$ versus $\logMstar$ correlation are quite stable within 20\%. However, the trend of the time evolution is significantly driven by the lack of constraining data at either low- or high-redshift: Excluding all CO-based galaxies leads to a factor of 10 higher gas fraction at $z\sim0$, because there is basically no constraint at $z<1$. While excluding all dust-based galaxies gives a factor of 2 higher gas fraction at $z\sim4$--6, due to little constraint at $z>3$.
Taking together the good consistency when fitting the non-redshift-dependent correlations, and the good agreement between CO- and dust-based gas mass calibrations in the literature (see beginning of Sect.~\ref{Section_Molecular_Gas_Mass_Calculation}), it is not only very reasonable to combine the CO- and dust-based samples and but also necessary to achieve sensible results.

\vspace{0.5truecm}
\section{Predictions from The Fitted Gas Evolution Functions}
\label{Section: Predictions from Galaxy Molecular Gas Evolution Functions}

\begin{figure*}[htbp]
\centering
\includegraphics[width=0.835\textwidth]{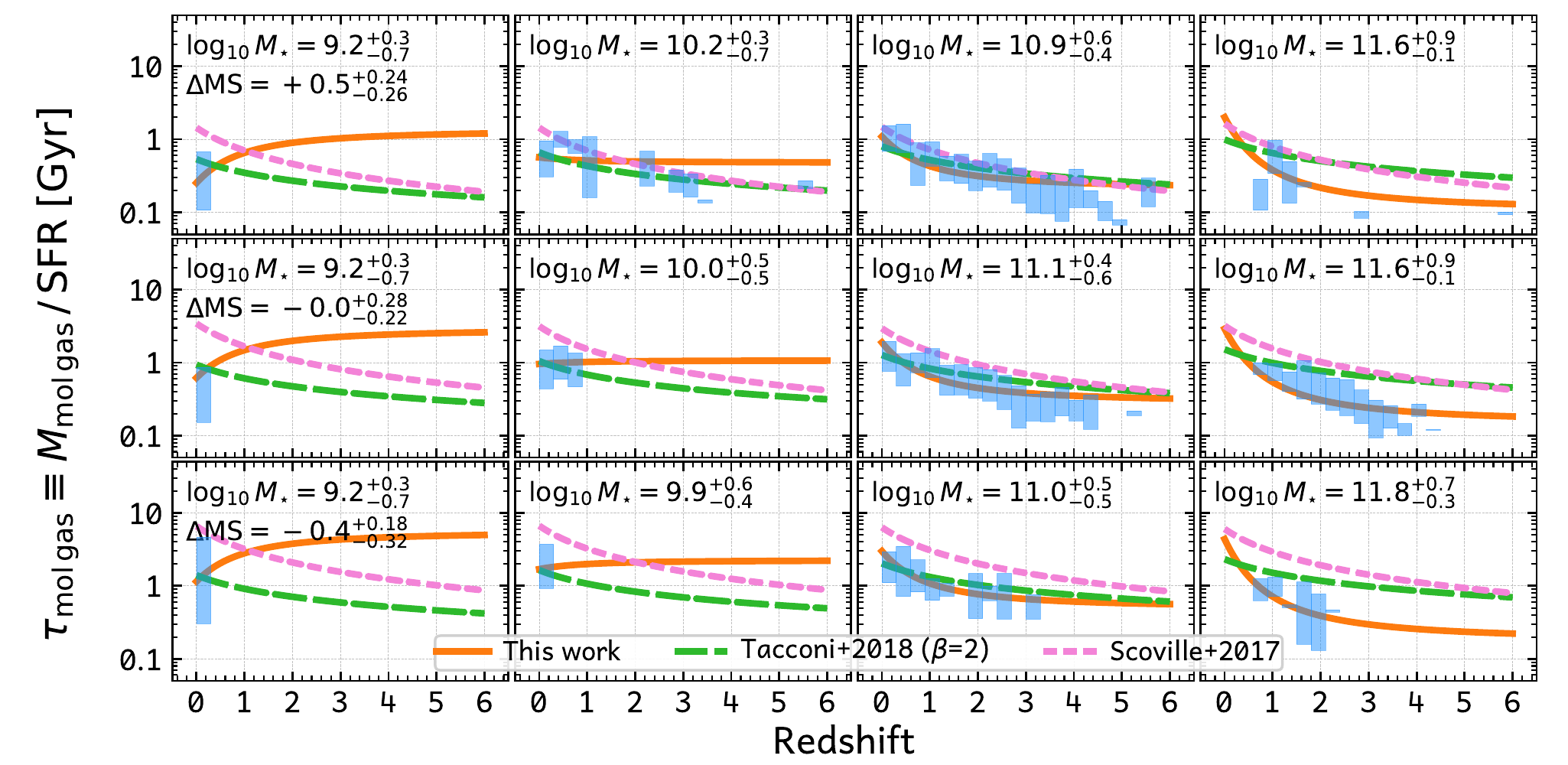}
\vspace{0.5ex}
\caption{%
Evolution of the molecular gas depletion time $\tauDepl \equiv \Mmolgas/\SFR$ (in units of $\mathrm{[Gyr]}$) with redshift in $4\times3$ bins of $\DeltaMS$ and $\logMstar$. From {\it left} to {\it right}, $\logMstar$ increases from 9.0 to 12.0 with a step of 1.0 and bin width of 1.0, and from {\it bottom} to {\it top} $\DeltaMS$ increases from $-0.5$ to $+0.5$ with a step of 0.5 and bin width of 0.5. 
We show the evolution function Eq.~\ref{Equation_tauDepl_deltaGas} from this work and those from \citetalias{Tacconi2018} and \citetalias{Scoville2017} (i.e., their best-fits to Eqs.~\ref{Equation_Tacconi2018} and \ref{Equation_Scoville2017}, respectively) in each panel (see the labels at the bottom). 
These functions are calculated with the mean $\DeltaMS$ and $\logMstar$ of the subsample data available within each bin. 
Blue rectangles represent the $\mathrm{mean}(\tauDepl)\pm1\sigma$ ranges of all galaxies from A$^3$COSMOS and the literature in bins of redshift in each panel.
We caution that this figure does not show the quality of data fitting because data still have variation in $\DeltaMS$ and $\logMstar$ even within each panel. See Fig.~\ref{Plot_multi_variate_function_fitting} for the fitting quality, and see Sect.~\ref{Section_Results_z_tauDepl} for the discussion of this figure. 
}
\label{Plot_z_tauDepl_comparison}
\includegraphics[width=0.835\textwidth]{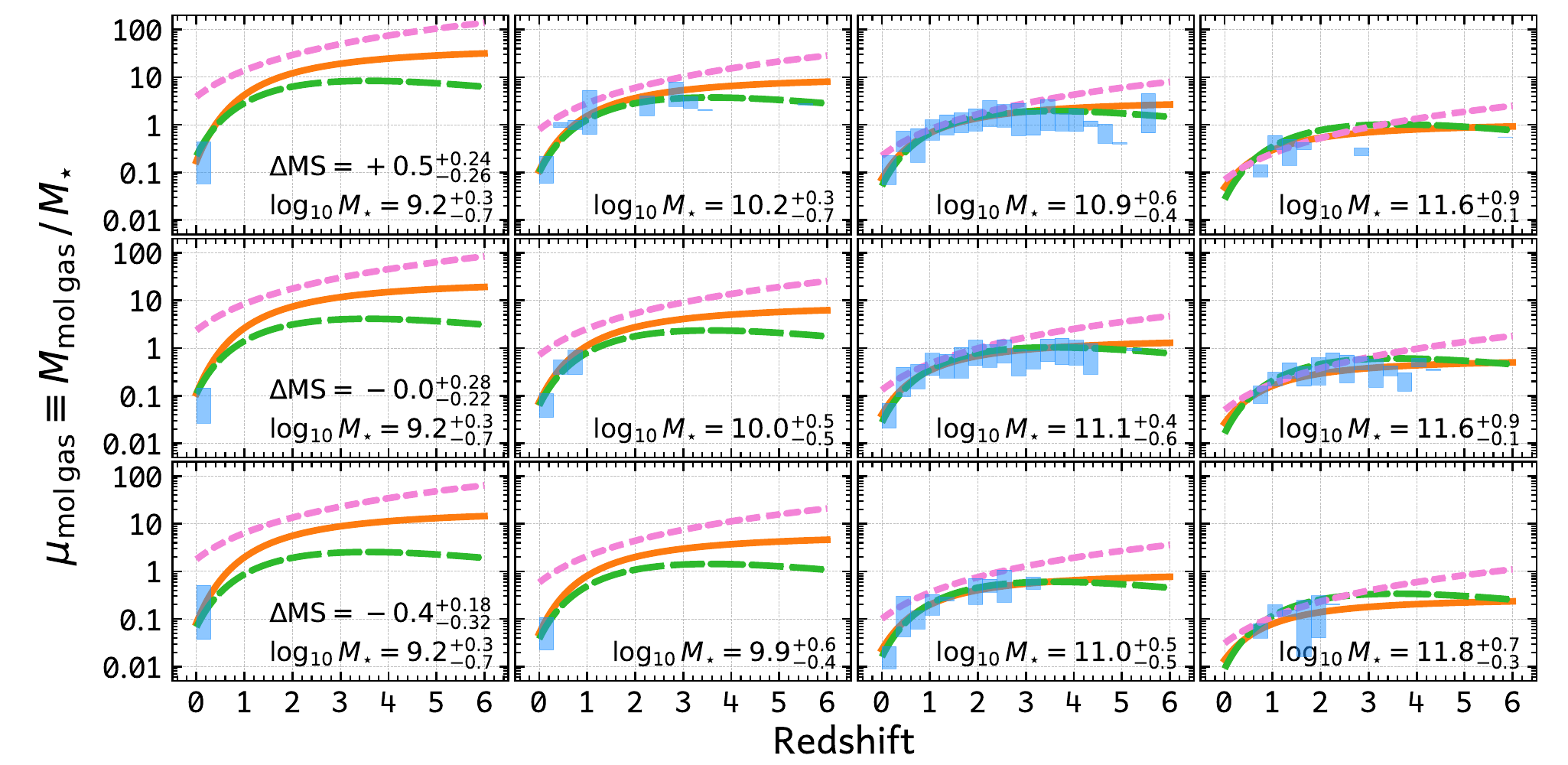}
\vspace{0.5ex}
\caption{%
Analogous to Fig.~\ref{Plot_z_tauDepl_comparison}, but for the evolution of the molecular gas to stellar mass ratio $\deltaGas$. The parameters for each panel are the same as in Fig.~\ref{Plot_z_tauDepl_comparison}. See Fig.~\ref{Plot_z_tauDepl_comparison} caption for the description and Sect.~\ref{Section_Results_z_deltaGas} for the discussion of this figure. 
}
\label{Plot_z_deltaGas_comparison}
\vspace{2ex}
\end{figure*}

\vspace{0.25truecm}
\subsection{Evolution of molecular gas depletion time}
\label{Section_Results_z_tauDepl}

Here we discuss the cosmic evolution of the molecular gas depletion time $\tauDepl$ as predicted by our best-fit function (Eq.~\ref{Equation_tauDepl_deltaGas}) for galaxies in bins of stellar mass and MS offset. 
In Fig.~\ref{Plot_z_tauDepl_comparison}, we bin all our 1,653 galaxies into $4\times3$ bins, with the $\logMstar$ bin center ranging from $9.0$ to $12.0$ (bin width 1.0) and the $\DeltaMS$ bin center ranging from $-0.5$ to $+0.5$ (bin width 0.5). The predictions of our best-fit function are shown as solid lines, while the predictions from \citetalias{Tacconi2018} and \citetalias{Scoville2017} functions (with their fitting) are shown as long- and short-dashed lines, respectively, for comparison. Galaxies in each panel are also binned in small redshift interval so as to show the mean and scatter at each redshift. 

From the figure, we can see that our function behaves differently than the other two functions. The evolution of $\tauDepl$ exhibits a much stronger dependency on stellar mass in our function. For very massive ($\logMstar\sim12.0$) galaxies, our function predicts a factor of about 20 increase in $\tauDepl$ from very early cosmic time to the present, while the \citetalias{Tacconi2018} and \citetalias{Scoville2017} functions predict only a factor of 5--8 increase. Data from this work favors our function in these bins. 
Meanwhile, for low-mass ($\logMstar\sim9.0$) galaxies, our function predicts a reversed evolutionary trend than the \citetalias{Tacconi2018} and \citetalias{Scoville2017} functions. That means, a galaxy with a stellar mass as low as $\logMstar\sim9.0$ has a longer depletion time at an earlier cosmic time, and its star formation speeds up with cosmic age.  Current data in these bins are not sufficient to clearly distinguish which function is better. 
The few CO observations available for local dwarf galaxies (e.g., \citealt{Bolatto2011}, \citealt{Cormier2014}) show that $\tauDepl$ ranges from $0.1$ to $1.0\;\mathrm{Gyr}$ (with SFR$\sim20-30$ to $0.04\;\Msyr$, i.e., from high to low $\DeltaMS$, respectively). These observations still agree with the predictions of our function.

We caution that this figure does not track the evolution of individual galaxies, as they grow in stellar mass and may have rapidly changed $\DeltaMS$ with time. Thus, for example, the flat $\tauDepl$ versus redshift trend for less-massive ($\logMstar\sim10.0$) galaxies seen in the middle columns of the figure does not imply a constant $\tauDepl$ for an individual galaxy across its evolution history --- its stellar mass growth will move it into a higher stellar mass $\tauDepl$ evolution track.
In the $\logMstar\sim10.0$ and $\DeltaMS\sim0.5$ bin, our function does not fit well the $z\sim3-5$ galaxies while the \citetalias{Tacconi2018} and \citetalias{Scoville2017} functions do. This is mainly driven by the small number of low-mass starburst galaxies in this redshift range. As already discussed in Sect.~\ref{Section_Results_SF_Law}, our sample within this range is sparse, biased and the statistics is expected to be less significant. 

If only looking at the function predictions, our function actually provides a coherent picture of galaxy ``down-sizing'' (e.g., \citealt{Cowie1996}; \citealt{Thomas2005}), i.e., more-massive galaxies (possibly in more-massive dark matter halos) evolve earlier than less-massive galaxies. 
Meanwhile, the star formation in the most massive galaxies quickly slows down at redshift 2--3, which probably points to the ``mass-quenching'' effect (e.g., \citealt{PengYingJie2010_SMF}).

Below we also compare the predictions of our functions with other works in the literature.
Our formula predicts that, for local galaxies with 
stellar mass $3 \times 10^{9}$, $3 \times 10^{10}$, $3 \times 10^{11}$ and $3 \times 10^{12} \, \Msun$, 
their $\tauDepl=0.7$, 1.3, 2.6 and 5.0~Gyr, respectively.
In comparison, \cite{HuangMeiLing2014,HuangMeiLing2015} studied about 600 local galaxies from the HERACLES (\citealt{Leroy2009}), ATLAS3D (\citealt{Cappellari2011}; \citealt{Alatalo2013}) and COLD GASS (\citealt{Saintonge2011a,Saintonge2011b}) surveys, and found $\tauDepl = -0.36 \log_{10} \mathrm{sSFR} - 0.14 \log_{10} (\Sigma_{\star}) + 5.87$\,\footnote{Note that we are using their $\mathrm{sSFR}$ function instead of the $\Sigma_{\SFR}$ function in their abstract.}. 
This translates into $\tauDepl=1.2$, 1.5, 2.0 and 2.9~Gyr for the four aforementioned stellar masses, assuming that the galaxy size follows the \cite{FernandezLorenzo2013} and \cite{Shen_2003_Galaxy_Sizes} size--mass relation. 
Thus the predictions agree within 20\% for the two intermediate stellar mass ranges, or $\sim50\%$ for all ranges.
We note that the $3 \times 10^{12} \, \Msun$ case is an extrapolation of their function as their data only probe galaxies with $10^{10} < \Mstar/\Msun < 10^{11.5}$.

New observations are needed in the future to clearly distinguish which function is better, and confirm whether our function can reproduce ``down-sizing'' and ``mass-quenching''. Such observations should prioritize low-mass galaxies at high redshift (with enough sensitivity and integration time), as well as highest-mass but below-MS galaxies at the early cosmic time (though such galaxies are still rarely found).

\vspace{0.25truecm}
\subsection{Evolution of molecular gas fraction}
\label{Section_Results_z_deltaGas}

Similar to the previous section, we show in Fig.~\ref{Plot_z_deltaGas_comparison} the binned view of the evolution of $\deltaGas$ as predicted by our best-fit function Eq.~\ref{Equation_tauDepl_deltaGas}. 
The three evolution functions in Fig.~\ref{Plot_z_deltaGas_comparison}, i.e., from our Eq.~\ref{Equation_tauDepl_deltaGas}, \citetalias{Tacconi2018} and \citetalias{Scoville2017} consistently show that $\deltaGas$ has a strong dependency on stellar mass. More-massive galaxies have a lower gas fraction at the same redshift. 
These functions are also very close to each other for $\logMstar \gtrsim 11$ galaxies at all redshifts below $\sim3$. 
For lower-mass galaxies, our function locates between the \citetalias{Tacconi2018} and \citetalias{Scoville2017} ones. 
\citetalias{Scoville2017}'s function does not fit well local galaxies because they do not include local samples in their analysis. But at high redshift our function in this work predicts similarly high gas fraction as the \citetalias{Scoville2017} function, which are a factor of ten higher than those expected from the \citetalias{Tacconi2018} function (for $\logMstar\sim9$ galaxies). 

The dependency of $\deltaGas$ or $\fgas$ on stellar mass has also been found much earlier for local galaxies \citep[e.g.,][]{Young_and_Scoville_1991_ARAA,Kennicutt1998ARAA,McGaugh1997,Schombert2001}.
\cite{Young_and_Scoville_1991_ARAA} reported an increase in gas fraction by two orders of magnitude from early-type to late-type galaxies (along the Hubble sequence) in the local Universe. This is equivalent to similar orders of magnitude increase in their IR to $H$-band luminosity ratio, i.e., $\propto$~sSFR \citep{Kennicutt1998ARAA}. \cite{McGaugh1997} and \cite{Schombert2001} also found strong decreases in the gas fraction with brighter $B$-band magnitude (higher stellar mass) and higher stellar surface density including low surface brightness local galaxies. This is in agreement with our function's prediction. 

More recently, \cite{JiangXueJian2015} reported a similarly strong decrease in gas fraction versus stellar mass as reported here, down to a stellar mass of $10^{8.5-9.0}\;\Msun$ (see also \citealt{CaoTianWen2017}; \citealt{Saintonge2017}) with a non-linear behavior. In their sample, $\deltaGas$ is about 0.08--0.3 for $\logMstar\sim9-10$ galaxies, then decreases to about 0.02--0.1 for $\logMstar\sim10-11$ galaxies, 
which is slightly below this work at $z\sim0$ and is likely caused by their use of a constant $\alphaCO$ while the true $\alphaCO$ might be higher for low-mass metal-poor galaxies.

In summary, the predictions from this work, \citetalias{Scoville2017} and \citetalias{Tacconi2018} only obviously differ in those regimes where not much data are currently available, i.e. at low stellar masses across cosmic time and for all stellar masses at $z>5-6$. 
This work's predictions agree with other individual observations in the literature, and our evolution function has the physical implications of ``down-sizing'' and ``mass-quenching'' in galaxy evolution. In general our analysis also raises the need for future CO and RJ-dust observations of below-MS and/or less-massive ($\logMstar\lesssim10$) galaxy samples.

\begin{figure*}[htb!]
\centering
\includegraphics[width=0.75\textwidth]{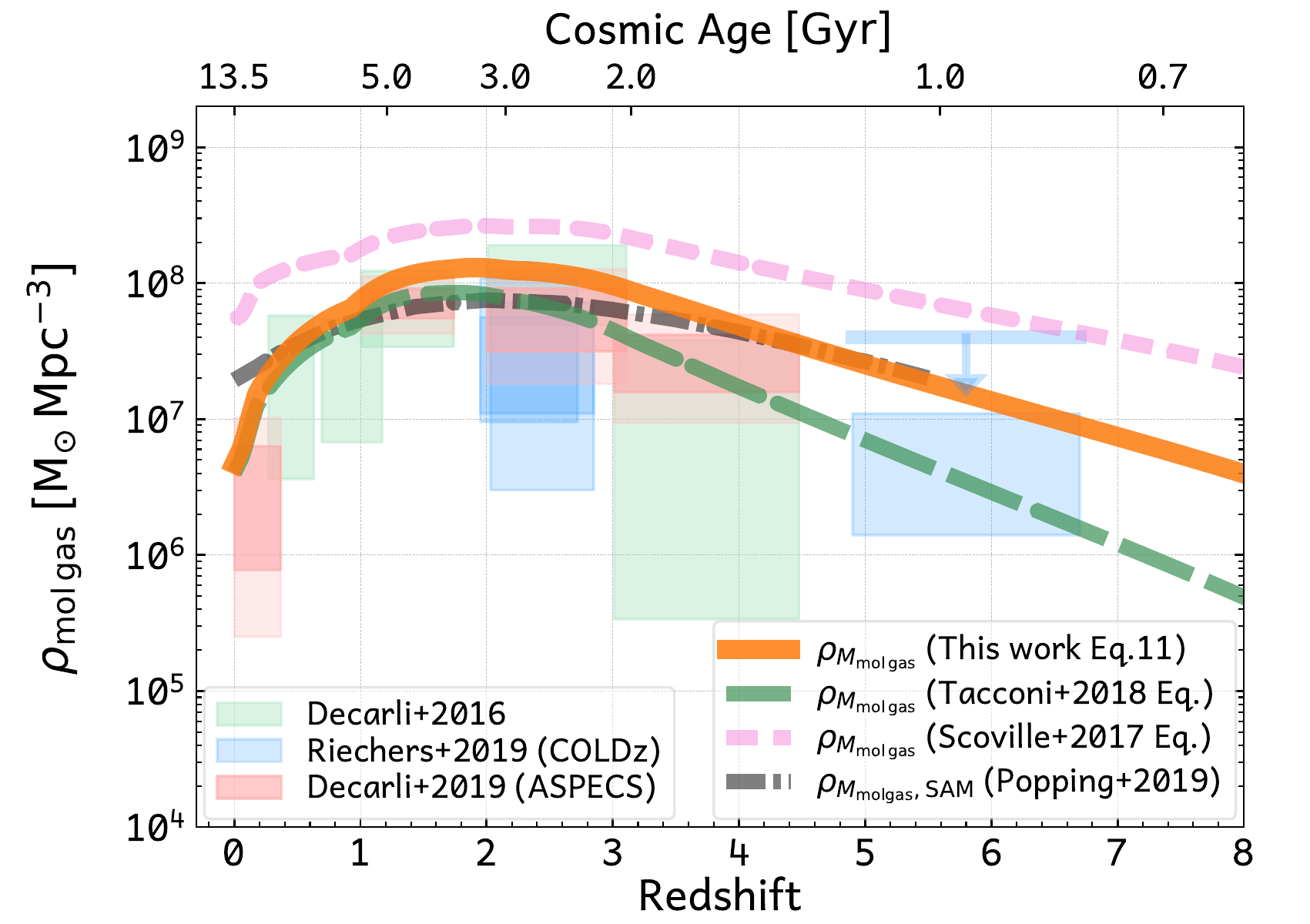}
\vspace{0.5ex}%
\caption{%
Cosmic evolution of the cold molecular gas mass density. Results from high-$z$ CO blind deep field studies from \citet{Decarli2016}, \citet{Riechers2019} and \citet{Decarli2019} are shown as green, blue and red boxes, with the X-sides (Y-sides) indicating the observed redshift range ($5^{\mathrm{th}}$ and $95^{\mathrm{th}}$ percentiles). The $z\sim6$ arrow is an upper limit from \citealt{Riechers2019}. 
The orange solid, green long-dashed and pink short-dashed lines are the SMF-integrated molecular gas mass density based on the SMFs presented in Sect.~\ref{Section_Adopting_the_SMFs} (see also Appx.~\ref{Section_Appendix_SMF_CSFRD}; integrated down to $\Mstar=10^{9.0}\;\Msun$) and gas fraction function from this work (Eq.~\ref{Equation_tauDepl_deltaGas}), \citetalias{Tacconi2018} and \citetalias{Scoville2017}, respectively. 
The black dash-dot line is from the Semi-Analytic Model (SAM) simulation of \cite{Popping2019} (based on \cite{Popping2014}). 
}
\label{Plot_cosmic_evolution_of_rho_molgas_with_data}
\vspace{2ex}
\end{figure*}

\begin{figure*}[ht]
\centering
\includegraphics[width=0.75\textwidth]{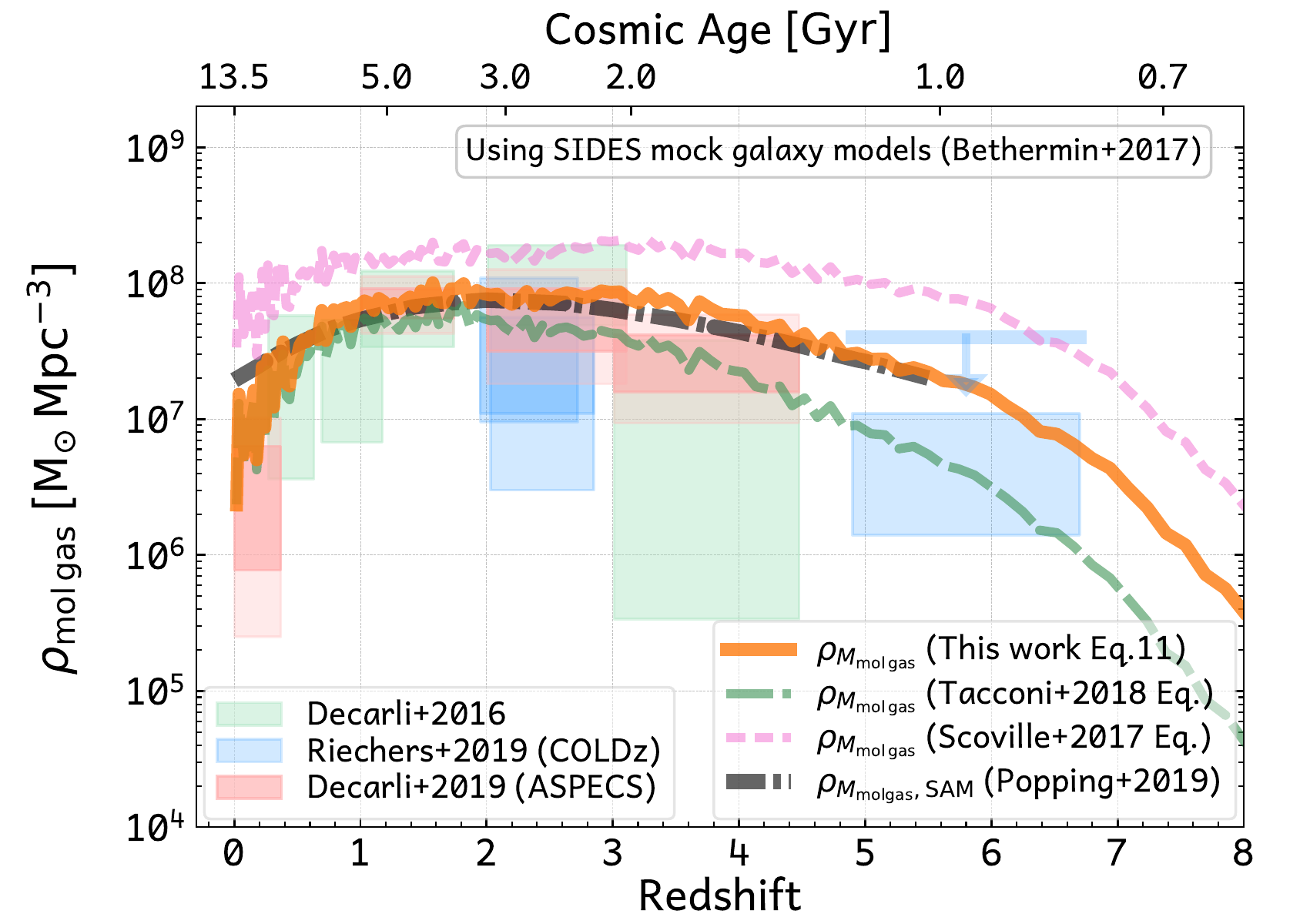}
\vspace{0.5ex}%
\caption{%
Analogous to Fig.~\ref{Plot_cosmic_evolution_of_rho_molgas_with_data}, now using the 2-SFM galaxy model based ``SIDES'' mock galaxy catalog (\citealt{Bethermin2017Model}) to derive the molecular gas mass density. See Sect.~\ref{Section_Results_z_rhomolgas_2SFM} for details. See Fig.~\ref{Plot_cosmic_evolution_of_rho_molgas_with_data} caption for symbols, curves and labels. 
}
\label{Plot_cosmic_evolution_of_rho_molgas_via_galaxy_model}
\vspace{2ex}
\end{figure*}

\vspace{0.5truecm}
\section{Implication for the Cosmic Evolution of Cold Molecular Gas Density}
\label{Section_Implication}

In this section, we study the implication of our cold molecular gas fraction function (Eq.~\ref{Equation_tauDepl_deltaGas}) for the cosmic molecular gas mass density evolution. This requires us to know: (a) the number density of star-forming galaxies at each redshift; (b) their stellar mass distribution at each redshift; and (c) their SFRs. 

The number density and stellar mass distribution evolution of star-forming galaxies have been reasonably well measured through star-forming galaxies' stellar mass function (SMF) studies. We discuss them in detail in Sect.~\ref{Section_Adopting_the_SMFs}. 

Then, by either simply assuming that all star-forming galaxies are MS galaxies (Sect.~\ref{Section_Results_z_rhomolgas}), or more realistically adopting the aforementioned 2-SFM galaxy modeling (\citealt{Sargent2014}; \citealt{Bethermin2017Model}) as we do later in Sect.~\ref{Section_Results_z_rhomolgas_2SFM}, we obtain a SFR for each galaxy corresponding to its stellar mass and redshift. 
With the SFR and $\DeltaMS$, the stellar mass is further converted to gas mass by applying our gas fraction function. Finally, by integrating over all star-forming galaxies, we obtain the cosmic molecular gas mass density at each redshift as presented in Sects.~\ref{Section_Results_z_rhomolgas}~and~\ref{Section_Results_z_rhomolgas_2SFM}.

Such a method is also used by \cite{Maeda2017}, who fitted molecular gas fraction versus stellar mass correlations at two redshift bins ($z\sim0$ and $z=1-1.5$), and then integrated the cosmic molecular gas mass density using stellar mass functions. Other earlier works (\citealt{Sargent2013}; see also \citealt{Carilli2013Review}) instead fitted a molecular gas mass versus SFR correlation (i.e., SF law; independent of redshift and stellar mass) to infer gas mass and integrate over stellar mass functions to obtain the cosmic molecular gas mass density.

\vspace{0.25truecm}
\subsection{Adopting the stellar mass functions (SMFs)}
\label{Section_Adopting_the_SMFs}

In recent years, deep \textit{HST}, \textit{Spitzer} and ground-based near-IR observations in deep fields have pushed the accurate measurements of the star-forming galaxies' SMFs out to $z\sim4-5$ (e.g., \citealt{Marchesini2009}; \citealt{PengYingJie2010_SMF}; \citealt{Baldry2012_SMF}; \citealt{Santini2012_SMF}; \citealt{Moustakas2013_SMF}; \citealt{Muzzin2013}; \citealt{Ilbert2013}; \citealt{Grazian2015}; \citealt{Song2016}; \citealt{Davidzon2017}; \citealt{Wright2017_SMF,Wright2018_SMF}). Similarly, deep \textit{Herschel} far-infrared/sub-mm and ground-based sub-mm surveys pushed the accurate measurements of cosmic SFR density (CSFRD) out to $z\sim3-4$ as well (e.g., \citealt{Madau2014Review}; \citealt{Liudz2017}; and references therein). The CSFRD represents the SFR at each cosmic epoch, thus by integrating the SFR across all the previous cosmic times, we will be able to obtain the total stellar mass density at that time. Meanwhile, the integration of the (star-forming galaxies') SMF at that cosmic time should in principle equal to the total stellar mass integrated from the CSFRD. In Appx.~\ref{Section_Appendix_SMF_CSFRD}, we verify that they are in good agreement for the redshift bins where empirical SMFs are available. 

Note that the CSFRD has been described as a function of redshift (double-powerlaw; \citealt{Madau2014Review}), while it is still difficult to characterize the SMF as a contiguous function of redshift. \cite{Wright2018_SMF} provide such a functional form, however, their function exhibits certain deviations from direct measurements (see Appx.~\ref{Section_Appendix_SMF_CSFRD}). Therefore we construct our SMFs by adopting the $z$-evolving shape of the SMFs in the literature and normalize them according to the integrated CSFRD. The full description of this procedure and its verification on observational data can be found in 
Appx.~\ref{Section_Appendix_SMF_CSFRD}. 
Thus our assumed (star-forming galaxies') SMFs and CSFRDs are consistent with each other at each redshift.

\vspace{0.25truecm}
\subsection{Integrating cosmic molecular gas mass density}
\label{Section_Results_z_rhomolgas}

Based on the assumption that ``all'' star-forming galaxies exactly follow our gas fraction function (Eq.~\ref{Equation_tauDepl_deltaGas}), and their number density obeys the SMF at each redshift, we can compute the molecular gas mass density by integrating the product of gas fraction, stellar mass and SMF in each stellar mass bin at each redshift: 
\begin{equation}
\begin{split}
&\rho_{\mathrm{molgas}} (z) = \\[1ex] 
&\ \sum\limits_{\Mstar\;\mathrm{bins}} \Phi_{\mathrm{SMF}} (z,\Mstar) \times \Mstar \times \deltaGas (z, \Mstar,\DeltaMS)
\end{split}
\label{Equation_integrating_SMF}
\end{equation}

In Fig.~\ref{Plot_cosmic_evolution_of_rho_molgas_with_data} we present the integrated cosmic cold molecular gas mass density versus redshift, using three different gas fraction functions $\deltaGas (z, \Mstar,\DeltaMS)$, our Eq.~\,\ref{Equation_tauDepl_deltaGas} (orange solid line), \citetalias{Tacconi2018} (green long-dashed line) and  \citetalias{Scoville2017} (pink short-dashed line). The same SMFs are used for the three gas fraction functions. 

Note that the result is sensitive to the lower stellar mass limit down to which the integration is performed. \cite{Davidzon2017} adopt a lower limit of $\Mstar=10^{8.0}\;\Msun$ when integrating SMFs to compute the cosmic stellar mass density. To match the CO blind deep field data (e.g., \citealt{Riechers2019}; \citealt{Decarli2019}), we integrate only down to $\Mstar=10^{9.0}\;\Msun$,
i.e., an order of magnitude shallower. 

In Fig.~\ref{Plot_cosmic_evolution_of_rho_molgas_with_data}, we compare results from three recent CO blind deep field surveys (\citealt{Decarli2016}, \citealt{Riechers2019} and \citealt{Decarli2019}, from the ASPECS-pilot, COLDz and ASPECS-LP surveys, respectively), 
to the gas evolution curves derived from our, \citetalias{Tacconi2018} and \citetalias{Scoville2017} functions. The form of the function $\deltaGas (z, \Mstar,\DeltaMS)$ significantly impacts the resulting cosmic cold gas mass density curve. Both our and the \citetalias{Tacconi2018} functions provide very reasonable fits to the data without any tuning (except for the integration limit). Due to the fact that observationally CO luminosity detection limit varies with redshift and sample (or excitation ``correction''), and is in general higher than the integration limits we chose, the currently available data can not sufficiently constrain these functions.

\vspace{0.25truecm}
\subsection{Alternative method: Cosmic molecular gas mass density with mock galaxy models}
\label{Section_Results_z_rhomolgas_2SFM}

The drawback of the SMF$\times\deltaGas$ integration in the previous section is that it only accounts for galaxies located exactly on the MS. In order to account for starburst galaxies as well as the scatter of the MS, we adopt here an alternative approach to derive the cosmic cold molecular gas mass density --- we calculate for each mock galaxy \citep[simulated under the 2-SFM framework by][]{Bethermin2017Model} the cold molecular gas mass using our $\deltaGas$ function before summing them up within each redshift bin. 

The ``SIDES'' simulation \citep[Simulated Infrared Dusty Extragalactic Sky\,\footnote{Available at \url{http://cesam.lam.fr/sides}.};][]{Bethermin2017Model} generated 1,489,629 mock galaxies within a 2~deg$^2$ lightcone from redshift 0.02 to 9.95. 
Different sets of SMFs were adopted according to redshift (\citet{Kelvin2014} for local galaxies, \citet{Moutard2016} at $z<1.5$, \citet{Davidzon2017} at $1.5<z<4$ and \citet{Grazian2015} at $z>4$). Stellar masses were assigned to dark matter halos via abundance matching, and a certain recipe for the star-forming galaxy fraction was assumed at each redshift. The modeling also accounts for the scatter of the star-forming MS coming from both the MS population itself and starburst galaxies, thus it reasonably reproduces true galaxy distributions. 

We use the SIDES mock galaxy catalog and select $\log_{10}\Mstar>9.0$ star-forming galaxies (as in the previous section to match CO luminosity function studies), then apply Eq.~\ref{Equation_tauDepl_deltaGas} (as well as the \citetalias{Tacconi2018} and \citetalias{Scoville2017} functions) to each galaxy to obtain its gas fraction, and hence to derive its molecular gas mass. We integrate the molecular gas mass for all galaxies in a given redshift bin, then divide it by the corresponding comoving volume to obtain the cosmic cold gas mass density $\rho_{\mathrm{mol\,gas}}$. We sample the redshift range from 0 to 15 with 500 bins (i.e., bin size $\sim$0.00253 in $\log_{10}(1+z)$; as in the previous section). 

The results are presented in Fig.~\ref{Plot_cosmic_evolution_of_rho_molgas_via_galaxy_model}. The wiggling at the low-redshift end is likely due to the cosmic variance. At higher redshifts ($z>0.5$) our curve coincidentally agrees with the Semi-Analytic Model (SAM) simulation by \cite{Popping2019}. Other simulations, \cite{Obreschkow2009} and \cite{Lagos2011}, can be seen in Fig.~5 of \cite{Riechers2019}: at $z\sim5$, the \cite{Popping2019} simulation exhibit a 0.2~dex lower $\rho_{\mathrm{mol\,gas}}$ than that of \cite{Lagos2011}, and the \cite{Obreschkow2009} $\rho_{\mathrm{mol\,gas}}$ is 0.1~dex lower than \cite{Lagos2011}; while the three are reversed at $z\sim0.5$, but still within 0.2~dex. Thus in general the simulations and the predictions with the functional form derived here are in good agreement.

When using the \citetalias{Tacconi2018} and \citetalias{Scoville2017} functions for the computation, the corresponding cold molecular gas mass density curves show large difference. The \citetalias{Scoville2017} function leads to a much higher cold molecular gas mass density at all redshifts, which is likely because their function predicts significantly higher $\deltaGas$ (see Fig.~\ref{Plot_z_deltaGas_comparison}).%
\,\footnote{%
We caution that \citetalias{Scoville2017} used a different MS function than this work and \citetalias{Tacconi2018}. Our test in Appx.~\ref{Section_MS} shows that their MS can explain half of the discrepancy seen in Figs.~\ref{Plot_cosmic_evolution_of_rho_molgas_with_data}~and~\ref{Plot_cosmic_evolution_of_rho_molgas_via_galaxy_model}. The other major contributor to the discrepancy is the functional form. As shown in Figs.~\ref{Plot_z_deltaGas_comparison}, their gas fraction's functional form is too high at both low- and high-redshift ($z<1$ and $z>4$) and for less-massive ($\logMstar<10$) galaxies. While integrating all galaxies to compute the cosmic gas mass density, such a difference in the functional forms causes a large discrepancy.%
}
The \citetalias{Tacconi2018} function results in fully (marginally) consistent cold molecular gas mass densities as our function at $z \lesssim 1$ ($z\sim2-3$), however, it predicts 0.4--0.9 dex lower values at $z>4$. This is mainly driven by the downturn of their $\deltaGas$ function at $z\gtrsim4$ (as mentioned in Sect.~\ref{Section_Fitting_Multi_Variate_Function}) and probably also affected by their systematic lower $\deltaGas$ for low-mass galaxies (see Fig.~\ref{Plot_z_deltaGas_comparison}). Nevertheless, due to the large uncertainties in the CO blind deep field data, it is still hard to distinguish whether our function is statistically better than the \citetalias{Tacconi2018} function. We will further investigate this with simulated galaxies (\citealt{Popping2019}) in future work.

In summary, the above comparisons indicate that our knowledge on galaxy star-forming MS (e.g., the two-star-formation model (2-SFM); \citealt{Sargent2014}), stellar mass functions (see references in \citealt{Bethermin2017Model}) and molecular gas fraction parametrization (using our functional form of $\deltaGas$ in Eq.~\ref{Equation_tauDepl_deltaGas}) are moving towards a coherent picture.

\vspace{0.5truecm}
\section{Summary}
\label{Section_Summary}

In this work, we present a comprehensive analysis of galaxy molecular gas scaling relations and their evolution using a robust ALMA-detected galaxy catalog from our paper~\romup{1} (A$^3$COSMOS). Each galaxy in the catalog has a redshift, stellar mass, SFR and dust mass from far-infrared spectral energy distribution (SED) fitting including the ALMA data and rich multi-wavelength data from the literature (see paper~\romup{1} for the details). We compared four methods of molecular gas mass calibration using SED-fitted dust mass and/or Rayleigh-Jeans (RJ)-tail dust continuum (Sect.~\ref{Section_comparing_gas_mass_calibration}), 
from which we determine that the RJ-dust continuum method (with \citealt{Hughes2017} luminosity-dependent calibration) better infers the gas mass. 
Meanwhile, we also comprehensively discuss several related topics in the gas mass calibration, i.e., $\alphaCO$, $\deltaGDR$, molecular-to-atomic fraction and metallicity, and their biases to this work in the Appendix (Appx.~\ref{Section_ZACO}~to~\ref{Section_MZR}).

Due to the sample inhomogeneity, higher-redshift (e.g., $z>4$) galaxies do not always have RJ-tail wavelength coverage. Thus we investigated the effect of band conversion with \textsc{MAGPHYS} high-$z$ SED fitting for galaxies whose longest-wavelength ALMA data do not cover RJ-tail wavelengths. We found that it potentially results in a factor of 2--6 underestimation of gas mass at $z>4$ (see Sect.~\ref{Section_Band_conversion} and Appx.~\ref{Section_Appendix_SED_band_conversion}). 

We combine our A$^3$COSMOS sample with \NumberOfComplementarySamples{} complementary samples in the literature from local to high redshift (see 
Table~\ref{Table_galaxy_samples}) to study the scaling relations and cosmic evolution of molecular gas depletion time $\tauDepl$ and molecular gas to stellar mass ratio $\deltaGas$. 
We parameterize the $\tauDepl$ and $\deltaGas$ as functions of galaxy's cosmic age, stellar mass and SFR. We tested both \citet[][\citetalias{Tacconi2018}]{Tacconi2018} and \citet[][\citetalias{Scoville2017}]{Scoville2017} functions (shown in Eqs.~\ref{Equation_Tacconi2018}~and~\ref{Equation_Scoville2017} respectively), meanwhile also propose a new functional form in Eq.~\ref{Equation_tauDepl_deltaGas} which accounts for the galaxies' different evolution driven by their stellar masses. Then, by applying the gas fraction scaling relation to galaxies' stellar mass functions and integrating over all stellar masses, we obtain the evolution of cosmic cold molecular gas mass density, which is in a coherent picture with the known cosmic SFR density evolution and the semi-analytic modeling of galaxies in the cosmological simulations (Figs.~\ref{Plot_cosmic_evolution_of_rho_molgas_with_data}~and~\ref{Plot_cosmic_evolution_of_rho_molgas_via_galaxy_model} respectively). 

Further more, we emphasize the following points:

\begin{itemize}
\item The distribution of our sample's redshifts, stellar masses and SFRs are consistent with previous studies where they overlap in the parameter space (e.g., see contours in Figs.~\ref{Plot_z_tauDepl_multi_panel}~and~\ref{Plot_z_deltaGas_multi_panel}). Given our total sample of more than 1,600 galaxies, we see that the composite sample selection is biased to strong starbursts with $\DeltaMS\sim0.5-1.5$ and $\logMstar\sim10-11$ at $z\sim0.08-1.0$ (see Fig.~\ref{Plot_z_DeltaMS}), and biased to the most massive galaxies with $\DeltaMS\sim0.0$ and $\logMstar\sim12$ at $z>3$ (see Figs.~\ref{Plot_z_deltaGas_3D}~and~\ref{Plot_z_sSFR_gas_evolution}). In particular at $z>4$ the dust continuum observations are mainly probing rest-frame wavelengths shorter than 250\,$\mu$m, for which the SED-fitting-extrapolated RJ-tail flux might be under-predicted. 
However, they do not statistically affect our functional fitting due to their low number. 
\item The parametrizations of $\tauDepl$ and $\deltaGas$ with the functions in this work, \citetalias{Tacconi2018} and \citetalias{Scoville2017} are roughly consistent where the data are commonly sampled in the parameter space, i.e., $z\sim1-3$, $\DeltaMS>0$ and $\logMstar>10.5$ (see Figs.~\ref{Plot_z_tauDepl_comparison}~and~\ref{Plot_z_deltaGas_comparison}). They differ significantly for low-mass and/or main-sequence or below-main-sequence galaxies, which, however, could not be verified with the current dataset. 
The chi-square statistics for these parametrization show that our new functional form and the \citetalias{Tacconi2018} are similarly good, and are better than the \citetalias{Scoville2017} functional form which has one (two) less free parameter(s) than the \citetalias{Tacconi2018} one (ours). 
We emphasize that our new functional form implicitly leads to a {\it ``down-sizing''} in galaxy evolution and probably a {\it ``mass-quenching''} effect. Although further data are needed to verify these effects (see Sects.~\ref{Section_Results_z_tauDepl}~and~\ref{Section_Results_z_deltaGas} as well as Figs.~\ref{Plot_z_tauDepl_comparison}~and~\ref{Plot_z_deltaGas_comparison}), the results are promising to build a most comprehensive picture of gas evolution. 
\item The integration of galaxies' stellar mass function with the application of gas fraction scaling relation involves many assumptions. Noticeable differences are found between the simpler assumption that all star-forming galaxies exactly follow the main sequence (Fig.~\ref{Plot_cosmic_evolution_of_rho_molgas_with_data}) and the more realistic 2-SFM galaxy modeling (Fig.~\ref{Plot_cosmic_evolution_of_rho_molgas_via_galaxy_model}) which accounts for the starburst/main-sequence dichotomy and uses different stellar mass functions than in this work (Appx.~\ref{Section_Appendix_SMF_CSFRD}). 
The realistic galaxy modeling has a better agreement with semi analytic models (\citealt{Popping2014,Popping2019}). 
Among the three functional forms discussed in this work, only our new functional form (Eq.~\ref{Equation_tauDepl_deltaGas}) of the gas fraction scaling relation could achieve such a high consistency. 
\item Compared to CO blind deep field surveys, our analytically-derived cold molecular gas mass densities agree within their upper boundary. This is understandable as the current CO surveys usually could not sample well enough the faint-end of the CO line luminosity function, thus the integration of CO luminosity functions is usually down to only $\log_{10} (L^{\prime}_{\mathrm{CO}}/[\mathrm{K\,km\,s^{-1}\,pc^2}])\sim 9.5$ (to avoid extrapolating the faint-end; see e.g., \citealt{Riechers2019}; \citealt{Decarli2019}). 
\item Finally, our large, robust dataset strongly supports a coherent picture of the evolution galaxies' gas, stellar and SFR which can be parameterized by the main sequence functions (e.g., \citealt{Speagle2014}; Leslie et al. submitted; Appx.~\ref{Section_MS}), stellar mass functions (e.g., \citealt{Davidzon2017}; Appx.~\ref{Section_Appendix_SMF_CSFRD}) and gas scaling functions (Eq.~\ref{Equation_tauDepl_deltaGas}). The integration of stellar mass function times the main sequence function (over stellar mass at each redshift) gives the cosmic SFR density, and the integration of stellar mass function times the molecular gas fraction function (over stellar mass at each redshift) results in the cosmic molecular gas mass density. Integrating the cosmic SFR density curve (across cosmic time) further leads to the cosmic stellar mass density growth curve, which in return is consistent with the integration of stellar mass functions across cosmic time. 
\end{itemize}

\vspace{-0.25cm}

\acknowledgments

\vspace{-0.75cm}%
DL, ES and PL acknowledge funding from the European Research
Council (ERC) under the European Union's Horizon 2020 research and innovation programme (grant agreement No. 694343).
B.M., E.F.J.A. and F.B. acknowledge support of the Collaborative Research Center 956, subproject A1 and C4, funded by the Deutsche Forschungsgemeinschaft (DFG) -- project ID 184018867. 
SL acknowledges funding from Deutsche Forschungsgemeinschaft (DFG) Grant SCH 536/9-1.
D.R. acknowledges support from the National Science Foundation under grant numbers AST-1614213 and AST-1910107 and from the Alexander von Humboldt Foundation through a Humboldt Research Fellowship for Experienced Researchers. 
GEM acknowledges support from the Villum Fonden research grant 13160 ``Gas to stars, stars to dust: tracing star formation across cosmic time'', the Cosmic Dawn Center of Excellence funded by the Danish National Research Foundation and  the ERC Consolidator Grant funding scheme (project ConTExt, grant number No. 648179). 
YG's research is supported by National Key Basic Research and Development Program of China (grant No. 2017YFA0402704), 
National Natural Science Foundation of China (grant Nos. 11861131007, 11420101002), and Chinese Academy of Sciences Key
Research Program of Frontier Sciences (grant No. QYZDJSSW-SLH008). 
We thank the anonymous referee for constructive comments. We thank Matthieu B\'{e}thermin for a very helpful discussion on galaxy modeling.

This paper makes use of the following ALMA data: 
\path{ADS/JAO.ALMA#2011.0.00064.S}, 
\path{ADS/JAO.ALMA#2011.0.00097.S}, 
\path{ADS/JAO.ALMA#2011.0.00539.S}, 
\path{ADS/JAO.ALMA#2011.0.00742.S}, 
\path{ADS/JAO.ALMA#2012.1.00076.S}, 
\path{ADS/JAO.ALMA#2012.1.00323.S}, 
\path{ADS/JAO.ALMA#2012.1.00523.S}, 
\path{ADS/JAO.ALMA#2012.1.00536.S}, 
\path{ADS/JAO.ALMA#2012.1.00919.S}, 
\path{ADS/JAO.ALMA#2012.1.00952.S}, 
\path{ADS/JAO.ALMA#2012.1.00978.S}, 
\path{ADS/JAO.ALMA#2013.1.00034.S}, 
\path{ADS/JAO.ALMA#2013.1.00092.S}, 
\path{ADS/JAO.ALMA#2013.1.00118.S}, 
\path{ADS/JAO.ALMA#2013.1.00151.S}, 
\path{ADS/JAO.ALMA#2013.1.00171.S}, 
\path{ADS/JAO.ALMA#2013.1.00208.S}, 
\path{ADS/JAO.ALMA#2013.1.00276.S}, 
\path{ADS/JAO.ALMA#2013.1.00668.S}, 
\path{ADS/JAO.ALMA#2013.1.00815.S}, 
\path{ADS/JAO.ALMA#2013.1.00884.S}, 
\path{ADS/JAO.ALMA#2013.1.00914.S}, 
\path{ADS/JAO.ALMA#2013.1.01258.S}, 
\path{ADS/JAO.ALMA#2013.1.01292.S}, 
\path{ADS/JAO.ALMA#2015.1.00026.S}, 
\path{ADS/JAO.ALMA#2015.1.00055.S}, 
\path{ADS/JAO.ALMA#2015.1.00122.S}, 
\path{ADS/JAO.ALMA#2015.1.00137.S}, 
\path{ADS/JAO.ALMA#2015.1.00260.S}, 
\path{ADS/JAO.ALMA#2015.1.00299.S}, 
\path{ADS/JAO.ALMA#2015.1.00379.S}, 
\path{ADS/JAO.ALMA#2015.1.00388.S}, 
\path{ADS/JAO.ALMA#2015.1.00540.S}, 
\path{ADS/JAO.ALMA#2015.1.00568.S}, 
\path{ADS/JAO.ALMA#2015.1.00664.S}, 
\path{ADS/JAO.ALMA#2015.1.00704.S}, 
\path{ADS/JAO.ALMA#2015.1.00853.S}, 
\path{ADS/JAO.ALMA#2015.1.00861.S}, 
\path{ADS/JAO.ALMA#2015.1.00862.S}, 
\path{ADS/JAO.ALMA#2015.1.00928.S}, 
\path{ADS/JAO.ALMA#2015.1.01074.S}, 
\path{ADS/JAO.ALMA#2015.1.01105.S}, 
\path{ADS/JAO.ALMA#2015.1.01111.S}, 
\path{ADS/JAO.ALMA#2015.1.01171.S}, 
\path{ADS/JAO.ALMA#2015.1.01212.S}, 
\path{ADS/JAO.ALMA#2015.1.01495.S}, 
\path{ADS/JAO.ALMA#2015.1.01590.S}, 
\path{ADS/JAO.ALMA#2015.A.00026.S}, 
\path{ADS/JAO.ALMA#2016.1.00478.S}, 
\path{ADS/JAO.ALMA#2016.1.00624.S}, 
\path{ADS/JAO.ALMA#2016.1.00735.S}. 
ALMA is a partnership of ESO (representing its member states), NSF (USA) and NINS (Japan), together with NRC (Canada), MOST and ASIAA (Taiwan), and KASI (Republic of Korea), in cooperation with the Republic of Chile. The Joint ALMA Observatory is operated by ESO, AUI/NRAO and NAOJ. 

\facility{ALMA}

\clearpage




\begin{table*}[htb]
\vspace{3cm}
\caption{%
	Best-Fit Coefficients for the Molecular Gas to Stellar Mass Ratio and Molecular Gas Depletion Time Functions
	\label{Table_functions}
}
%
%
\begin{tabularx}{1.0\textwidth}{
		l @{\extracolsep{4pt}}
		@{\extracolsep{\fill}}
		*4{>{\centering\arraybackslash}c@{}} 
	}
\hline
\hline

$\log_{10} ( \deltaGas )$ & $\mathsf{a}$ ($+ \mathsf{ak}$) & $\mathsf{b}$ & $\mathsf{c}$ ($+ \mathsf{ck}$) & $\mathsf{d}$ \\

$(\equiv \log_{10} (\Mmolgas / \Mstar))$ & ($\DeltaMS$) & ($\log_{10} M_{\star,10}$)$^{\dagger}$ & ($\log_{10} (1+z)$ or $t$) & (norm.) \\[2pt]

\cline{2-5}\\[-10pt]


This work, Eq.~\ref{Equation_tauDepl_deltaGas} & $+0.4195+0.1195{\times}\log_{10}M_{\star,10}$ & $-0.6907$ & $(-0.1543+0.0320{\times}\log_{10}M_{\star,10}){\times}t$ & $+0.9339$ \\

This work, Eq.~\ref{Equation_Tacconi2018} & $+0.54$ & $-0.39$ & $-4.42{\times}(\log_{10}(1+z)-0.58)^{2}$ & $+0.309$ \\

\citetalias{Tacconi2018} best-fit, Eq.~\ref{Equation_Tacconi2018} & $+0.53$ & $-0.35$ & $-3.62{\times}(\log_{10}(1+z)-0.66)^{2}$ & $+0.365$ \\

\citetalias{Tacconi2018} best-fit, Eq.~\ref{Equation_Tacconi2018} & $+0.53$ & $-0.35$ & $-3.62{\times}(\log_{10}(1+z)-0.66)^{2}$ & $+0.120$ \\

This work, Eq.~\ref{Equation_Scoville2017} & $+0.57$ & $-0.28$ & $+2.27{\times}\log_{10}(1+z)$ & $-1.106$ \\

\citetalias{Scoville2017} best-fit, Eq.~\ref{Equation_Scoville2017} & $+0.32$ & $-0.70$ & $+1.84{\times}\log_{10}(1+z)$ & $-0.149$ \\

\hline

$\log_{10} (\tauDepl / \mathrm{[Gyr]})$ &  &  &  &  \\

$(\equiv \log_{10} (\Mmolgas / \SFR))$ &  &  &  &  \\[2pt]

\cline{2-5}\\[-10pt]

This work, Eq.~\ref{Equation_tauDepl_deltaGas} & $-0.5724+0.1120{\times}\log_{10}M_{\star,10}$ & $-0.5174$ & $(-0.0030+0.0568{\times}\log_{10}M_{\star,10}){\times}t$ & $+0.0269$ \\

This work, Eq.~\ref{Equation_Tacconi2018} & $-0.40$ & $+0.08$ & $-0.91{\times}\log_{10}(1+z)$ & $+0.022$ \\

\citetalias{Tacconi2018} best-fit, Eq.~\ref{Equation_Tacconi2018} & $-0.44$ & $+0.09$ & $-0.62{\times}\log_{10}(1+z)$ & $+0.027$ \\

This work, Eq.~\ref{Equation_Scoville2017} & $-0.39$ & $+0.08$ & $-0.91{\times}\log_{10}(1+z)$ & $+0.021$ \\

\citetalias{Scoville2017} best-fit, Eq.~\ref{Equation_Scoville2017} & $-0.70$ & $-0.01$ & $-1.04{\times}\log_{10}(1+z)$ & $+0.509$ \\


\hline

\end{tabularx}

$^{\dagger}$
$\log_{10} M_{\star,10}$ represents $\log_{10} (M_{\star}/10^{10}\,\mathrm{M_{\odot}})$. \\

%
%
%
%
\end{table*}



\begin{table*}[htb]
\caption{%
	Statistics of Fitting the Functions of Molecular Gas to Stellar Mass Ratio ($\deltaGas\equiv\Mmolgas/\Mstar$) with Subsamples
	\label{Table_function_fittings}
}
%
%
\begin{tabularx}{1.0\textwidth}{
		l
		@{\extracolsep{\fill}}
		l
		l
		l
		l
		l
		l
		l
		l
		l 
		@{}
	}
\hline
\hline

Fitting Function & \multicolumn{3}{c}{All data points} & \multicolumn{3}{c}{Without $z>4$ data} & \multicolumn{3}{c}{Without $z>1$ CO} \\

\cline{2-4}
\cline{5-7}
\cline{8-10}

(For $\deltaGas$) & $N$ & $\chi^2$ & $\chi^2_{\mathrm{redu.}}$ & $N$ & $\chi^2$ & $\chi^2_{\mathrm{redu.}}$ & $N$ & $\chi^2$ & $\chi^2_{\mathrm{redu.}}$ \\

\hline


Eq.~\ref{Equation_tauDepl_deltaGas} (This work)
& 1663 & 1426.98 & 0.86 & 1617 & 1390.16 & 0.86 & 1554 & 1349.56 & 0.87 \\

Eq.~\ref{Equation_Tacconi2018} (This work)
& 1663 & 1444.83 & 0.87 & 1617 & 1404.74 & 0.87 & 1554 & 1369.32 & 0.88 \\

Eq.~\ref{Equation_Tacconi2018} (\citetalias{Tacconi2018}'s fit)
& 1663 & 1503.94 & 0.91 & 1617 & 1478.11 & 0.92 & 1554 & 1427.91 & 0.92 \\

Eq.~\ref{Equation_Scoville2017} (This work)
& 1663 & 2086.55 & 1.26 & 1617 & 1863.42 & 1.16 & 1554 & 1921.61 & 1.24 \\

Eq.~\ref{Equation_Scoville2017} (\citetalias{Scoville2017}'s fit)
& 1663 & 10230.35 & 6.17 & 1617 & 10029.8 & 6.22 & 1554 & 10141.08 & 6.54 \\


\hline

\end{tabularx}
%
%
%
%
\end{table*}

\vspace{2cm}

\makeatletter\onecolumngrid@push\makeatother
\FloatBarrier
\makeatletter\onecolumngrid@pop\makeatother
\clearpage

\appendix
\counterwithin{figure}{section}
\counterwithin{table}{section}

\FloatBarrier

\section{Empirical Galaxy Scaling Relations Used in This Work}
\label{Section_Galaxy_Scaling_Relations}

Scaling relations describe how galaxy properties correlate with each other and are important for understanding galaxy populations and their evolution over cosmic time. As this work studies the molecular gas evolution in galaxies, the four scaling relations below are relevant and sometimes needed for our analysis. Calibrations of these correlations are widely studied in the literature, however, their validity for different types of galaxies (i.e., different $z$, $\Mstar$ and SFR) are rarely studied. Here we compare a number of empirical calibrations and discuss their biases. This comparison guides our choice of the most suitable correlations to use in the analysis described in the main body of the paper.

\vspace{0.25truecm}
\subsection{CO-to-H$_2$ conversion factor ($\alphaCO$) versus metallicity}
\label{Section_ZACO}

\begin{figure*}[ht]
\centering
\includegraphics[width=0.497\linewidth]{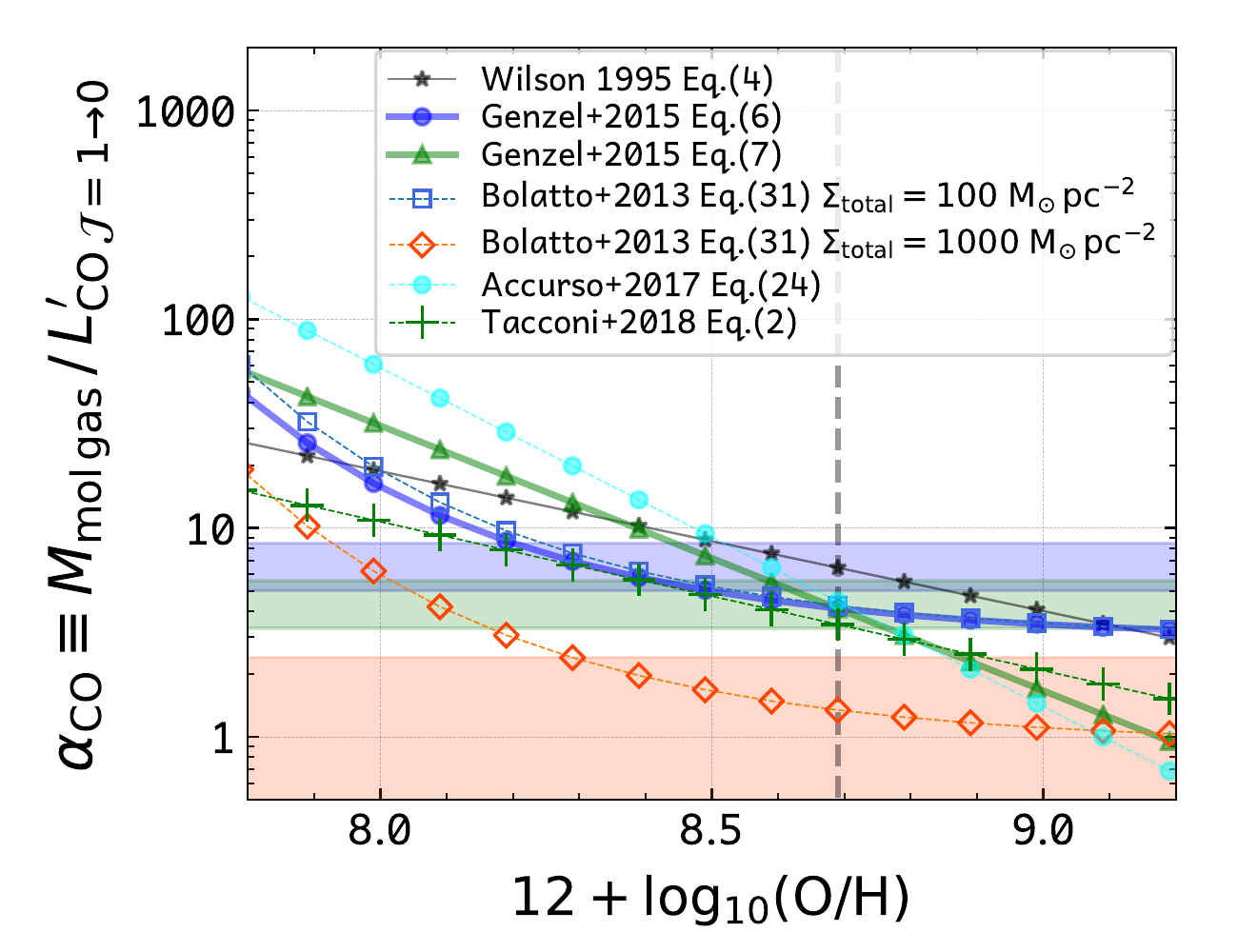}
\includegraphics[width=0.497\linewidth]{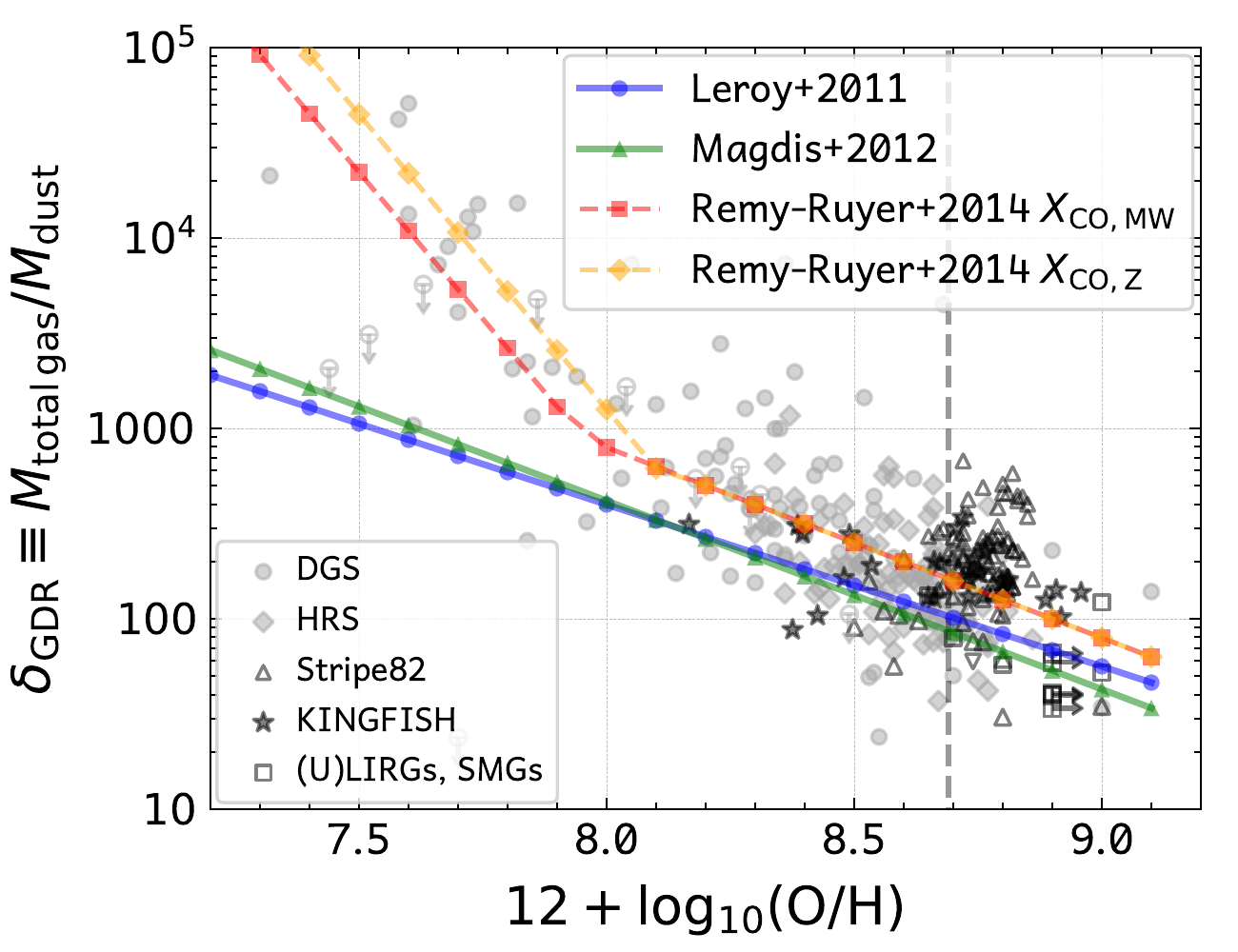}
\caption{%
{\it\normalsize Left panel:} 
The metallicity dependency of the CO-to-H$_2$ conversion factor $\alphaCO$. 
Symbols/lines are from \cite{Wilson1995}, \cite{Genzel2015}, \cite{Bolatto2013ARAA}, \cite{Accurso2017}, and \citetalias{Tacconi2018} as labeled. 
The horizontal blue-, green- and orange-shaded regions correspond to $\alphaCO=6.5$, $4.3$, and $0.8 \; \mathrm{M_{\odot}\,(K\,km\,s^{-1}\,pc^2)^{-1}}$, representing Galactic GMCs, the inner Galactic disk and ULIRGs, respectively. 
See Appx.~\ref{Section_ZACO} for more details. 
The vertical dashed line indicates solar metallicity ($\metalZOH_{\odot}=8.69$; \citealt{Asplund2009}). 
{\it\normalsize Right panel:} 
The metallicity dependency of the gas-to-dust mass ratio $\deltaGDR$. 
Symbols/lines are from \cite{Leroy2011GDR}, \cite{Magdis2012SED} and \cite{RemyRuyer2014} as labeled. 
See Appx.~\ref{Section_ZGDR} for more details. 
See Table~\ref{Table_galaxy_samples} for references of the data points in both panels. 
}
\label{Plot_metalZ_deltaGD}
\end{figure*}

The CO-to-H$_2$ conversion factor, $\alphaCO$, is an empirical ratio converting CO line luminosity to total molecular gas mass. It has been found to be relatively constant in the inner Galactic Giant Molecular Clouds (GMCs), 
being around $4.6 \; \mathrm{\Msun \, (K\,km\,s^{-1}\,pc^{2})^{-1}}$ ($X_{\mathrm{CO}} = 2.1 \times 10^{20} \; \mathrm{cm^{-2}\,(K\,km\,s^{-1}\,pc^2)^{-1}}$) (\citealt{Solomon1987}; \citealt{Solomon1991}; \citealt{Solomon2005}), or $6.5 \; \mathrm{\Msun \, (K\,km\,s^{-1}\,pc^2)^{-1}}$ ($X_{\mathrm{CO}} = 3.0 \times 10^{20} \; \mathrm{cm^{-2}\,(K\,km\,s^{-1}\,pc^2)^{-1}}$) when including heavy elements which are mostly helium. 
While it is as low as $0.8 \; \mathrm{\Msun \, (K\,km\,s^{-1}\,pc^{2})^{-1}}$ ($X_{\mathrm{CO}} = 2.0 \times 10^{20} \; \mathrm{cm^{-2}\,(K\,km\,s^{-1}\,pc^2)^{-1}}$; $\pm0.5\,\mathrm{dex}$) in local ULIRGs (\citealt{Solomon1997}; \citealt{Downes1998}; \citealt{Solomon2005}). 
The calibration of $\alphaCO$ relies on a number of other (molecular) gas mass tracers, including virial mass, optically-thin CO isotopologues, dust extinction, dust emission (via the gas-to-dust ratio, e.g., Sect.~\ref{Section_ZGDR}), and diffuse $\gamma$-ray radiation. 
More details are given in the recent review by \cite{Bolatto2013ARAA}. 
In this work, we only focus on the established $\alphaCO$--metallicity relations presented in \cite{Genzel2015} and \citetalias{Tacconi2018}, as shown in the left panel of Fig.~\ref{Plot_metalZ_deltaGD}, to homogenize the molecular gas mass calculation for our complementary samples.

\vspace{0.25truecm}
\subsection{Gas-to-dust ratio ($\deltaGDR$) versus metallicity}
\label{Section_ZGDR}

The gas-to-dust mass ratio, $\deltaGDR \equiv \Mtotalgas \, / \, \Mdust$, describes the correlation between the total amount of gas (molecular plus atomic, compositing almost all of the ISM) and dust. As dust grains are usually assumed to be well-mixed within the ISM, $\deltaGDR$ should be predictable by ISM chemical models (e.g., see recent review by \citealt{Galliano2018ARAA}). We will skip the physical mechanism behind this and refer the reader to \citet{Galliano2018ARAA}. Here we aim at understanding how $\deltaGDR$ can be applied for molecular gas mass estimation for high-redshift galaxies. 

The calibration of $\deltaGDR$ is usually based on observations of CO and H\,{\sc i} emission lines plus multi-wavelength photometry to which SED fitting is performed (e.g., \citealt{Leroy2011GDR}; \citealt{Sandstrom2013}; \citealt{RemyRuyer2014}; \citealt{Lisenfeld2000}; \citealt{Magdis2011SED}; \citealt{Tan2014}). 
These works found that $\deltaGDR$ is correlated with galaxies' gas phase metallicity, as illustrated by data from our sample compilation (see Tab.~\ref{Table_galaxy_samples}) in Fig.~\ref{Plot_metalZ_deltaGD} ({\it right panel}). 
$\deltaGDR$ is around 100 for galaxies with solar- and super-solar-metallicity,
while it increases non-linearly toward lower metallicity, reaching over 1000 in extremely metal-poor ($<10\%\,Z_{\odot}$) galaxies (e.g., \citealt{Elmegreen2013}; \citealt{Shi2014,Shi2016}). The difference between the derived relations of \cite{Leroy2011GDR} and \cite{RemyRuyer2014} is about 0.1\,dex in the super-solar metallicity regime, increases to 0.2\,dex at 0.2 solar metallicity, and then quickly becomes much larger at even lower metallicity. 

As $\deltaGDR$ is calibrated with total gas mass instead of molecular gas mass, a molecular-to-total gas mass ratio, $\fmol \equiv \Mmolgas/\Mtotalgas$, needs to be considered. It is discussed in the next section (Appx.~\ref{Section_ZFMOL}).

\begin{figure}[b]
\centering
\includegraphics[width=0.487\linewidth]{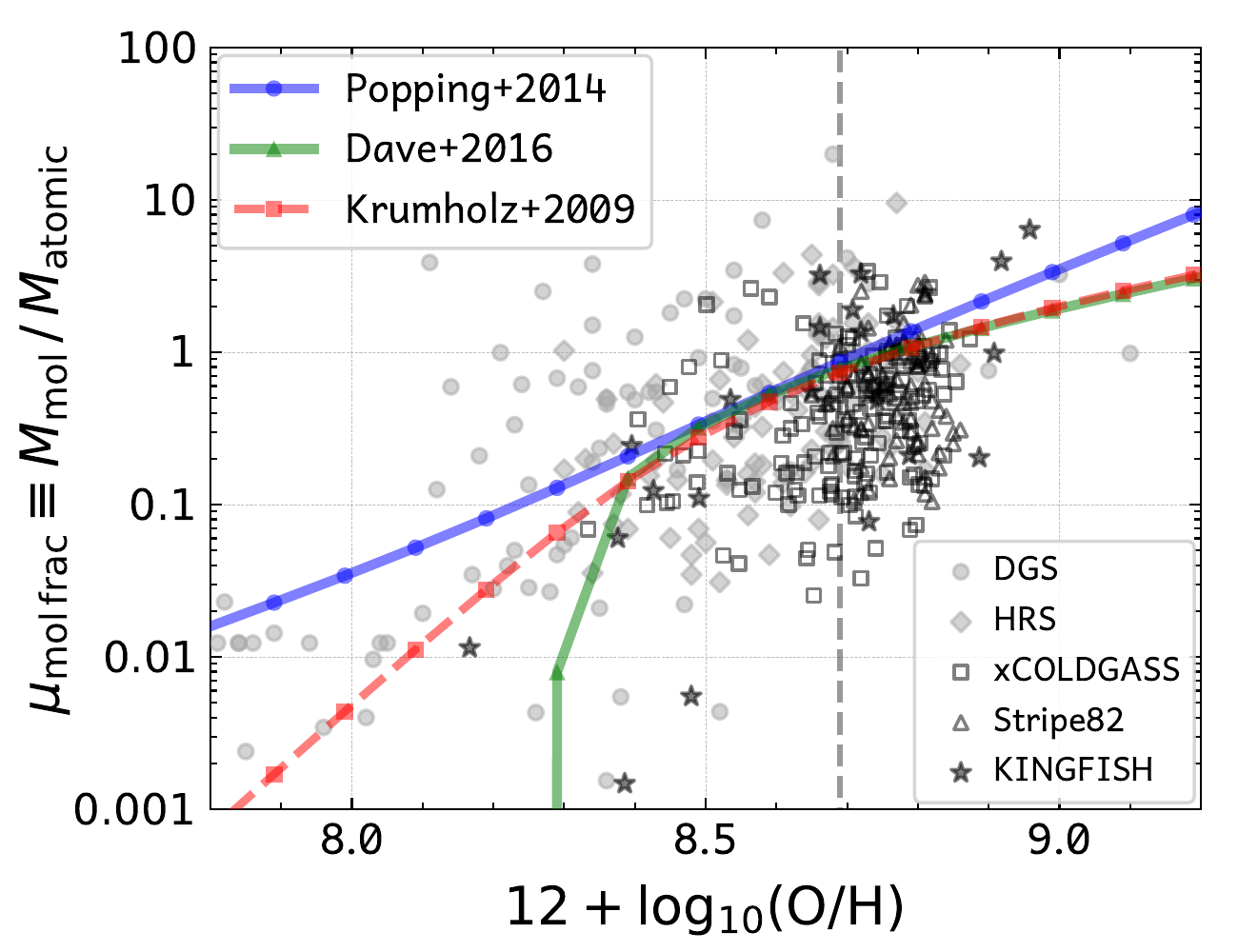}
\includegraphics[width=0.487\linewidth]{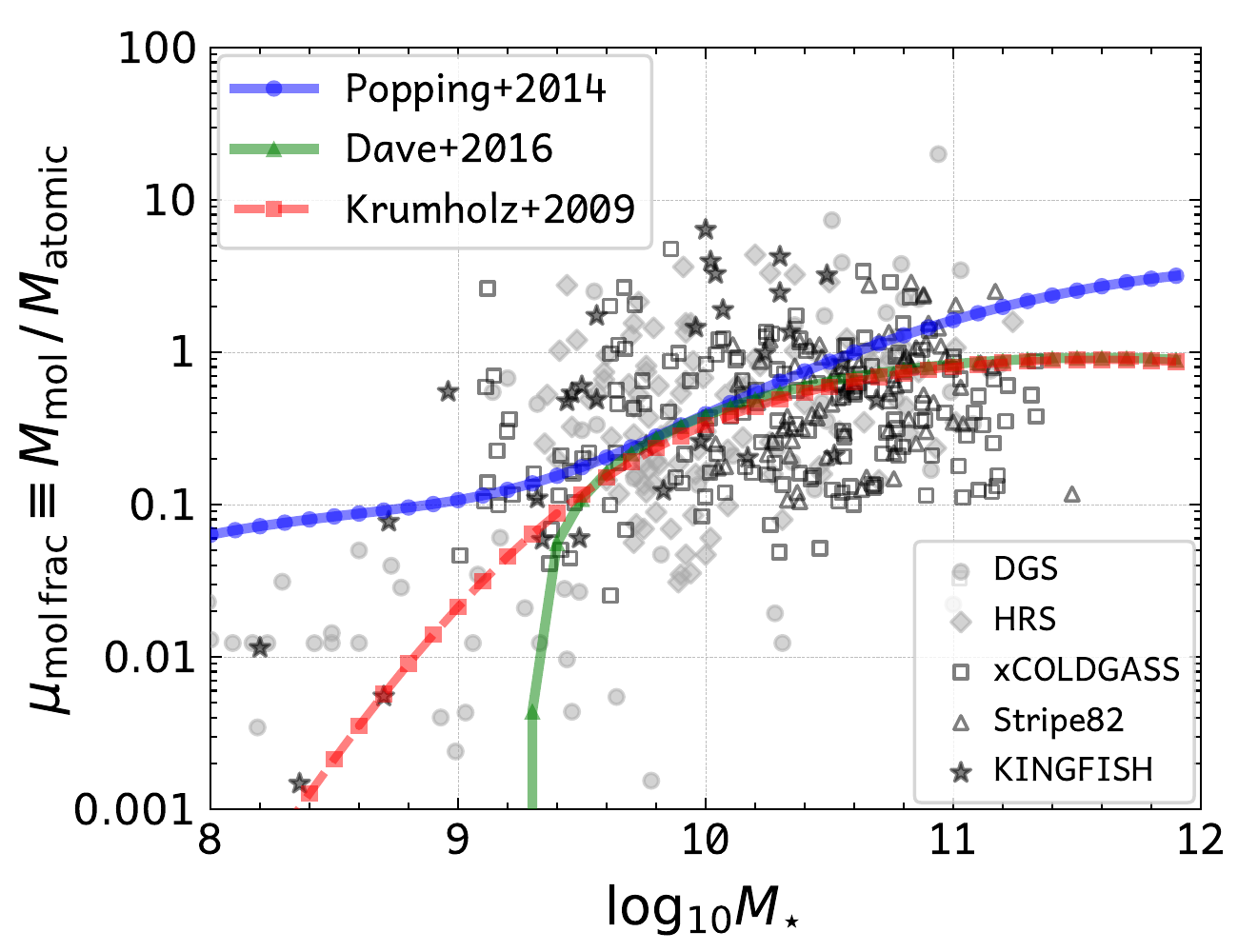}
\caption{%
Molecular-to-atomic gas mass ratio $\deltaMol\equiv \Mmolgas/\Matomicgas$ versus gas phase metallicity ($12 + \log_{10} (\mathrm{O/H})$; {\it left panel}; using the \citet[][PP04]{PP04} calibration) and stellar mass ({\it right panel}; using a \citet{Chabrier2003} initial mass function). 
Data points are compiled from the literature: see Table~\ref{Table_galaxy_samples} for the references of DGS, HRS, Stripe82 and KINGFISH surveys; in addition we used the atomic gas mass $M_{\mathrm{H\small{I}}}$ from \cite{Catinella2018} for the xCOLDGASS survey \citep{Saintonge2017}. 
Theoretical models from \citet[][their Eq.~8]{Popping2014}, \citet[][their Eqs.~1~and~2]{Dave2016} and \citet[][their Eq.~2]{KMT09} are overlaid as colored lines. 
\label{Plot_metalZ_molecular2atomic}
}
\end{figure}

\vspace{0.25truecm}
\subsection{Molecular hydrogen fraction ($\fmol$) versus metallicity}
\label{Section_ZFMOL}

The molecular hydrogen fraction, $\fmol \equiv \Mmolgas/\Mtotalgas$, is the ratio between molecular gas and molecular+atomic gas. In the following, we use $\deltaMol \equiv \Mmolgas/\Matomicgas$ for the molecular-to-atomic gas mass ratio. 

$\fmol$ (or $\deltaMol$) also correlates with metallicity, e.g., the amount of dust grains, as Hydrogen molecules form mainly on the surface of dust grains (e.g., \citealt{Hollenbach1971}), and the abundance of dust grains depends on the metal enrichment by recent star formation activities (e.g., \citealt{Draine2003ARAA}). 
The correlations between $\fmol$ and $\metalZOH$ and $\Mstar$ are illustrated in Fig.~\ref{Plot_metalZ_molecular2atomic}, where theoretical models from \cite{KMT09}, \cite{Popping2014} and \cite{Dave2016} are compared to the data. 

In Fig.~\ref{Plot_metalZ_molecular2atomic}, we show $\deltaMol$ versus metallicity and stellar mass with a large compilation of 
524 
galaxies from the literature (see labels and figure caption). All galaxies have $\Mmolgas$ from CO observations, $\Matomicgas$ from H{\sc i} observations, $\metalZOH$ from optical spectroscopy and $\Mstar$ from multi-wavelength optical/near-infrared data. The data points exhibit a large scatter in both panels, which is probably caused by the uncertainties in $\Mmolgas$, $\Matomicgas$ and metallicity. The metallicity-dependent CO-to-H$_2$ conversion factor has on average a $\sim38\%$ uncertainty in the \cite{Saintonge2017} catalog, where the conversion factor is computed based on \cite{Accurso2017}, and the observed CO line flux has a $\sim5-30\%$ uncertainty. H{\sc i} line flux has a $\sim2-20\%$ uncertainty in their catalog, and in addition the conversion from H{\sc i} line flux to $\Matomicgas$ may have a $30\%$ or higher uncertainty due to the assumption of optically thin H{\sc i} (e.g., \citealt{Fukui2018}). These uncertainties add up in total to at least $\sim50-60\%$ uncertainty for the Y-axis. 

Three theoretical models from \cite{Popping2014}, \cite{Dave2016} and \cite{KMT09} are overlaid as colored lines. Comparing with the data, the \cite{KMT09} model provides the best fit at the low-metallicity end. While at high metallicities, it seems the data is not statistically meaningful and all three models provide reasonable predictions. 

The figure shows that for local galaxies with solar-abundant metallicity and $\Mstar>10^{10}\;\Msun$, molecular gas nearly dominates the total gas mass ($\left<{\fmol}\right> > 50\%$). 
At higher redshifts, however, there is no observational constraint. 
We can only assume such scaling relations are still valid at higher redshifts. 
In principle, higher-redshift galaxies have higher SFRs and gas density at the same stellar mass (see Appx.~\ref{Section_MS}), $\fmol$ should be at least as high as those of similar stellar mass and metallicity local analogs. 
Thus it is common to ignore the atomic gas contribution in high-redshift galaxies with $\Mstar>10^{10}\;\Msun$ (e.g., \citealtalias{Tacconi2018}).

\vspace{0.25truecm}
\subsection{Mass metallicity relation (MZR)}
\label{Section_MZR}

The Fundamental Metallicity Relation (FMR; the correlation between metallicity, stellar mass and SFR; \citealt{Mannucci2010,Mannucci2011}) and Mass-Metallicity Relation (MZR; the correlation between metallicity and stellar mass; e.g., \citealt{Kewley2008}) are usually used to infer the metallicity and metallicity-related properties (e.g., $\alphaCO$, $\deltaGDR$) of high-redshift galaxies when no sufficient optical nebular emission line information is available. 
A number of FMR and MZRs exist in the literature (see below), with metallicity ($12+\log_{10} (\mathrm{O/H})$) parameterized as a function of $\Mstar$ and/or $z$ or SFR. However, whether the FMR and MZRs are valid across cosmic time or within a given stellar mass range is less studied.

\begin{figure}[htb]
\centering
\includegraphics[width=0.325\linewidth]{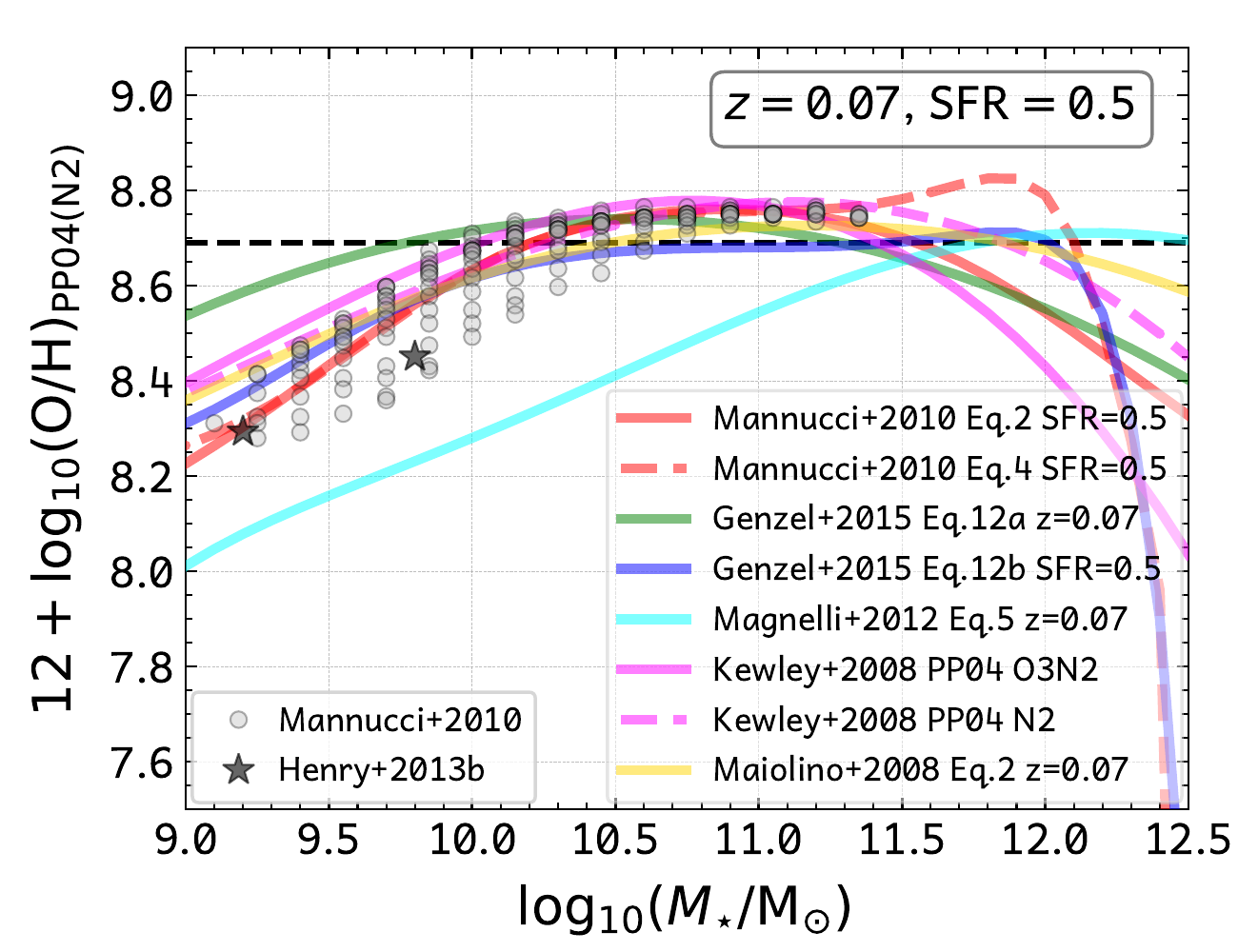}
\includegraphics[width=0.325\linewidth]{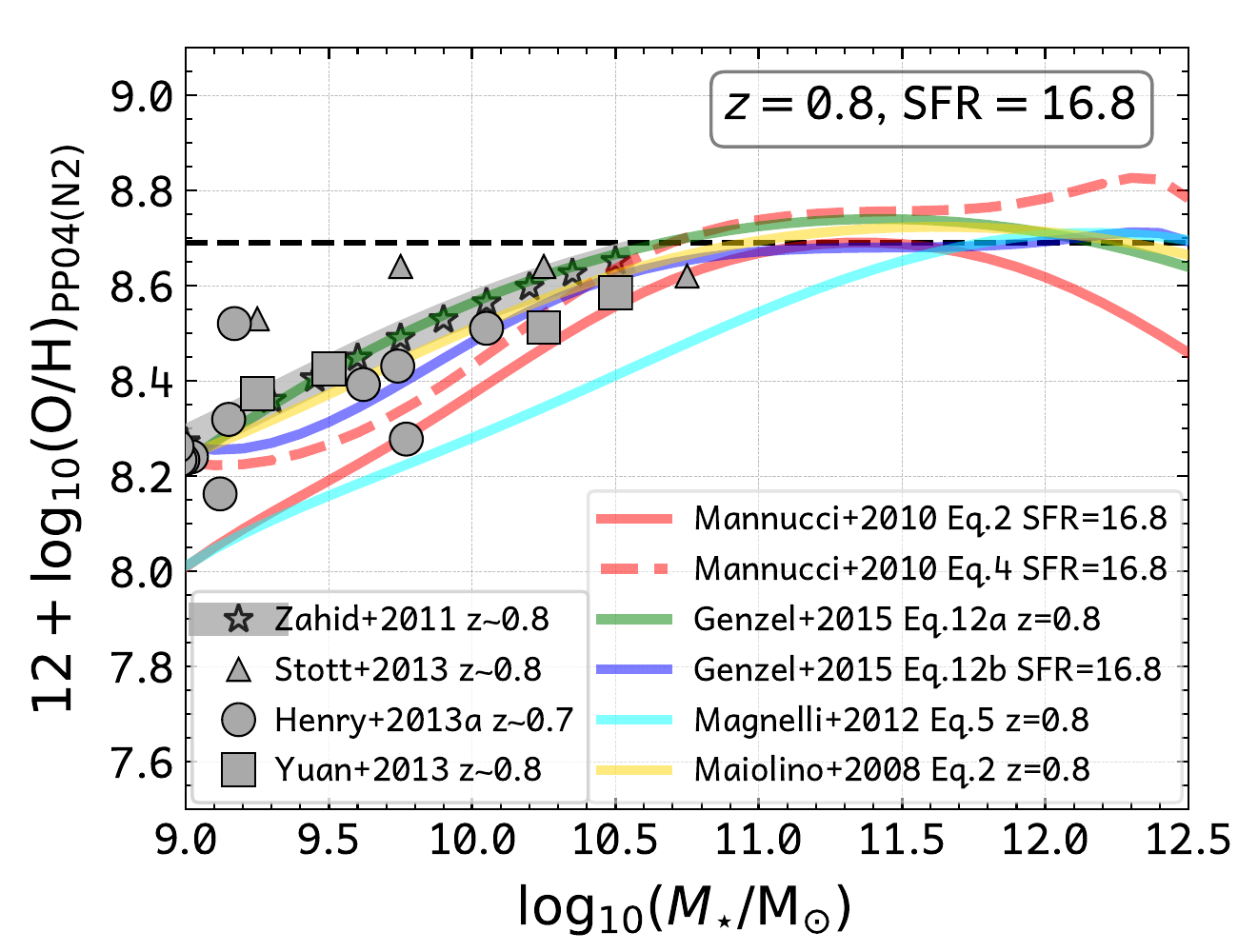}
\includegraphics[width=0.325\linewidth]{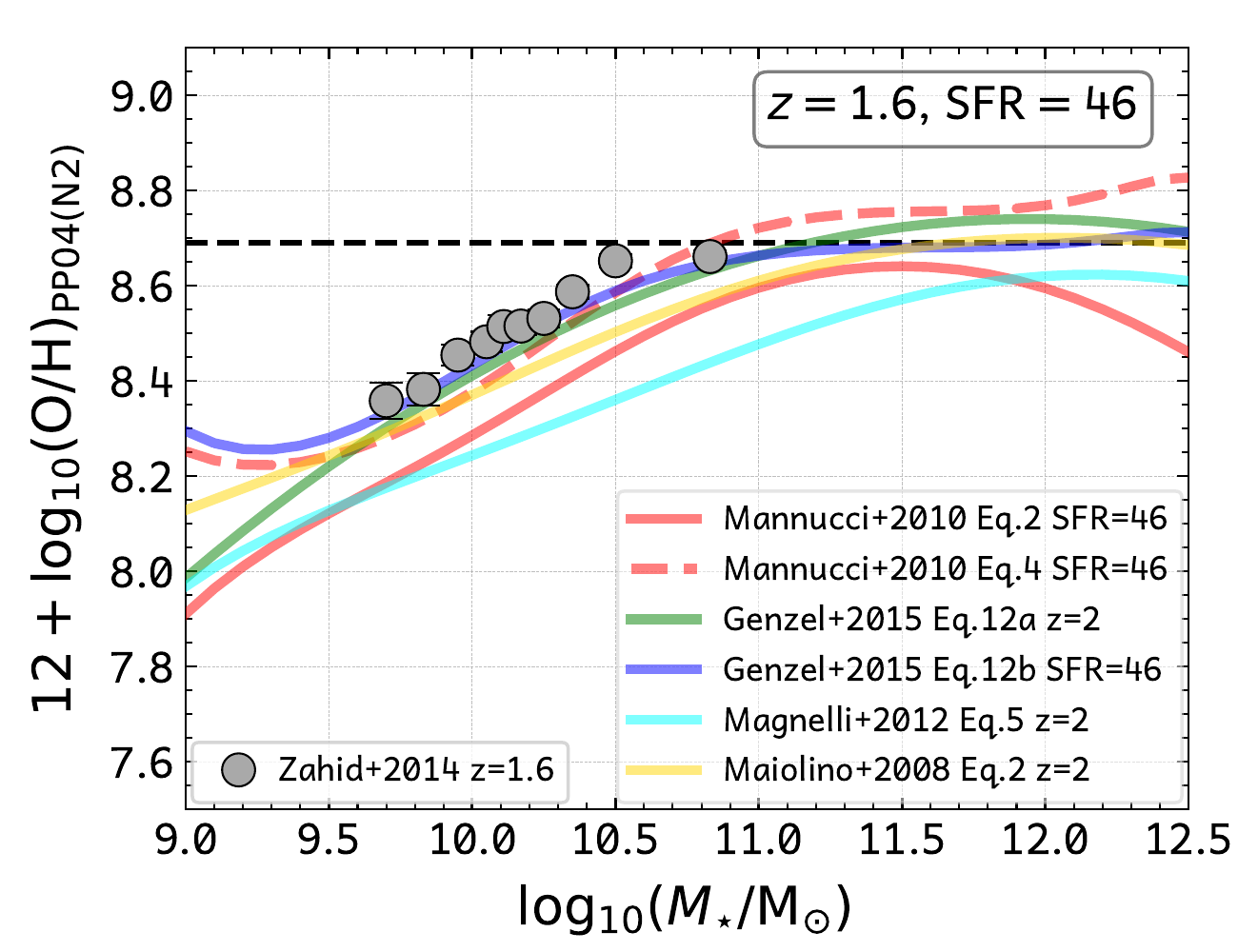}
\includegraphics[width=0.325\linewidth]{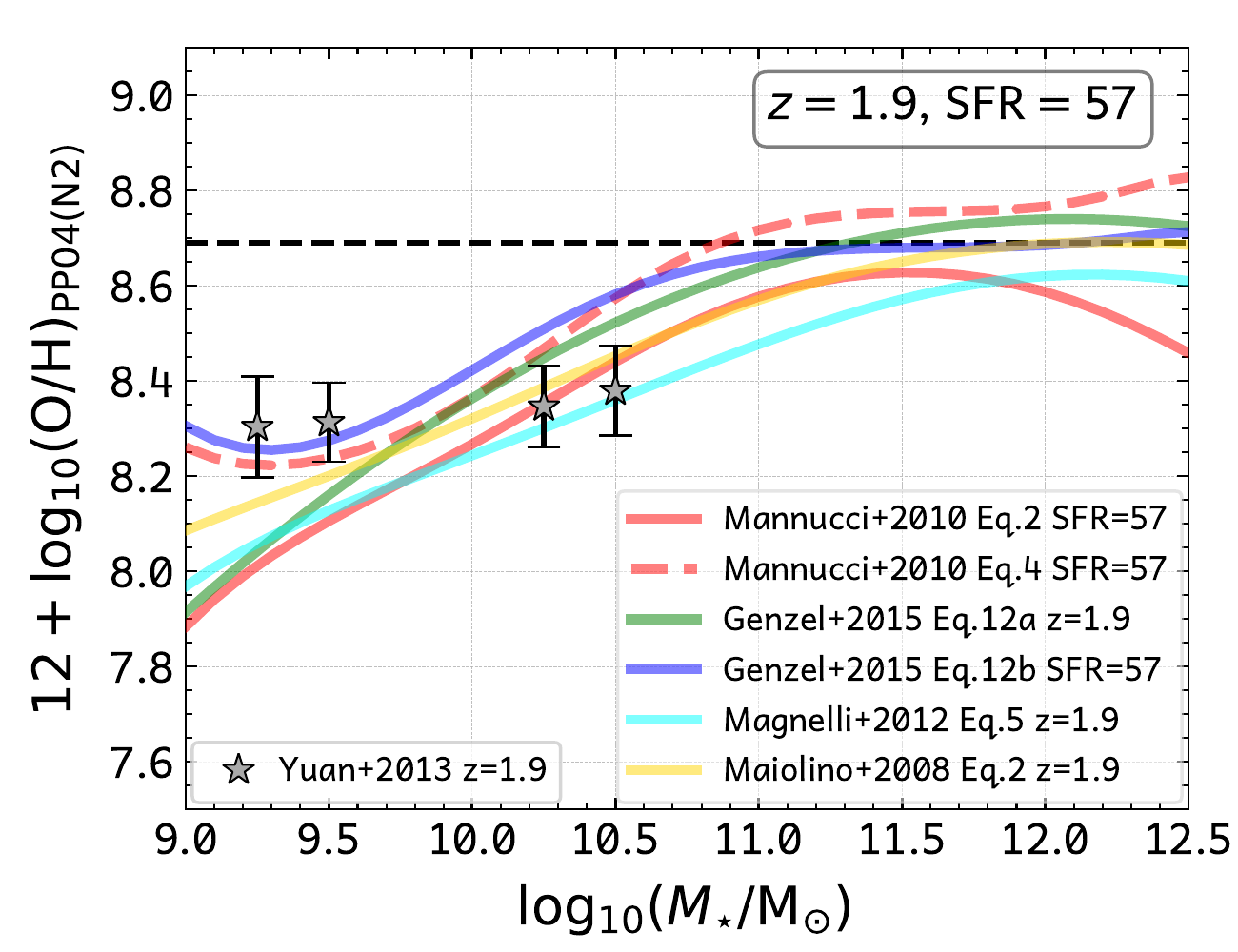}
\includegraphics[width=0.325\linewidth]{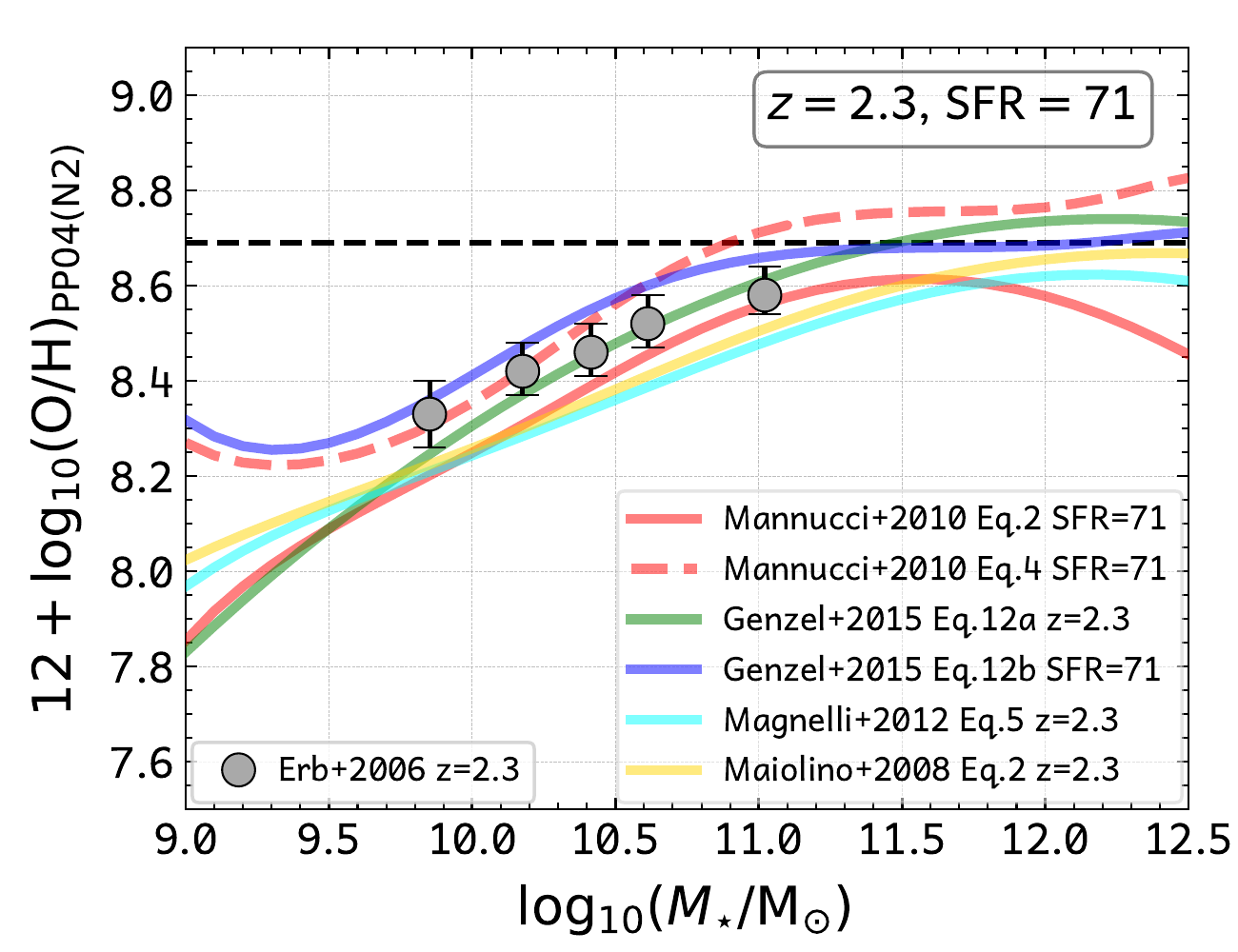}
\includegraphics[width=0.325\linewidth]{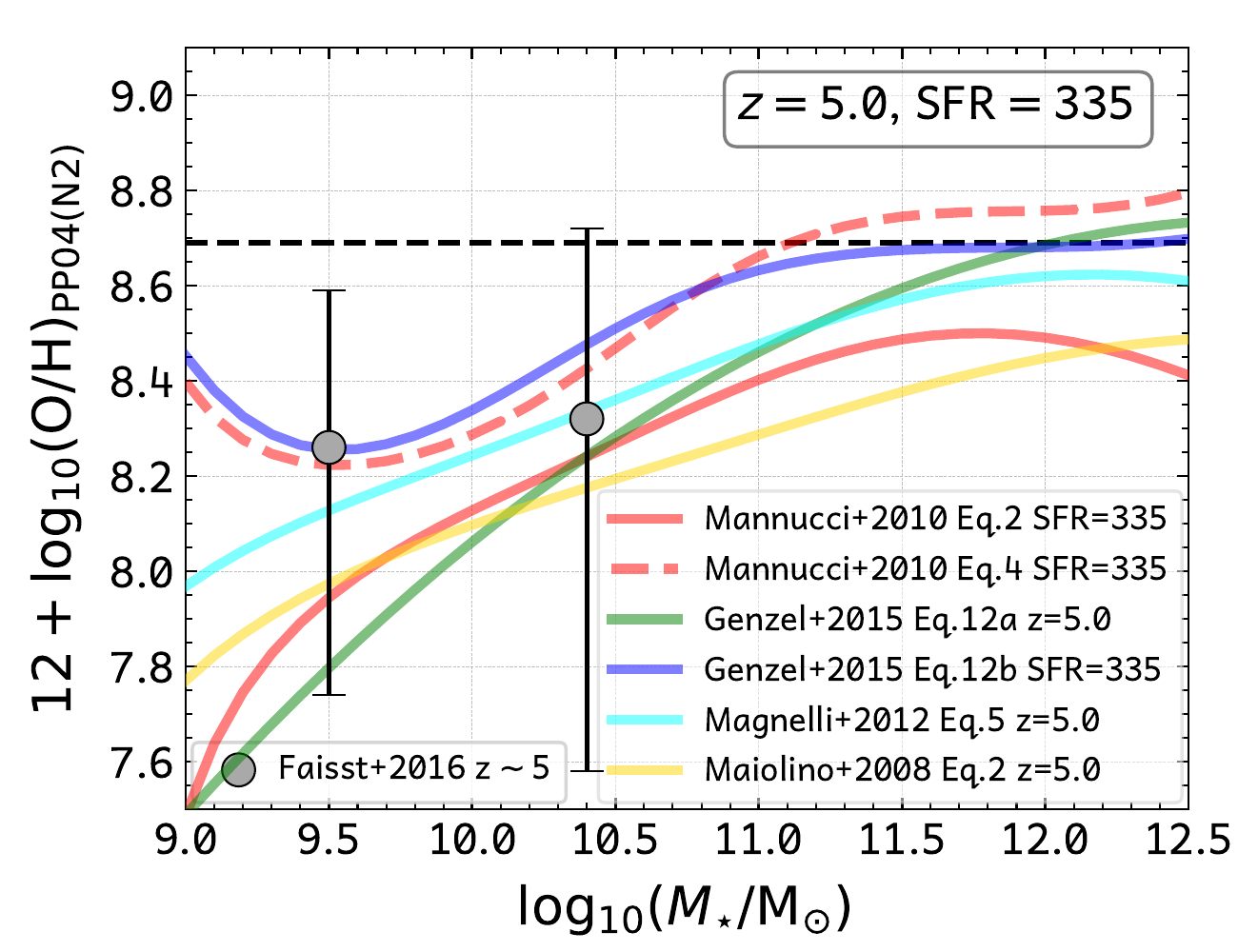}
\includegraphics[width=0.325\linewidth]{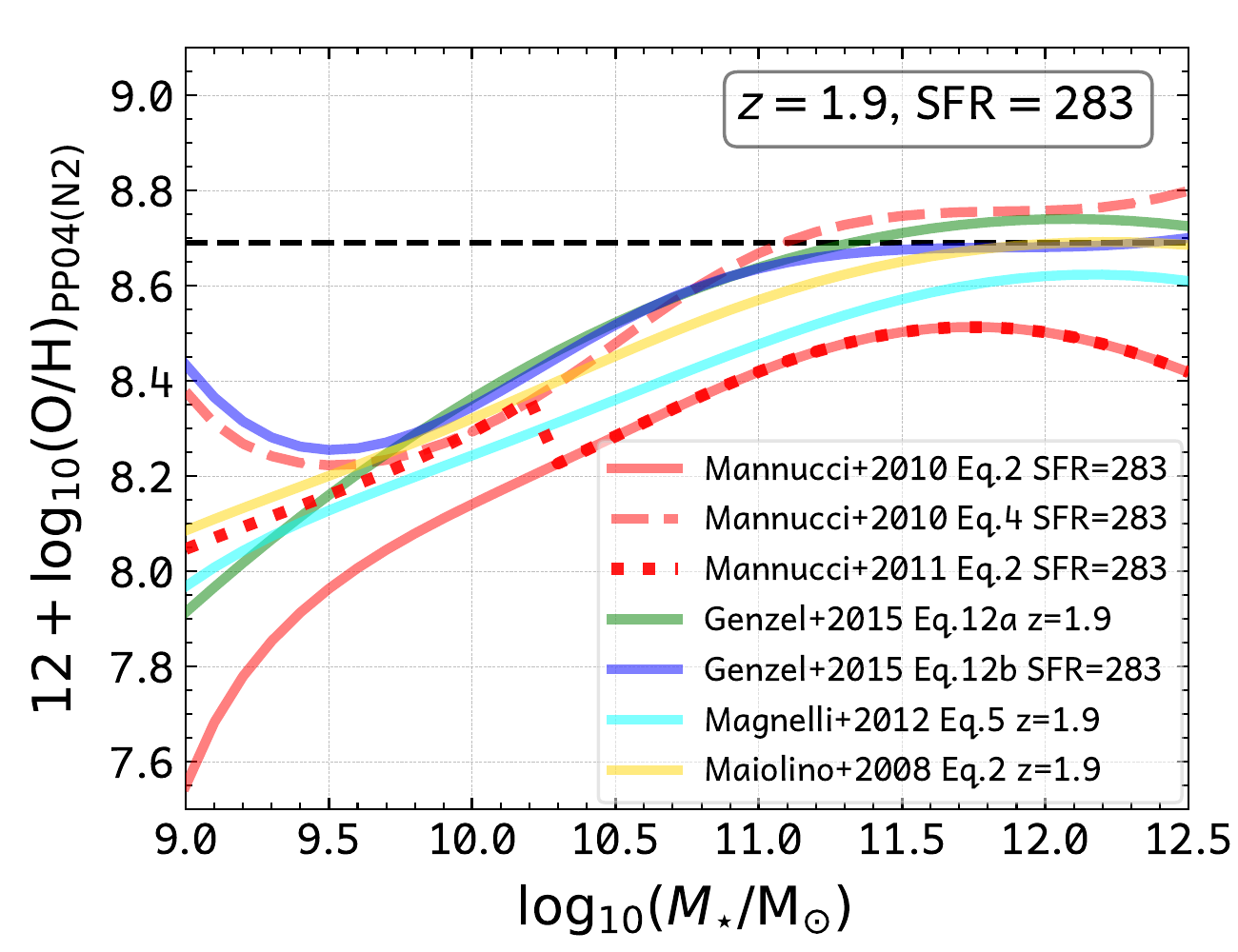}
\includegraphics[width=0.325\linewidth]{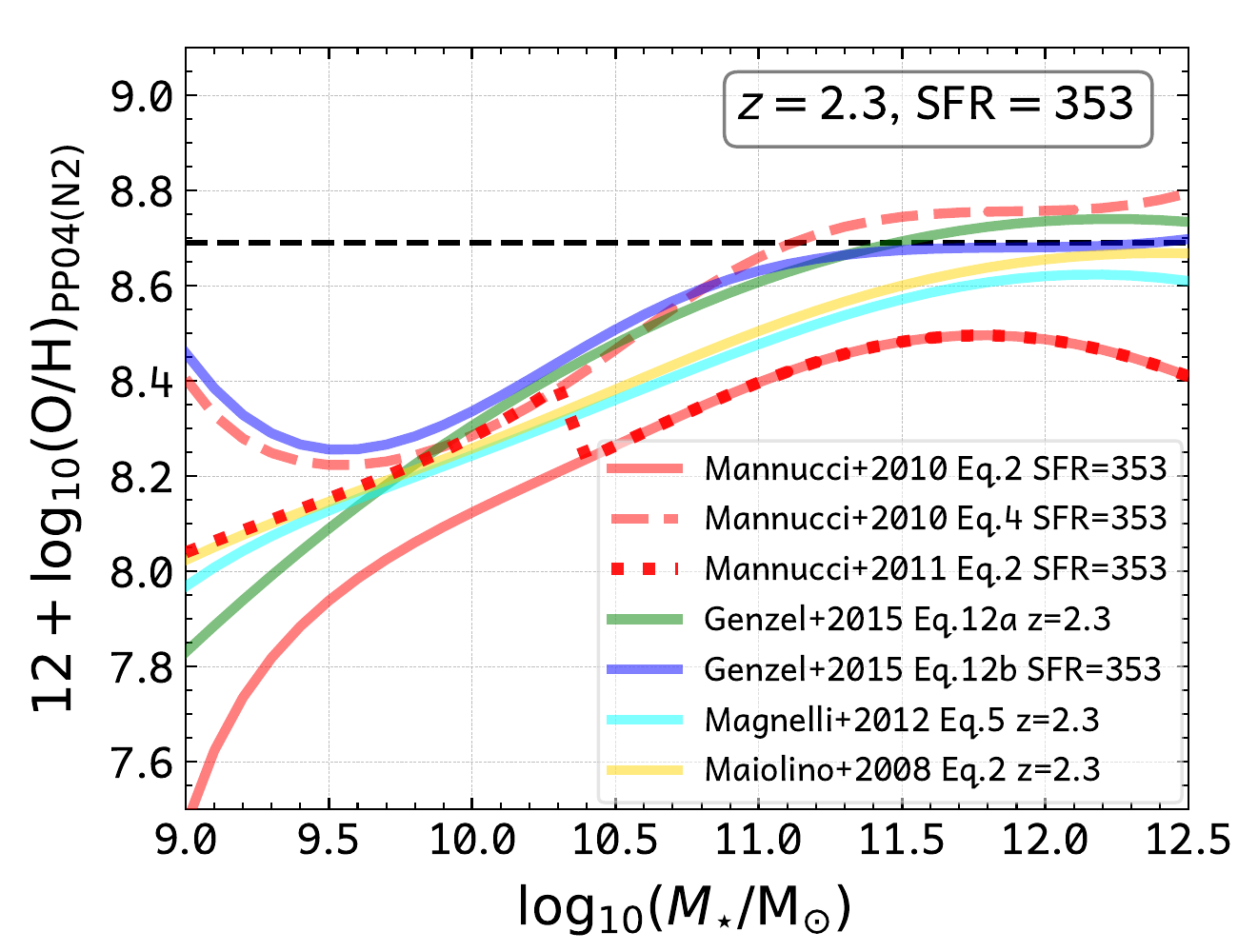}
\includegraphics[width=0.325\linewidth]{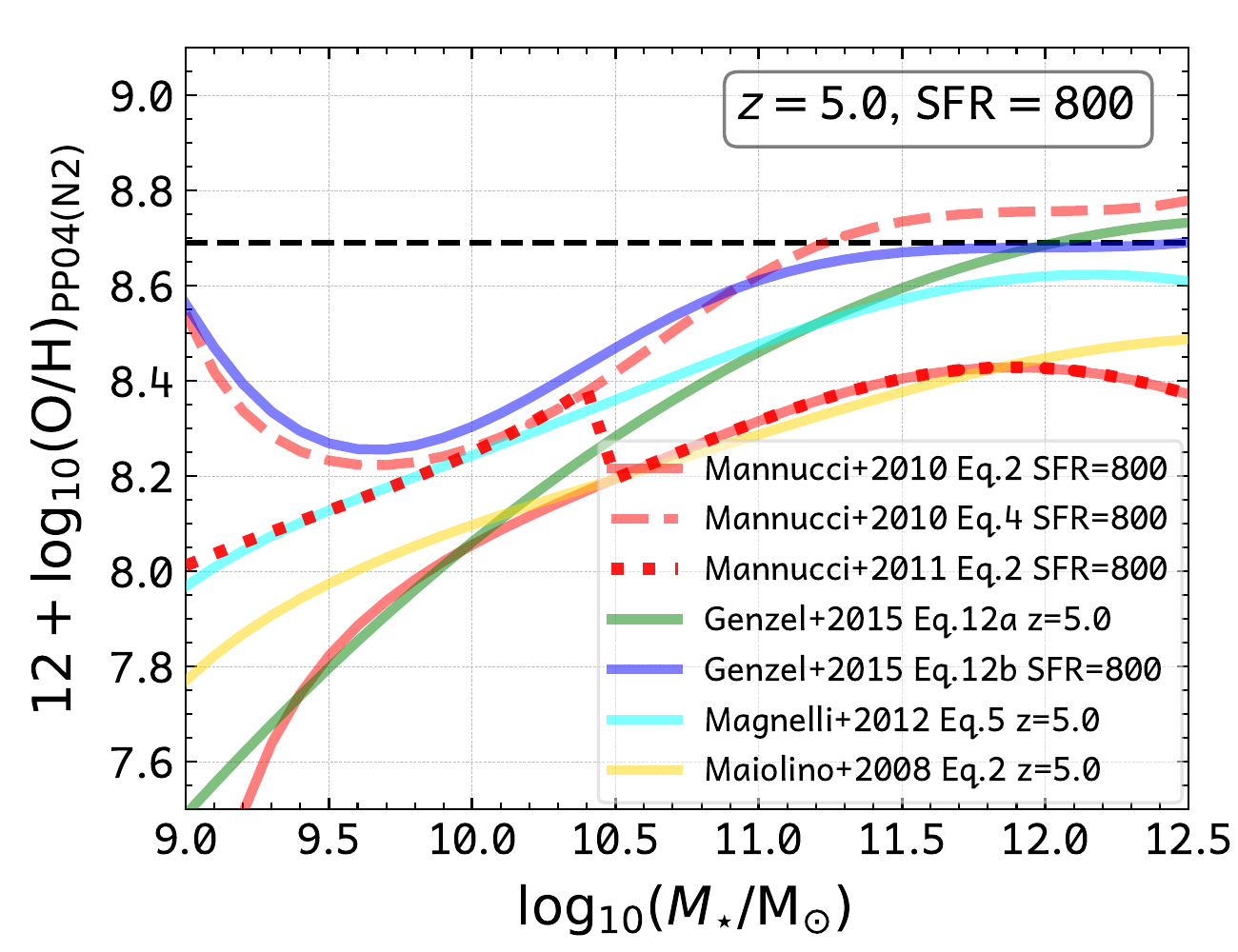}
\vspace{1ex}
\caption{Comparison of different metallicity relations in the literature for main-sequence galaxies in six redshift bins ({\it the top two rows}) and for starburst galaxies in three redshift bins ({\it the bottom row}). Metallicity relations are equations that compute metallicity from stellar mass, SFR, and/or redshift. The curves in each panel are the metallicity computed with each equation in bins of stellar mass, but at a given redshift and SFR as labeled in the top right corner. In each panel, equations are as labeled at the bottom right and described in Appx.~\ref{Section_MZR}, while data points are taken from observations in the literature as labeled in the bottom left. The horizontal black dashed line indicates solar metallicity. }
\label{Plot_metalZ_Mstar_SFR_z_bins}
\end{figure}

Here we take the following seven most widely used FMR and MZRs for high-redshift studies and compared them in Fig.~\ref{Plot_metalZ_Mstar_SFR_z_bins} so that their validities can be more clearly seen in bins of redshift:
\begin{itemize}
\item \cite{Mannucci2010} presented their FMR in their Eqs.~2~and~4 for local galaxies. Their metallicity values are originally calibrated from optical emission lines following the \citet[][M08]{Maiolino2008} prescription instead of \citet[][PP04]{PP04} therefore we converted their derived metallicity to the PP04 calibration by solving both \cite{Mannucci2010} Eq.~1 (the polynomial form, instead of their Eq.~2) and \cite{Maiolino2008} Eq.~1 (with their Table~4's second row coefficients). 
Both their Eqs.~2~and~4 are shown in Fig.~\ref{Plot_metalZ_Mstar_SFR_z_bins}. 
The caveat of their Eq.~2 includes that: (a) at a given redshift and SFR, it first increases with stellar mass then drops quickly when $\logMstar>11.2$; and (b) it predicts the lowest metallicity for starburst galaxies at $z>1$. 
And the caveat of their Eq.~4 is the nonphysical extrapolation for (a) main-sequence galaxies at all redshift with $\logMstar>11.2$; and (b) $z\gtrsim1$ main-sequence galaxies with $\logMstar<9.5$. 
\item \cite{Mannucci2011} provided an updated version of their FMR in \cite{Mannucci2010} for lower-mass galaxies. The update is only for $\logMstar - 0.32 \times \log_{10}\mathrm{SFR} < 9.5$ case, i.e., low mass and/or high SFR. 
Therefore we show this equation only in the bottom panels for starburst galaxies. Note that there is a nonphysical jump in metallicity at $\logMstar\sim10.2-10.5$. 
\item \cite{Genzel2015} Eq.(12a) is originally from \cite{Wuyts2014} and also adopted by \citetalias{Tacconi2018} in the identical form. This formula considers both redshift and stellar mass as the parameters determining metallicity. It predicts reasonable metallicities except at $\logMstar\gtrsim11$ at local (or $\logMstar\gtrsim11.5-12$ at $z\gtrsim1$). This motivates our modification of this equation as described in Eq.~\ref{Equation_metallicity}. 
\item \cite{Genzel2015} Eq.(12b) is based on \cite{Mannucci2010}'s Eq.~4, with a M08-to-PP04 conversion applied by solving both \cite{Mannucci2010} Eq.~2 and \cite{Maiolino2008} Eq.~1 (see their Table~4's second row coefficients). Therefore the curve of this equation is very similar to the \cite{Mannucci2010} Eq.~4 curve, yet we caution that the M08-to-PP04 conversion is different in their work than here, and our conversion (by solving \cite{Mannucci2010} Eq.~2 and \cite{Maiolino2008} Eq.~1) should be more precise. 
\item \cite{Magnelli2012} Eq.~5 uses the \cite{Denicolo2002} calibration. Thus we convert this calibration to the PP04 N2 calibration following \cite{Kewley2008}. They assumed two different MZRs distinguished by redshift at 1.5. 
We caution that it is much lower at low-redshift ($z\lesssim1$). It also always predicts sub-solar metallicity for galaxies at $z>1$. 
\item \cite{Kewley2008} PP04 O3N2 and N2 MZRs as listed in their Table~2\,\footnote{Note that their equation in the Table~2 caption should be $y = a + b x + c x^2 + d x^3$.}. Their equation only depends on stellar mass and has no redshift evolution, therefore we only show their curve in the first panel. It predicts too high metallicity for high-redshift galaxies with $\logMstar<10.5$, and like \cite{Mannucci2010} Eq.~2, it also has a nonphysical drop with increasing stellar mass when $\logMstar>11.2$. 
\item \cite{Maiolino2008} Eq.~2 with the coefficients listed in their Table~5. They fitted five different MZRs at five redshifts they analyzed. Here we linearly interpolate their coefficients in redshift so as to plot their curves in Fig.~\ref{Plot_metalZ_Mstar_SFR_z_bins}. The equation seems reasonable at low-$z$ ($z<2-3$) but predicts significantly sub-solar metallicity at $z\gtrsim5$ even for starbursts. 
\end{itemize}

Fig.~\ref{Plot_metalZ_Mstar_SFR_z_bins} shows that most of these formulae are consistent (within 0.2\,dex) only for main-sequence galaxies at $z\lesssim2$ and with $\logMstar\sim10-11.2$. Subtle differences exist among these curves and the reader should consider the proper choice of MZR or FMR to use. A small difference, e.g., a 0.1\,dex lower/higher metallicity, could translate into a factor of 1.6 higher/lower $\deltaGDR$ when assuming the $\deltaGDR$--metallicity relation in Fig.~\ref{Plot_metalZ_deltaGD}.
Also note that the prescription for deriving $\metalZOH$ from optical emission lines is important, as can be seen by comparing the first and third formulae at the high-mass end. Finally, we caution that these formulae do not agree well at the low-mass regime ($\Mstar < 10^{10.0} \; \Msun$). But for the study in this work, although with such a large ALMA sample, we still do not probe such low-mass galaxies. Therefore these discrepancies are currently not an issue.

\vspace{0.25truecm}
\subsection{Stellar mass--SFR main sequence (MS)}
\label{Section_MS}

\begin{figure*}[htbp]
\centering
\includegraphics[width=\textwidth]{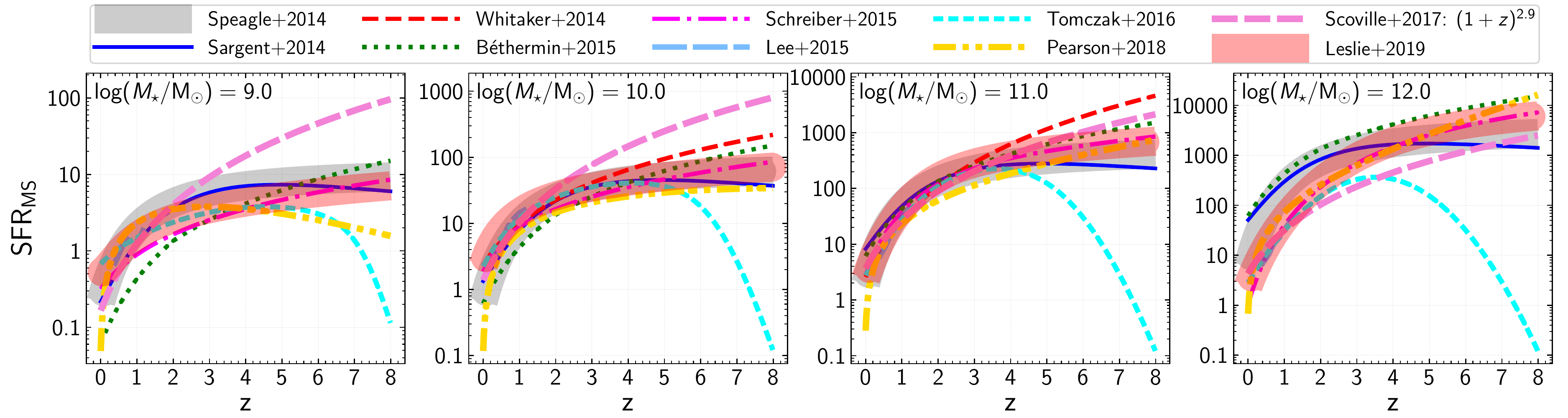}
\caption{%
Comparison of galaxy star-forming main-sequence functions in the literature as labeled at the top: \citet{Speagle2014}, \cite{Sargent2014}, \cite{Whitaker2014}, \cite{Bethermin2015}, \cite{Schreiber2015}, \cite{Lee2015}, \cite{Tomczak2016}, \cite{Pearson2018}, \citetalias{Scoville2017} and Leslie et al. (subm.). 
We show the MS evolution curves predicted by these functions at four representative stellar masses ({\it from left to right}): $\log_{10} (\Mstar/\Msun)=9.0$, $10.0$, $11.0$ and $12.0$. See discussion in Appx.~\ref{Section_MS}.
\label{Plot_SF_MS_bin_by_lgMstar_horizontal}
}
\end{figure*}

In Sect.~\ref{Section_Fitting_Multi_Variate_Function} we mentioned that we adopt the \cite{Speagle2014} MS (the \#49 fitting in their Table~7) with cosmic time as the variable. 
In Fig.~\ref{Plot_SF_MS_bin_by_lgMstar_horizontal} we compare a number of MS relations in the literature (see the labels therein). 
The \cite{Whitaker2014}, \cite{Lee2015} and \cite{Tomczak2016} MS are only valid at $z \lesssim 3$. The \cite{Sargent2014} MS predicts the lowest $\mathrm{sSFR_{MS}}$ at $z > 5$, while the \cite{Bethermin2015} MS predicts a factor of 2 higher $\mathrm{sSFR_{MS}}$ than average at $z > 5$ and is in general higher for $\log_{10} \Mstar > 10^{11}$ galaxies. The \cite{Pearson2018} MS is a factor of $<2$ lower than others at $z\sim2-3$ and in general lower for $\log_{10} \Mstar < 10^{10}$ galaxies. 
These MS calibrations have large scatter in the $\log_{10} \Mstar < 10^{9}$ and $\log_{10} \Mstar \gtrsim 10^{12}$ regimes which lack observational data. 
The \cite{Speagle2014} MS (with cosmic age) is closer to the average of all MS analyzed, therefore we adopt it for our work.

\begin{figure*}[htbp]
\centering
\includegraphics[width=0.825\textwidth]{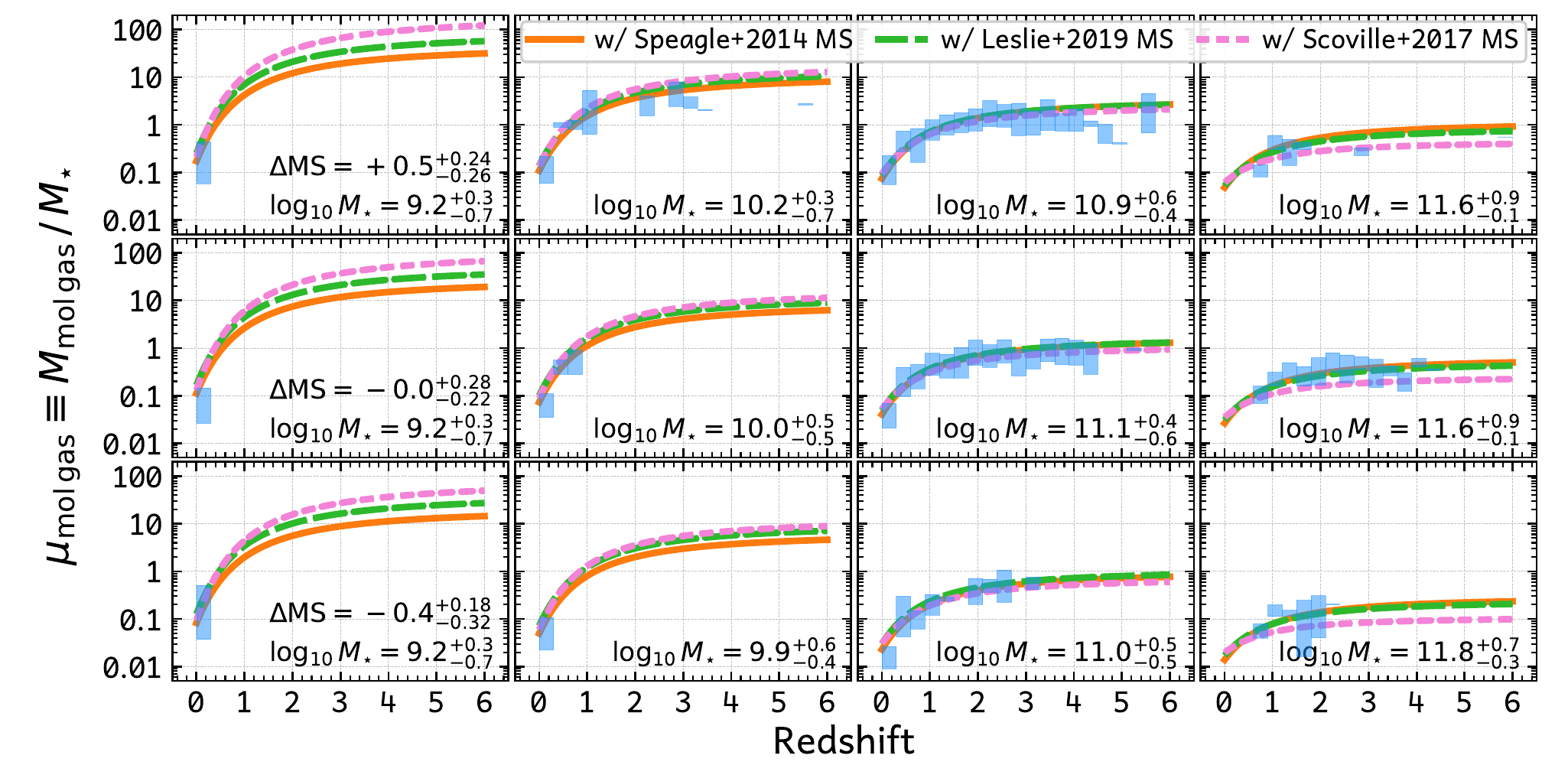}
\caption{%
Similar to Fig.~\ref{Plot_z_deltaGas_comparison}, comparing the evolution of gas fraction $\deltaGas$ curves obtained by using the Leslie et al. (subm.) MS (green long-dashed line), \citetalias{Scoville2017} MS (pink short-dashed line) and the \citet{Speagle2014} MS (orange solid line; same as in Fig.~\ref{Plot_z_deltaGas_comparison}). 
See description of the binning scheme and data boxes in Figs.~\ref{Plot_z_tauDepl_comparison} and~\ref{Plot_z_deltaGas_comparison} captions. 
And see discussion in Appx.~\ref{Section_MS}. 
\label{Plot_z_deltaGas_comparison_with_Leslie2019_MS}
}
\end{figure*}

The Leslie et al. (subm.) MS is potentially an alternative choice for a most reasonable MS to use. They derived the MS correlation from redshift $\sim0.2$ to $\sim6$ by stacking the VLA 3GHz large program data (\citealt{Smolcic2017a}; covering 2 sq. deg. COSMOS field with a sensitivity of $1\,\sigma\sim2.3\,\mu\mathrm{Jy}/\mathrm{beam}$ at a spatial resolution of 0.75'') using a large sample of $\sim300,000$ galaxies from the \citealt{Laigle2016} and \citealt{Davidzon2017} catalogs. The largest difference between the Leslie et al. MS and the \citealt{Speagle2014} one is that the former exhibits a flattening for a higher stellar mass, while the latter is a straight line at each redshift. Such a flattening, yet debated, has also been reported by other stacking studies (e.g., \citealt{Schreiber2015}; \citealt{Lee2015}; \citealt{Tomczak2016}). We refer the reader to these papers and Leslie et al. (subm.) for details on the shapes of different MS functions. Here we investigate how an MS with flattening affects our functional form by adopting the Leslie et al. MS and repeating our analysis from Sects.~\ref{Section_Fitting_Multi_Variate_Function}~to~\ref{Section_Results_z_deltaGas}. In Fig.~\ref{Plot_z_deltaGas_comparison_with_Leslie2019_MS}, we compare the obtained $\deltaGas$ evolution curves using the Leslie et al. MS to those using the \citet{Speagle2014} MS.

Fig.~\ref{Plot_z_deltaGas_comparison_with_Leslie2019_MS} shows that adopting the Leslie et al. MS results in an at most factor of two higher gas fraction for low-mass, high-redshift galaxies ($\logMstar\lesssim9.2$, $z\sim2-6$), while being indistinguishable from using the \cite{Speagle2014} MS for galaxies with $\logMstar\gtrsim10-11$. The difference at the low-mass end is likely because the Leslie et al. MS predicts two times higher sSFRs for low-mass galaxies at $z\sim0$ while a factor of two lower sSFR at $z\sim3$, as shown in the left panel of Fig.~\ref{Plot_SF_MS_bin_by_lgMstar_horizontal}. This leads to a systematically lower $\DeltaMS$ for low-mass galaxies at $z\sim0$, altering the slope of $\deltaGas$ versus $\DeltaMS$ to be shallower (by a small change of 0.04 in the coefficient $\mathsf{a}$ in Eq.~\ref{Equation_tauDepl_deltaGas}), meanwhile steepening the slope of $\deltaGas$ versus $\logMstar$ (by a change of 0.14 in the coefficient $\mathsf{b}$ in Eq.~\ref{Equation_tauDepl_deltaGas}). Thus it results in a $<2\times$ higher extrapolation for the gas fraction at the low-mass end. 
We note that the fits are indistinguishable where data are rich, i.e., using either Leslie et al. or \citet{Speagle2014} MS makes no obvious difference for $\logMstar\gtrsim10-11$ galaxies at all redshifts ($z\sim0-6$).

In Fig.~\ref{Plot_SF_MS_bin_by_lgMstar_horizontal}, we additionally show the MS relations used in \citetalias{Scoville2017} following the equations in their Sects.~2.1~and~2.2. Although their equation is not aimed for extrapolating out to $z>3-4$ (and exhibits a large excess compared to others), here we show their curve and use their MS to compute $\DeltaMS$ for our data out to $z\sim6$ for the sole purpose of evaluating the validity of their MS at these redshifts. In Fig.~\ref{Plot_z_deltaGas_comparison_with_Leslie2019_MS} we also repeated the fitting with our Eq.~\ref{Equation_tauDepl_deltaGas} functional form and used the \citetalias{Scoville2017} MS for $\DeltaMS$ normalization. The implied gas fraction evolution curve is $\sim2\times$ ($\sim4\times$) higher than that using the Leslie et al. MS (\citealt{Speagle2014} MS) at the low-mass end, meanwhile it also exhibits a $\sim2.5\times$ lower gas fraction at the massive end. This means that the difference in MS can indeed explain about half of the difference between our and \citetalias{Scoville2017} molecular gas mass density curves seen in Fig.~\ref{Plot_z_deltaGas_comparison} (the shape of the functional form is likely responsible for the other half of the difference).

\FloatBarrier

\section{Biases in Band Conversion with MAGPHYS SED Fitting}
\label{Section_Appendix_SED_band_conversion}

To verify the potential bias in using \textsc{MAGPHYS} SED fitting to predict the rest-frame RJ-tail (i.e., 850\,$\mu$m) dust continuum, we have done some tests using the multi-wavelength data from UV to submm for the JINGLE survey galaxy sample (\citealt{Saintonge2018}; \citealt{Smith2019}). JINGLE galaxies are main-sequence star-forming galaxies in the local Universe ($z<0.05$) with well-sampled SEDs including: JCMT/SCUBA2 850\,$\mu$m (\citealt{Smith2019}), \textit{Herschel} 70--500\,$\mu$m (\citealt{Pilbratt2010}; \citealt{Poglitsch2010}; \citealt{Griffin2010}), \textit{Spitzer} 3.6--24\,$\mu$m (\citealt{Werner2004}), \textit{WISE} 3.4--22\,$\mu$m (\citealt{Wright2010}), VISTA 1.2--2.2\,$\mu$m (Sutherland et al. 2015), 2MASS $J$, $H$ and $K$ (Skrutskie et al. 2006), SDSS optical (York et al. 2000; Eisenstein et al. 2011) and GALEX UV (Morrissey et al. 2007). 

We run the following \textsc{MAGPHYS} fitting tests: (a) fitting all data points at $\lambda \le 250\,\mu\mathrm{m}$, mimicking the $\lambda_{\mathrm{rest}} \le 250 \,\mu\mathrm{m}$ cases in Fig.~\ref{Plot_rest_frame_wavelength_vs_redshift} in the main text; (b) fitting all $\lambda < 8\,\mu\mathrm{m}$ photometry data points plus only one $\lambda = 160\,\mu\mathrm{m}$ data point, mimicking the cases where we have only one ALMA data point for fitting the whole dust SED (see Fig.~\ref{Plot_rest_frame_wavelength_vs_redshift}). 

\begin{figure}[h!]
\centering
\includegraphics[width=0.497\linewidth]{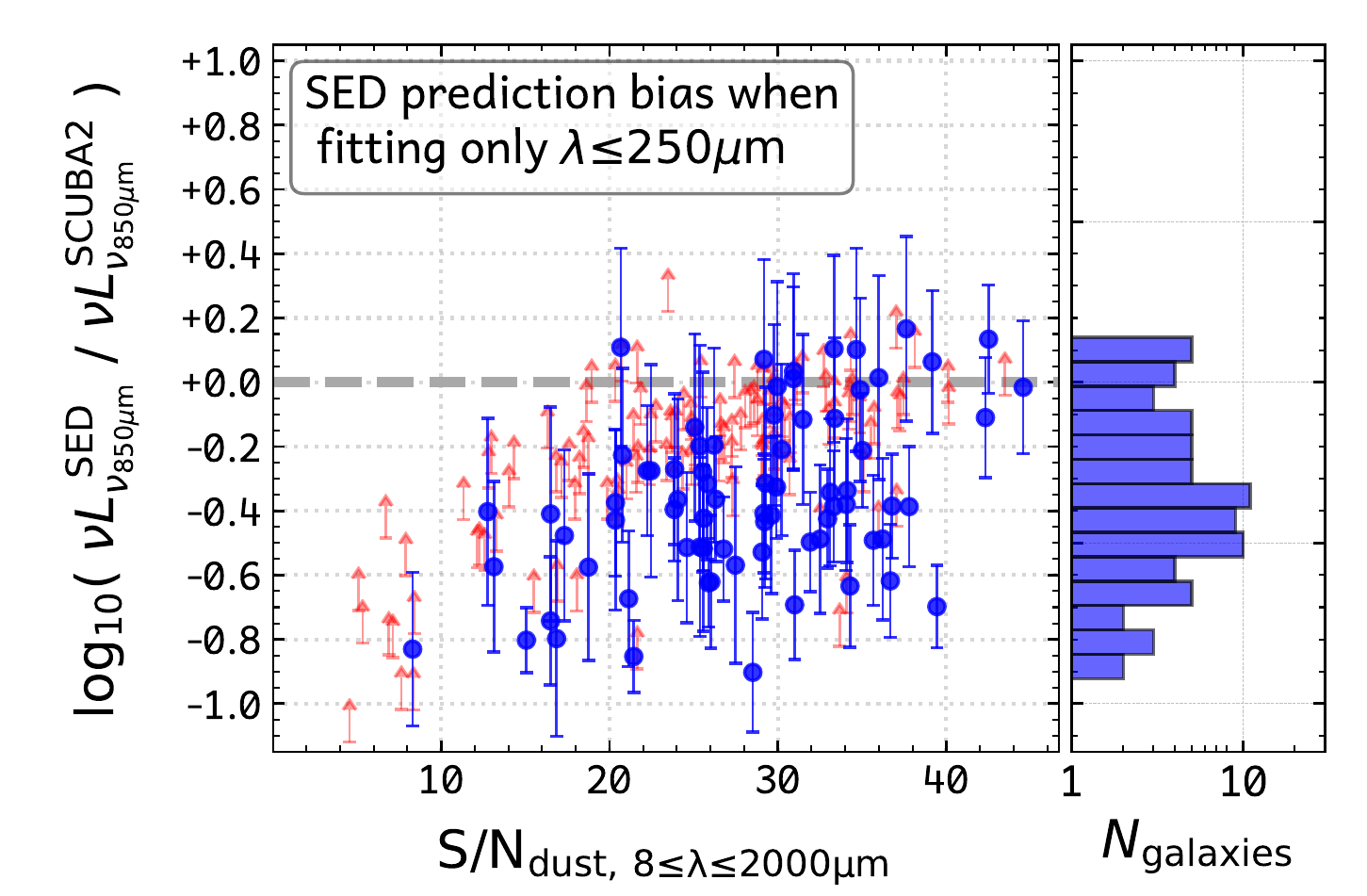}
\includegraphics[width=0.497\linewidth]{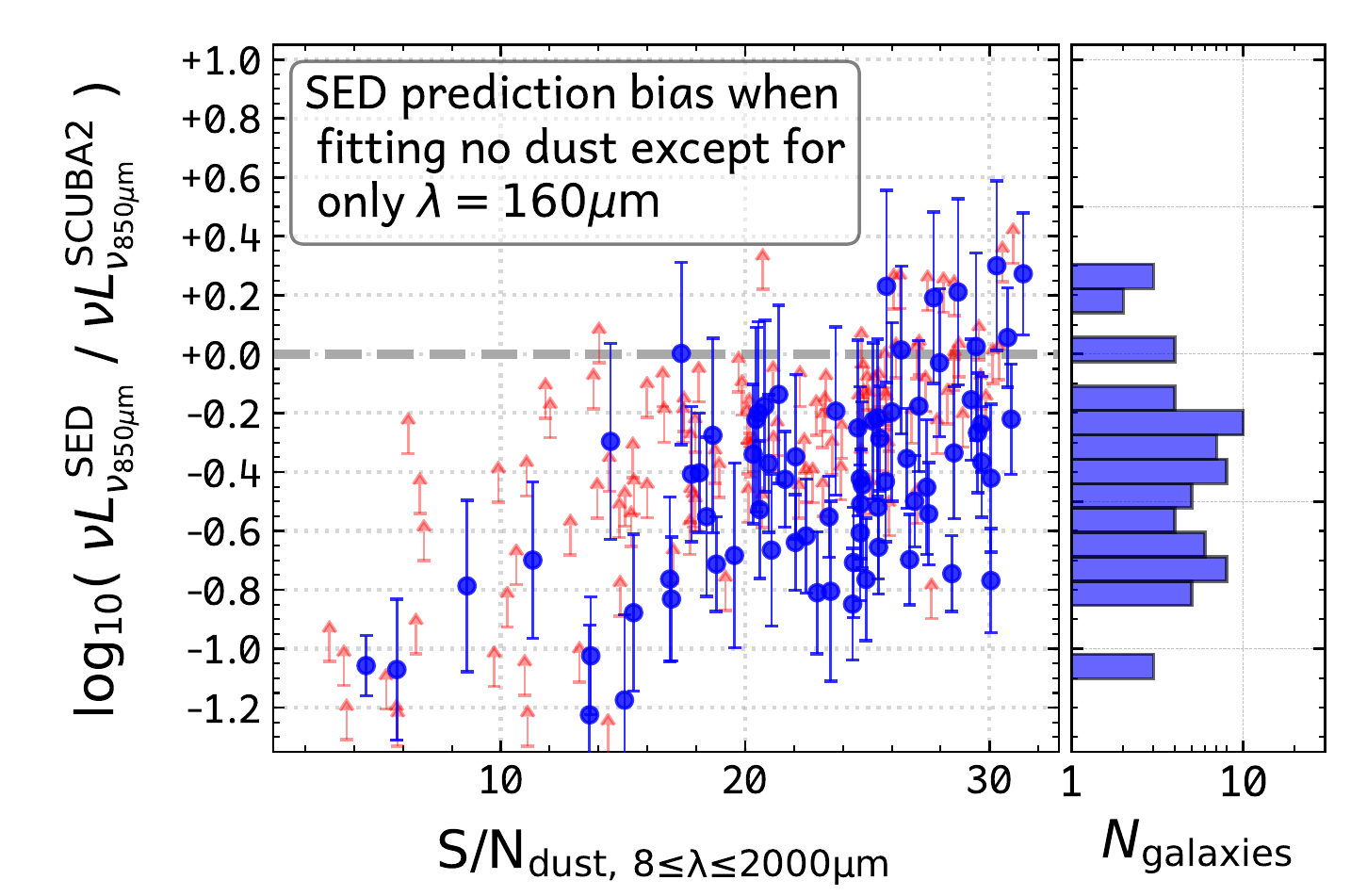}
\vspace{0.2ex}
\caption{%
Tests of \textsc{MAGPHYS} SED fitting with the JINGLE survey local ($z<0.05$) galaxy sample (\citealt{Saintonge2018}) and multi-wavelength data (from UV to SCUBA2 850\,$\mu$m; \citealt{Smith2019}), which show the biases of \textsc{MAGPHYS} SED fitting when ({\it left}): fitting all photometry data points at $\lambda \le 250\,\mu\mathrm{m}$ while no data point at longer wavelength is used; and ({\it right}) fitting photometry data points up to $\lambda < 8\,\mu\mathrm{m}$ plus only one $\lambda = 160\,\mu\mathrm{m}$ data point for the dust component. 
Galaxies with observed SCUBA2 850\,$\mu$m flux $\SNR \ge 3$ ($\SNR < 3$) are shown as blue circles with error bars indicating the observational errors (red arrows representing $3\,\sigma$ upper limits in the observed fluxes). 
See Appx.~\ref{Section_Appendix_SED_band_conversion} for details. 
}
\label{Plot_rest_frame_SED_flux_difference_no_250um}
\end{figure}

We compare the SED-predicted 850\,$\mu$m fluxes from both tests to the true observed 850\,$\mu$m fluxes in Fig.~\ref{Plot_rest_frame_SED_flux_difference_no_250um} (left and right panels, respectively). As also mentioned in Sect.~\ref{Section_Band_conversion}, the SED-predicted fluxes tend to be lower than the true fluxes, and the accuracy of predicting 850\,$\mu$m flux seems to depend on the $\SNR$ of all data points fitted for the dust component SED. When dust SED data points have a $\SNR\sim20$, the rest-frame 850\,$\mu$m fluxes tend to be underestimated by $\sim$0.5\,dex. If $\SNR>30$, it seems some galaxies have no bias while some still are under-predicted. For $\SNR<20$, the 850\,$\mu$m fluxes are significantly underestimated by $0.3-0.8\,$dex.

\vspace{0.5truecm}
\section{Markov chain Monte Carlo (MCMC) fitting}
\label{Section_Appendix_MCMC_Corner_Diagram}

\begin{figure}[H]
\centering
\includegraphics[width=0.497\linewidth]{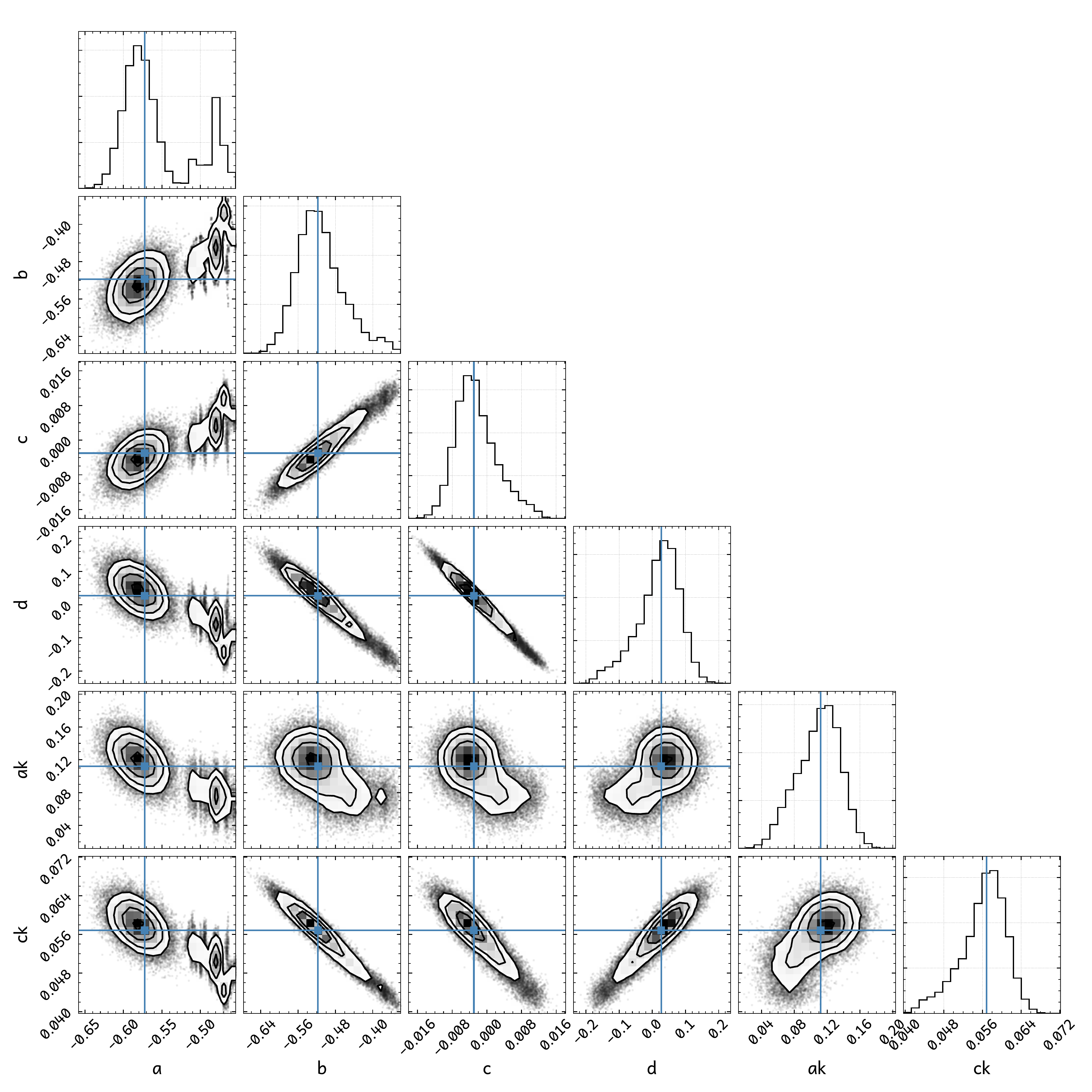}
\includegraphics[width=0.497\linewidth]{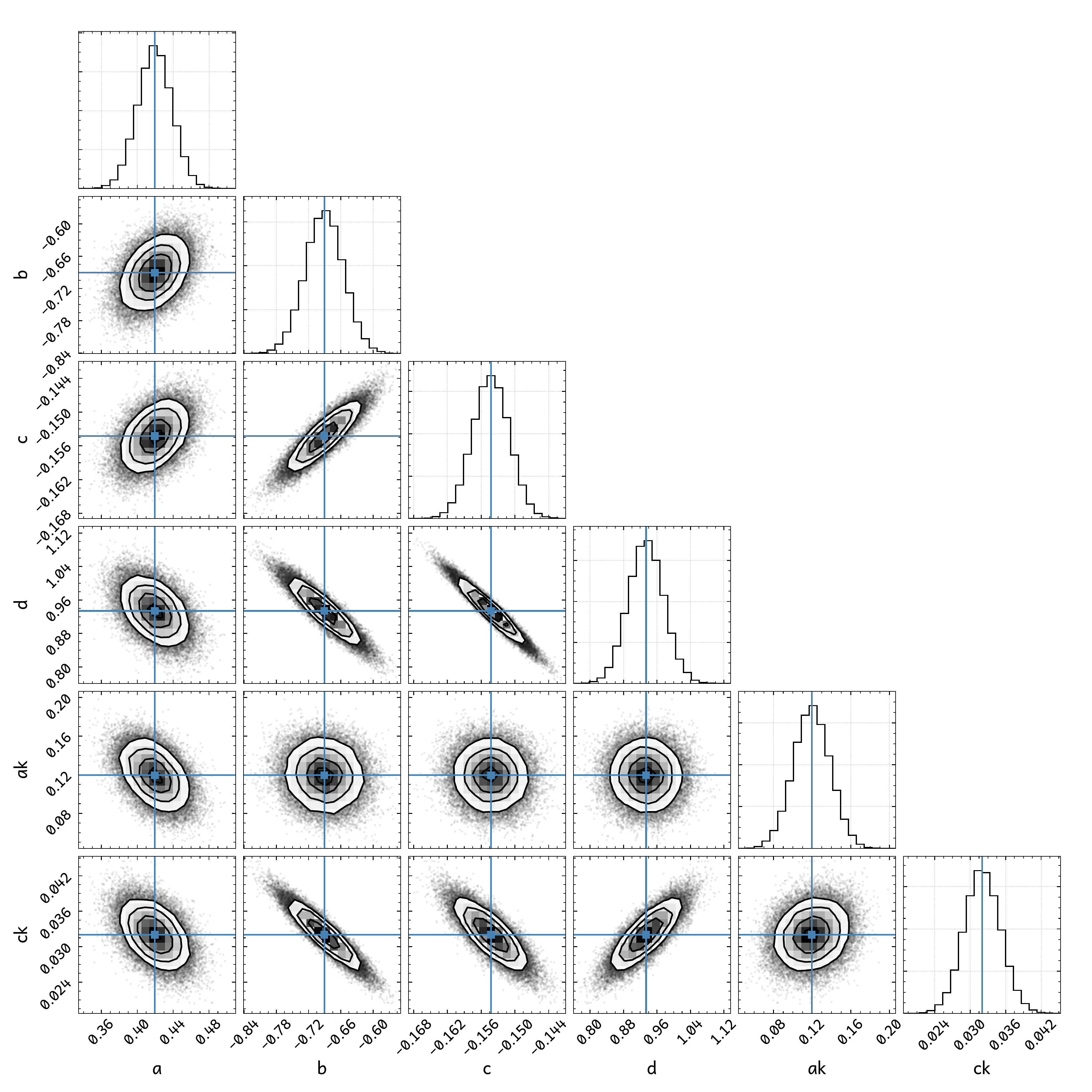}
\caption{%
Probability distributions of the coefficients in the $\tauDepl$ and $\deltaGas$ functional forms in Eq.~\ref{Equation_tauDepl_deltaGas} as fitted by our MCMC fitting to all our data. See the fitting in Sect.~\ref{Section_Fitting_Multi_Variate_Function}. 
The {\it left panel} is for the $\tauDepl$ function and the {\it right panel} is for the $\deltaGas$ function. 
}
\label{Plot_pymc3_corner_a3cosmos_function}
\end{figure}

Fig.~\ref{Plot_pymc3_corner_a3cosmos_function} shows the probability distributions of the coefficients in Eq.~\ref{Equation_tauDepl_deltaGas} obtained from our MCMC fitting in Sect.~\ref{Section_Fitting_Multi_Variate_Function}. In the fitting we allow the coefficients to vary within a relatively large range of $(-10,10)$. The probability distribution of each coefficient is shown to include the most probably areas as automatically determined by the \incode{Python} package \incode{corner}. All coefficients have a clear peak in their  probability distribution with a small width (uncertainty) of about 10\%. The $\tauDepl$ function's coefficients seem to have some second peaks which are probably due to the non-uniform, complicated sample biases (see Sect.~\ref{Section_Sample_and_Data}). The overall constraint of our fitting (to Eq.~\ref{Equation_tauDepl_deltaGas} and other fittings) is tight.

\vspace{0.5truecm}
\section{Stellar mass function (SMF)}
\label{Section_Appendix_SMF_CSFRD}

The galaxies' stellar mass function (SMF) at each redshift should in principle be consistent with the evolution of the cosmic SFR density (CSFRD). As mentioned in Sect.~\ref{Section_Adopting_the_SMFs}, at a given cosmic time, the integration of the CSFRD over previous cosmic times should be equal to the integration of the SMF at that cosmic time over all stellar masses. Note that the SMF is usually divided into two galaxy types: star-forming galaxies (SFGs) and quiescent galaxies (QGs), and the integration of the SMF should be the sum of both SFGs and QGs. 

Therefore we adopt SMFs for this work by adjusting known SMFs according to the integration of the CSFRD. For example at $z<0.085$, we adopt a SMF with the shape same as the SMF from \cite{PengYingJie2010_SMF}, for both SFG and QG types, and with the normalization of SFG+QG adjusted to the CSFRD-integrated total stellar mass at that redshift, while keeping the SFG and QG SMFs' relative normalization the same as in \cite{PengYingJie2010_SMF}. Similarly at higher redshifts, we adopt the shape of our SMF from an interpolation of the \cite{Davidzon2017} SMFs, as their SMFs are measured over multiple redshift bins ($0.2<z<5.0$). In the cases of $0.085<z<0.2$ and $z>4.0$, we adopt their $z=0.2$ and $z=4.0$ SMF shapes, respectively. The normalization is also adjusted such that SFG+QG SMFs' integrated total stellar mass equals the CSFRD-integrated total stellar mass. 

Note that during the integration of CSFRD over cosmic time, we have considered the loss of mass due to stellar evolution following \citet[][see their Eq.~11]{Conroy2009}. The choice of the mass losing timescale can be different, e.g., \cite{Ilbert2013} adopt 3\,Myr and \cite{Behroozi2015} adopt 1.4\,Myr. Compared to the \cite{Conroy2009} timescale, adopting 3\,Myr would lead to a 0.09\,dex higher integrated total stellar mass at $z=0$. 

\begin{figure*}[htbp]
\centering
\includegraphics[width=1\linewidth]{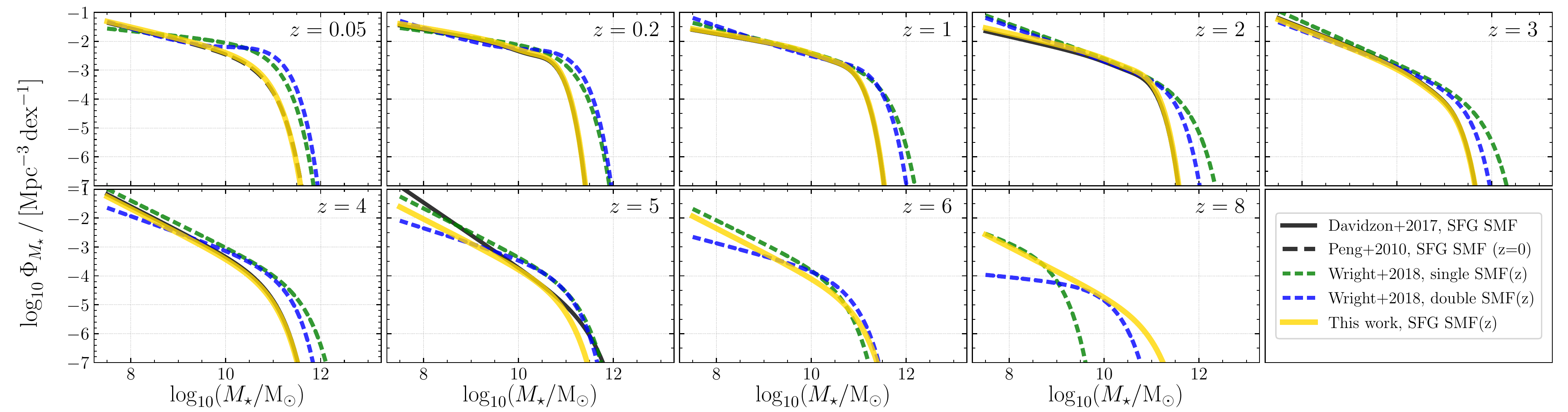}
\includegraphics[width=1\linewidth]{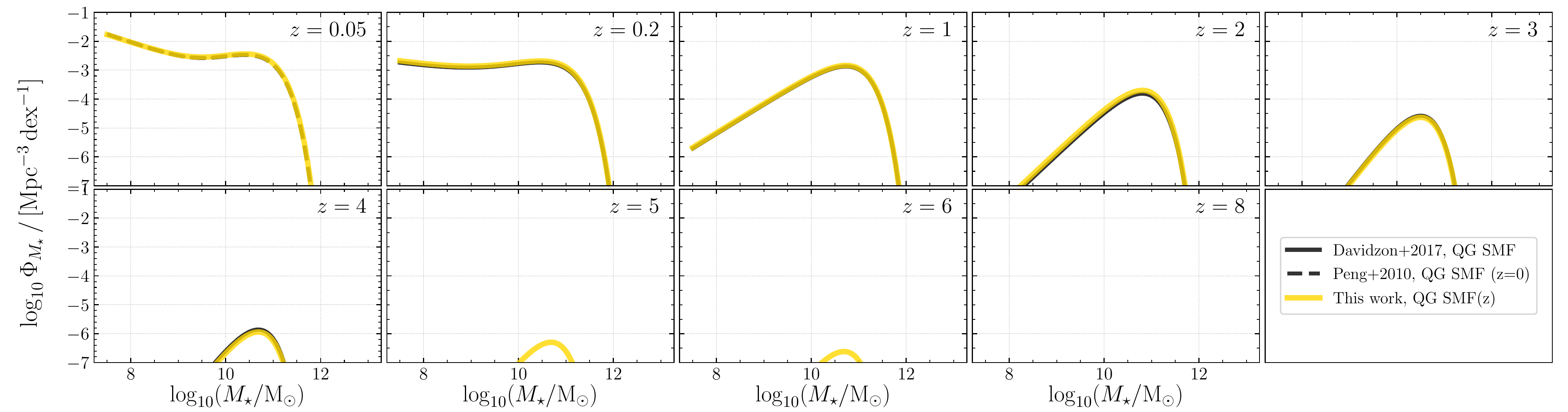}
\includegraphics[width=1\linewidth]{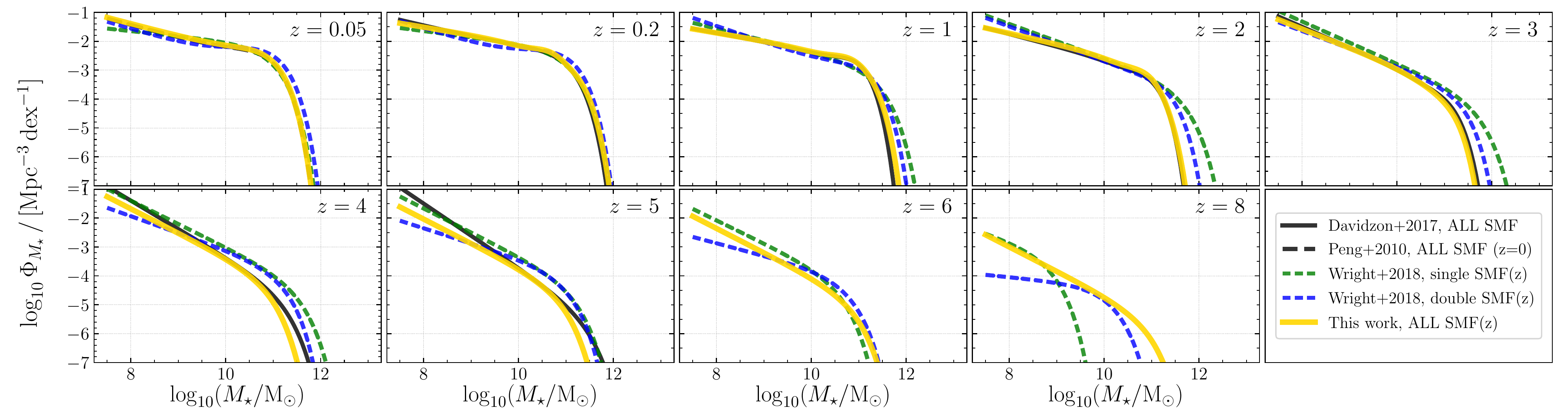}
\vspace{-2ex}
\caption{Comparison of stellar mass functions (SMFs) in 9 redshift bins. The top 9 panels are SMFs of star-forming galaxies (SFG); the middle 9 panels are SMFs of quiescent galaxies (QG); and the bottom 9 panels are the sum of SFG+QG, i.e., total galaxy SMFs. Labels represent the following references: \cite{Davidzon2017}, \cite{PengYingJie2010_SMF} and \cite{Wright2018_SMF}. The shape of the SMFs adopted in this work is from \cite{PengYingJie2010_SMF} if $z<0.08$, or the linear interpolation of \cite{Davidzon2017} SMFs at intermediate redshifts, or the shape of the \cite{Davidzon2017} SMFs at $z=4$ if $z>4$. And the normalization of the SMFs adopted in this work is set to be consistent with the integration of the cosmic SFR density (CSFRD; \citealt{Madau2014Review}). See Appx.~\ref{Section_Appendix_SMF_CSFRD} for details.}
\label{Plot_SMF}
\end{figure*}

In Fig.~\ref{Plot_SMF} we compare our adjusted SMFs with the measured SMFs from \cite{Davidzon2017} and \cite{PengYingJie2010_SMF}, which are recent measurements with the deepest data available at high redshift and in the local Universe, respectively. The very good agreement between our SMFs and theirs supports our knowledge of galaxy evolution characterized by SMFs and CSFRD. We also compare the SMFs from \cite{Wright2018_SMF}, who compiled a large number of data and performed a function fitting to characterize the evolution of SMF. We show both their single- and double-Schechter function fitting in Fig.~\ref{Plot_SMF}, however, their SMFs are too high at the massive end, while changing rapidly in shape at $z>6$. This perhaps shows the difficulty in obtaining a best fitting function, and is the reason that we adopt the CSFRD-adjusted SMFs rather than the function-characterized ones. 

\clearpage

\DeclareRobustCommand{\disambiguate}[3]{#1}
\bibliography{Biblio_all}

\end{document}